\documentclass[aip, pof, amsmath,amssymb, preprint, floatfix]{revtex4-2}

\usepackage{multirow}
\usepackage{array}

\usepackage{graphicx}
\usepackage{dcolumn}
\usepackage{bm}

\usepackage[utf8]{inputenc}
\usepackage[T1]{fontenc}
\usepackage{mathptmx}

\usepackage{nomencl}
\makenomenclature

\graphicspath{{Fig-R/}}

\usepackage{etoolbox}
\renewcommand\nomgroup[1]{%
  \item[\bfseries
  \ifstrequal{#1}{L}{Latin letters}{%
  \ifstrequal{#1}{G}{Greek letters}{}}%
]}
\newcommand{\g}{\dot{\gamma}}
\newcommand{\der}[2]{ \dfrac{\mathrm{d}#1}{\mathrm{d}#2}}
\newcommand{\lder}[2]{ \mathrm{d} #1 / \mathrm{d}#2 }
\newcommand{\hT}{\hat{\mathbb{T}}}
\newcommand{\hnen}{\hat{\mathbb{N}}_1}
\newcommand{\hnto}{\hat{\mathbb{N}}_2}
\newcommand{\hs}{\hat{\mathbb{S}}}

\begin{document}
\title{Full linear Phan-Thien--Tanner fluid model: Exact analytical solutions for steady, startup, and cessation regimes of shear and extensional flows}

\author{D. Shogin}
 \altaffiliation[Also at ]{The National IOR Centre of Norway, University of Stavanger, 4036 Ullandhaug, Norway} 
 \email{dimitri.shogin@uis.no}
\affiliation{Department of Energy Resources, University of Stavanger, 4036 Ullandhaug, Norway}

\date{\today}

\begin{abstract}
Exact, fully explicit, purely real analytical expressions for the material functions describing steady, startup, and cessation regimes of shear flows and of planar, uniaxial, and biaxial extensional flows of full linear Phan-Thien--Tanner fluids are obtained. These expressions, which have no analogs in the literature, are formulated in compact, beautiful  forms, partially due to the unique scaling procedure reducing the number of the model parameters from four to one. The properties of the material functions are investigated in detail. For steady extensional flows, the possible shapes of the extensional viscosity curves are described and the conditions for these shapes to occur are determined. For startup flows, it is found when exactly the stress dynamics is oscillatory, and, in this case, a detailed characterization of oscillations is given, which includes expressions for the position and magnitude of stress overshoots and undershoots. 
\end{abstract}

\maketitle
\nomenclature[La1]
{$a$}{Dimensionless coefficient encountered in forms $\mathfrak{T}$, $\mathfrak{H}$, and $\mathfrak{E}^+$
(defined in Table \ref{Tab:a})}
\nomenclature[La2]
{$A,B,C,K$}{Dimensionless coefficients encountered in forms $\mathfrak{T}$, $\mathfrak{H}$, and $\mathfrak{E}^+$
(defined in Table \ref{Tab:FormFunctions})}
\nomenclature[Ln1]
{$\hnen,\hnto$}{Dimensionless normal stress differences in steady shear and extensional flows 
(defined in Table \ref{Tab:DimensionlessVariables})}
\nomenclature[Ln2]
{$\hnen^\pm,\hnto^\pm$}{Dimensionless normal stress differences in startup ($+$) and cessation ($-$) of steady shear and extensional flows 
(defined in Table \ref{Tab:DimensionlessVariables})}
\nomenclature[Lv]
{$\bm{v}$}{Velocity vector 
[$\mathrm{m\;s^{-1}}$, first used in Eq. (\ref{Eq:Form:GSDerivative})]}
\nomenclature[Ls1]
{$\hs$}{Dimensionless shear stress in steady shear flow 
(defined in Table \ref{Tab:DimensionlessVariables})}
\nomenclature[Ls2]
{$\hs^\pm$}{Dimensionless shear stress in startup ($+$) and cessation ($-$) of steady shear flow 
(defined in Table \ref{Tab:DimensionlessVariables})}
\nomenclature[Lt1]
{$t$}{Time variable 
($\mathrm{s}$, first used in Table \ref{Tab:TensorFormsForDifferentFlowTypes})}
\nomenclature[Lt2]
{$\bar{t}$}{Dimensionless time variable 
[first used in Eq. (\ref{Eq:DF:DimensionlessTime})]}
\nomenclature[Lt3]
{$\hT$}{Dimensionless trace of the stress tensor in steady shear and extensional flows 
(defined in Table \ref{Tab:DimensionlessVariables})}
\nomenclature[Lt4]
{$\hT^\pm$}{Dimensionless trace of the stress tensor in startup ($+$) and cessation ($-$) of steady  shear and extensional flows 
(defined in Table \ref{Tab:DimensionlessVariables})}
\nomenclature[Ga]
{$\alpha$}
{Effective extensional flow parameter 
[dimensionless, defined by Eq. (\ref{Eq:DF:Alpha})]}
\nomenclature[Ge1]
{$\varepsilon$}{Extensibility parameter 
[dimensionless, first used in Eq. (\ref{Eq:Form:ConstitutiveEquation})]}
\nomenclature[Ge2]
{$\dot{\varepsilon}_0$}{Elongation rate/step elongation rate 
($\mathrm{s^{-1}}$, first used in Table \ref{Tab:TensorFormsForDifferentFlowTypes})}
\nomenclature[Gc1]
{$\dot{\bm{\gamma}}$}{Rate-of-strain tensor 
[$\mathrm{s^{-1}}$, first used in Eq. (\ref{Eq:Form:GSDerivative})]}
\nomenclature[Gc2]
{$\dot{\gamma}_0$}{Shear rate/step shear rate 
($\mathrm{s^{-1}}$, first used in Table \ref{Tab:TensorFormsForDifferentFlowTypes})}
\nomenclature[Gh1a]
{$\eta$}{Non-Newtonian viscosity 
($\mathrm{Pa\;s}$, defined in Table \ref{Tab:MaterialFunctions})}
\nomenclature[Gh1b]
{$\eta^\pm$}{Shear stress growth ($+$) and relaxation ($-$) functions 
($\mathrm{Pa\;s}$, defined in Table \ref{Tab:MaterialFunctions})}
\nomenclature[Gh2a]
{$\bar{\eta}$}{Uniaxial/biaxial extensional viscosity 
($\mathrm{Pa\;s}$, defined in Table \ref{Tab:MaterialFunctions})}
\nomenclature[Gh2b]
{$\bar{\eta}^\pm$}{Normal stress difference growth ($+$) and relaxation ($-$) functions in startup and cessation of steady uniaxial and biaxial extensional flows 
($\mathrm{Pa\;s}$, defined in Table \ref{Tab:MaterialFunctions})}
\nomenclature[Gh2c]
{$\eta_0$}{Zero-shear-rate viscosity 
[$\mathrm{Pa\;s}$, first used in Eq. (\ref{Eq:Form:ConstitutiveEquation})]}
\nomenclature[Gh3a]
{$\bar{\eta}_1,\bar{\eta}_2$}{Planar extensional viscosities 
($\mathrm{Pa\;s}$, defined in Table \ref{Tab:MaterialFunctions})}
\nomenclature[Gh3b]
{$\bar{\eta}_1^\pm,\bar{\eta}_2^\pm$}{Normal stress difference growth ($+$) and relaxation ($-$) functions in startup and cessation of steady planar extensional flow 
($\mathrm{Pa\;s}$, defined in Table \ref{Tab:MaterialFunctions})}
\nomenclature[Gl1]
{$\lambda$}{Time constant 
[$\mathrm{s}$, first used in Eq. (\ref{Eq:Form:ConstitutiveEquation})]}
\nomenclature[Gl2]
{$\Lambda$}{Dimensionless strain rate 
(defined in Table \ref{Tab:DimensionlessVariables})}
\nomenclature[Gn]
{$\xi$}
{Affinity parameter 
[dimensionless, first used in Eq. (\ref{Eq:Form:GSDerivative})]}
\nomenclature[Gt1]
{$\bm{\tau}$}
{Stress tensor 
[$\mathrm{Pa}$, first used in Eq. (\ref{Eq:Form:GSDerivative})]}
\nomenclature[Gt2]
{$\overset{\circ}{\bm{\tau}}$}
{Gordon-Schowalter derivative of the stress tensor 
[$\mathrm{Pa\;s^{-1}}$, defined by Eq. (\ref{Eq:Form:GSDerivative})]}
\nomenclature[Gt3]
{$\tau_{ij}$}
{Stress tensor components 
($\mathrm{Pa}$, first used in Table \ref{Tab:TensorFormsForDifferentFlowTypes})}
\nomenclature[Gw1]
{$\omega$}
{Dimensionless frequency 
(defined in Table \ref{Tab:FormFunctions})}
\nomenclature[Gw2]
{$\Omega$}
{Dimensionless damping factor 
(defined in Table \ref{Tab:FormFunctions})}
\nomenclature[Gx]
{$\chi$}
{Effective shear flow parameter 
[dimensionless, defined by Eq. (\ref{Eq:DF:Chi})]}
\nomenclature[Gy1]
{$\Psi_1,\Psi_2$}
{Normal stress coefficients 
($\mathrm{Pa\;s^2}$, defined in Table \ref{Tab:MaterialFunctions})}
\nomenclature[Gy2]
{$\Psi_1^\pm,\Psi_2^\pm$}
{Normal stress difference growth ($+$) and relaxation ($-$) functions in startup and cessation of steady shear flow 
($\mathrm{Pa\;s^2}$, defined in Table \ref{Tab:MaterialFunctions})}

\printnomenclature

\newpage 

\section{
\label{Sec:Intro}
Introduction}
The linear\cite{Phan-Thien1977} and exponential\cite{Phan-Thien1978} Phan-Thien--Tanner (PTT) models, derived in the early 1970s from the Lodge--Yamamoto network theory, are among the most used physical non-Newtonian fluid models. Nonlinear in stresses but compact and relatively simple, differential constitutive PTT equations capture a wide range of non-Newtonian phenomena in steady and transient flows. Over several decades, the PTT models have been used not only in rheological characterization of polymer solutions and melts,\cite{White1988, Quinzani1995, Li1998, Burghardt1999, Langouche1999, Solovyov1999, Dietz2015} resins,\cite{Hatzikiriakos1997, Christodoulou2000} pig liver and bread dough,\cite{Nasseri2004} and human blood,\cite{CampoDeano2013, Ramiar2017, Gudino2019} but also in analytical\cite{Oliveira1999, Pinho2000, Alves2001, Cruz2005, Ferras2016, Sarma2018, Housiadas2021, PerezSalas2021} and numerical\cite{Keunings1984, White1988, Baaijens1993, Carew1993, Azaiez1996, Baloch1996, Xue1998, Ngam2002, Majidi2009, Gudino2019, Ferras2020} studies of non-Newtonian flows.
\par 
The explicit formulas for the material functions of the linear PTT (LPTT) fluid in steady shear flows have been obtained by Alves \textit{et al.},\cite{Alves2001} who were also the first to investigate the monotonicity of these functions. At the same time, no analogs to these formulas have been derived so far for extensional flows. Xue \textit{et al.}\cite{Xue1998} have proposed an implicit expression for steady planar extensional viscosity, where an unknown function is defined as a solution of a certain cubic equation. Such formulation is not completely satisfactory because of its possible ambiguity: Three real solutions for the extensional viscosities of the PTT models, of which only one is physically reasonable, can exist.\cite{Keunings1984} This problem has been partially resolved by Shogin,\cite{Shogin2020SLPTT} who has obtained unambiguous analytical expressions for the material functions in steady uniaxial, biaxial, and planar extensional flows. Still, those expressions are not general: They apply exclusively to the so-called “simplified” version of the LPTT model and only at values of the model parameters in a certain range. Furthermore, the formulation of Shogin\cite{Shogin2020SLPTT} involves inverses of monotonic functions and, therefore, is not fully explicit.
\par 
Even fewer analytical results have been obtained for transient flows of PTT fluids, which is not surprising: In this case, systems of coupled nonlinear differential equations are typically involved. The only exact analytical solutions for startup and cessation flows are those for the simplified LPTT model, obtained recently by Shogin.\cite{Shogin2020SLPTT} Earlier analytical works on transient flows were focused on the qualitative analysis of the stress growth. In particular, Missaghi and Petrie\cite{Missaghi1982} have demonstrated that, somewhat unexpectedly, it is possible for PTT fluids to have a stress overshoot in uniaxial stress growth, and Stephenson\cite{Stephenson1986} has pointed out that the full PTT fluids may behave unphysically in startup of shear flows, providing that the step strain rate is considerably large.
\par 
Thus, despite the huge popularity of the PTT fluid models, some important aspects of their rheological properties remain unknown. The purpose of this work is to address this issue and provide a complete, detailed, and fully explicit analytical description of the LPTT fluid rheology in the most common steady and transient flows for all physically reasonable values of the model parameters.
\par 
The rest of this paper is organized as follows: In Sec. \ref{Sec:Formulation}, we define the rheometric flows to be considered in this work and review the constitutive equation of the LPTT model. In Sec. \ref{Sec:Eqs}, we introduce the dimensionless variables and formulate the systems of equations governing the stress--strain rate relations for steady flows and the evolution of stresses in time for transient flows. The analytical solutions for steady and transient flow regimes are obtained and discussed in Secs. \ref{Sec:SteSol} and \ref{Sec:TransientSolutions}, respectively, and conclusions are drawn in Sec. \ref{Sec:Conclusion}. When appropriate,  references to Appendixes \ref{App:Cubics}--\ref{App:FormShift} are made; most technical details, mathematical proofs, and lengthy derivations are placed there.
\par Finally, we should remark that every result obtained in this work has been verified through a comparison to the numerical solutions of the corresponding equations using Wolfram Mathematica. The Mathematica codes containing the verification of the main analytical results of this work are included as the supplementary material.
\section{Problem formulation
\label{Sec:Formulation}}
\subsection{The material functions
\label{Sec:Form:MaterialFunctions}}
In this work, we shall consider four important types of rheometric flows: simple shear flows and three cases of extensional flows [planar, uniaxial, and biaxial (sometimes called “equibiaxial”) extension]. These rheometric flows are conventionally defined through  assuming a fluid velocity field $(\bm{v})$ of particular form,\cite{Bird1987a} which also puts certain restrictions on the forms of the rate-of-strain tensor $\left[ \bm{\g} = (\nabla \bm{v}) + (\nabla \bm{v})^\top \right]$ and the stress tensor $(\bm{\tau})$; throughout this paper, lightface font, boldface Latin, and  boldface Greek shall be used to denote scalars, vectors, and second-rank tensors, respectively. 
\par 
The forms of $\bm{v}$, $\bm{\g}$, and $\bm{\tau}$ for shear flows and for planar, uniaxial, and biaxial extension are provided in Table \ref{Tab:TensorFormsForDifferentFlowTypes}. For convenience, Table \ref{Tab:TensorFormsForDifferentFlowTypes} also contains  expressions for the Gordon--Schowalter derivative, $\overset{\circ}{\bm{\tau}}$, of the stress tensor, calculated according to its definition,
\begin{table*}[htp]
\caption{
\label{Tab:TensorFormsForDifferentFlowTypes}
The velocity field ($\bm{v}$), the rate-of-strain tensor ($\bm{\dot{\gamma}}$), the stress tensor ($\bm{\tau}$), and its Gordon--Schowalter derivative ($\overset{\circ}{\bm{\tau}}$) in shear flows (A), planar extensional flows (B), and uniaxial and biaxial extensional flows (C, with $\dot{\epsilon}>0$ and $\dot{\epsilon}<0$, respectively), specified with respect to a Cartesian frame of reference $(x_1,x_2,x_3)$. Components of the rate-of-strain and stress tensors are functions of time.}
\begin{center}
{\begin{ruledtabular}
\begin{tabular}{c c c c c}
& $\bm{v}$ 
& $ \bm{\dot{\gamma}} $ 
& $ \bm{\tau} $ 
& $ 
\overset{\circ}{\bm{\tau}}$ \\[1pt]
\hline \rule{0pt}{3.5em}
A
& $\renewcommand*{\arraystretch}{0.7} \begin{bmatrix} 
\dot{\gamma}x_2 \\ 0 \\ 0
\end{bmatrix} $
& $\renewcommand*{\arraystretch}{0.7} \begin{bmatrix} 
0 & \dot{\gamma} & 0 \\
\dot{\gamma} & 0 & 0 \\
0 & 0 & 0
\end{bmatrix} $ 
& $ \renewcommand*{\arraystretch}{0.7} \begin{bmatrix} 
\tau_{11} & \tau_{12} & 0 \\
\tau_{12} & \tau_{22} & 0 \\
0 & 0 & \tau_{33}
\end{bmatrix} $ 
& $\dfrac{\mathrm{d} \boldsymbol{\tau}}{\mathrm{d}t}-\dot{\gamma}   \begin{bmatrix} 
(2-\xi)\tau_{12} & \dfrac{2-\xi}{2}\tau_{22}-\dfrac{\xi}{2}\tau_{11} & 0 \\
\dfrac{2-\xi}{2}\tau_{22}-\dfrac{\xi}{2}\tau_{11} & -\xi \tau_{12} & 0 \\
0 & 0 & 0
\end{bmatrix}$ \\ \rule{0pt}{3em}
B
& $ \renewcommand*{\arraystretch}{0.7} \begin{bmatrix} 
 -\dot{\epsilon}x_1 \\
0 \\
\dot{\epsilon}x_3
\end{bmatrix}$
& \renewcommand*{\arraystretch}{0.7} $\begin{bmatrix} 
- 2 \dot{\epsilon} & 0 & 0 \\
0 & 0 & 0 \\
0 & 0 & 2 \dot{\epsilon}
\end{bmatrix}$ 
& $\renewcommand*{\arraystretch}{0.7} \begin{bmatrix} 
\tau_{11} & 0 & 0 \\
0 & \tau_{22} & 0 \\
0 & 0 & \tau_{33}
\end{bmatrix} $
& $\dfrac{\mathrm{d} \boldsymbol{\tau}}{\mathrm{d}t}+ (1-\xi)\dot{\epsilon} \renewcommand*{\arraystretch}{0.7}  \begin{bmatrix} 
 2\tau_{11} & 0 & 0 \\
 0 & 0 & 0 \\
 0 & 0 & -2\tau_{33}
 \end{bmatrix}$  \\ \rule{0pt}{3em}
C
& $ \renewcommand*{\arraystretch}{0.7} \begin{bmatrix} 
 -\dot{\epsilon}x_1/2 \\
-\dot{\epsilon}x_2/2 \\
\dot{\epsilon}x_3
\end{bmatrix}$
& $ \renewcommand*{\arraystretch}{0.7} \begin{bmatrix} 
-\dot{\epsilon} & 0 & 0 \\
0 & -\dot{\epsilon} & 0 \\
0 & 0 & 2 \dot{\epsilon}
\end{bmatrix}$
& $ \renewcommand*{\arraystretch}{0.7} \begin{bmatrix} 
 \tau_{11} & 0 & 0 \\
 0 & \tau_{22} & 0 \\
 0 & 0 & \tau_{33}
 \end{bmatrix} $
& $\dfrac{\mathrm{d} \boldsymbol{\tau}}{\mathrm{d}t}+ (1-\xi) \dot{\epsilon} \renewcommand*{\arraystretch}{0.7}  \begin{bmatrix} 
\tau_{11} & 0 & 0 \\
0 & \tau_{22} & 0 \\
0 & 0 & -2\tau_{33}
\end{bmatrix}$ 
\end{tabular}
\end{ruledtabular}
}
\end{center}
\end{table*}
\begin{equation}
\label{Eq:Form:GSDerivative}
\overset{\circ}{\bm{\tau}} = \dfrac{\partial}{\partial t}\bm{\tau}+\bm{v}\cdot (\nabla \bm{\tau}) - \left\{(\nabla \bm{v})^\top \cdot \bm{\tau}+  \bm{\tau} \cdot (\nabla \bm{v})\right\}+\dfrac{\xi}{2} \left\{\bm{\g}\cdot \bm{\tau}+\bm{\tau}\cdot \bm{\g} \right\}.
\end{equation}
This special time derivative, with affinity parameter $\xi$ in the range $0\leq \xi \leq 1$, is encountered in the PTT constitutive equation (see the following Sec. \ref{Sec:Form:CE}) and contains the Oldroyd (upper-convected time) derivative and the corotational Jaumann derivative as special cases (at $\xi=0$ and $\xi=1$, respectively).
\par 
For the flow regimes considered in this work, the strain rates in Table \ref{Tab:TensorFormsForDifferentFlowTypes} ($\g$ and $\dot{\epsilon}$) can be written as
\begin{equation}
\label{Eq:Form:StrainRates}
\begin{bmatrix} 
\g (t) \\ \dot{\epsilon}(t)
\end{bmatrix} = \begin{bmatrix}
\g_0 \\ \dot{\epsilon}_0 \end{bmatrix} \Theta(t),
\end{equation}
where $\g_0$ and $\dot{\epsilon}_0$ are constants, and
\begin{equation}
\label{Eq:Form:Heaviside}
\Theta(t) = \left\{ \begin{array}{ll}
H(t) & \quad \text{for startup,} \\
1-H(t) & \quad \text{for cessation,} \\
1 & \quad \text{for steady flows.}
\end{array}\right.
\end{equation}
Here $H$ stands for the Heaviside step-function, with $H(t)=1$ at $t>0$ and $H(t)=0$ otherwise. Due to the flow symmetry, it is sufficient to consider $\g_0>0$ for shear flows and $\dot{\epsilon}_0>0$ for planar extension, while uniaxial and biaxial extensional flows are kinematically similar and can be described with the same equations but assuming elongation rates of opposite signs ($\dot{\epsilon}_0>0$ for uniaxial extension and $\dot{\epsilon}_0<0$ for biaxial extension).\cite{Bird1987a, Shogin2020SLPTT}
\par 
The material functions for startup, cessation, and steady regimes of shear and extensional flows are defined in Table \ref{Tab:MaterialFunctions}. Note that two independent normal stress differences are needed to describe planar extension; following Shogin,\cite{Shogin2020SLPTT} we choose $\tau_{33}-\tau_{11}$ and $\tau_{33}-\tau_{22}$. In contrast, one normal stress difference is sufficient for uniaxial and biaxial extension, since $\tau_{11}=\tau_{22}$ due to the axial symmetry of such flows.
\begin{table*}[htp]
\caption{
\label{Tab:MaterialFunctions}
Material functions describing startup ($+$) and cessation ($-$) of steady shear and extensional flows (top row) and their steady-flow analogs (bottom row). The stress tensor components are functions of time. The sign convention of Bird \textit{et al.}\cite{Bird1987a} is adopted for the stress tensor. }
\begin{center}
\begin{ruledtabular}
\begin{tabular}{c c c}
Shear flow & Planar extension & Uniaxial and biaxial extension \\ 
\hline \rule{0pt}{4.em}
$\begin{bmatrix}
\eta^\pm(t,\g_0)\\
\Psi_1^\pm (t,\g_0) \\
\Psi_2^\pm (t,\g_0)
\end{bmatrix}
=-\dfrac{1}{\g_0^2}
\begin{bmatrix}
\g_0 \tau_{12} \\
\tau_{11}-\tau_{22} \\
\tau_{22}-\tau_{33}
\end{bmatrix}$
& $ \begin{bmatrix}
\bar{\eta}_1^\pm(t,\dot{\epsilon}_0) \\
\bar{\eta}_2^\pm(t,\dot{\epsilon}_0)
\end{bmatrix}
=
-\dfrac{1}{\dot{\epsilon}_0}
\begin{bmatrix}
\tau_{33}-\tau_{11} \\
\tau_{33}-\tau_{22}
\end{bmatrix}$
& $\bar{\eta}(t,\dot{\epsilon}_0)
=
-\dfrac{\tau_{33}-\tau_{11}}{\dot{\epsilon}_0}$ \\
\rule{0pt}{4.em}
$\begin{bmatrix}
\eta(\g_0)\\
\Psi_1(\g_0) \\
\Psi_2(\g_0)
\end{bmatrix} = \displaystyle{\lim_{t\to \infty}}
\begin{bmatrix}
\eta^\pm(t,\g_0)\\
\Psi_1^\pm (t,\g_0) \\
\Psi_2^\pm (t,\g_0)
\end{bmatrix}$
&
$\begin{bmatrix}
\bar{\eta}_1(\dot{\epsilon}_0) \\
\bar{\eta}_2(\dot{\epsilon}_0)
\end{bmatrix} = \displaystyle{\lim_{t\to \infty}}
\begin{bmatrix}
\bar{\eta}_1^\pm(t,\dot{\epsilon}_0) \\
\bar{\eta}_2^\pm(t,\dot{\epsilon}_0)
\end{bmatrix}$
&
$\bar{\eta}(\dot{\epsilon}_0) = 
\displaystyle{\lim_{t\to \infty}}
\bar{\eta}^\pm(t,\dot{\epsilon}_0)$

\end{tabular}
\end{ruledtabular}
\end{center}
\end{table*}
\subsection{The constitutive equation
\label{Sec:Form:CE}}
The stress and rate-of-strain tensors of a single-mode LPTT fluid are related through a nonlinear differential constitutive equation,
\begin{equation}
\label{Eq:Form:ConstitutiveEquation}
\left(1-\dfrac{\epsilon \lambda}{\eta_0} \mathrm{tr} \bm{\tau}\right) \bm{\tau} + \lambda \overset{\circ}{\bm{\tau}} = -\eta_0 \bm{\g}.
\end{equation}
In addition to $\xi$ in the Gordon--Schowalter derivative [see Eq. (\ref{Eq:Form:GSDerivative})], the three other model parameters encountered in Eq. (\ref{Eq:Form:ConstitutiveEquation}) are the zero-shear-rate viscosity of the fluid $(\eta_0)$, the time constant $(\lambda)$, and the extensional parameter $(\epsilon)$, all of them being positive constants.
\par 
The dimensionless extensional parameter, $\epsilon$, also called the PTT factor,\cite{Nasseri2004} is related to the nonlinearity of the model and is known to strongly affect the fluid rheology in extensional flows while having a much smaller impact on the shear-flow properties. In the original work of Phan-Thien and Tanner,\cite{Phan-Thien1977} it was assumed that $\epsilon$ must be small (the authors suggested $\epsilon \sim 10^{-2}$). Later, however, it was shown that larger values ($0<\epsilon<2$) are also relevant for rheological characterization.\cite{Dietz2015, Hatzikiriakos1997, Christodoulou2000, Nasseri2004, CampoDeano2013}
\par  
For the affinity parameter, we shall restrict our consideration to $0\leq \xi <1$, with $\xi \neq 0$ corresponding to the “full” (non-affine) PTT model and $\xi=0$ yielding the “simplified” (affine) version of the model. The case when $\xi=1$ is not of actual rheological interest: It is easy to show that a LPTT fluid with $\xi=1$ behaves as a corotational Maxwell fluid in shear flows and as a standard Maxwell fluid in extensional flows.
\par 
Finally, we should note that the material functions of multimode PTT fluids are simply the sums of the corresponding material functions of separate modes.\cite{Shogin2020SLPTT} Therefore, a single-mode version of the LPTT fluid model shall be considered in this work without loss of generality.
\section{Equations for the stresses
\label{Sec:Eqs}}
The equations governing the rheological behavior of the fluid in startup and cessation regimes of shear and extensional flows are obtained by substituting the expressions for $\bm{\g}$, $\bm{\tau}$, and $\overset{\circ}{\bm{\tau}}$ from the appropriate row of Table \ref{Tab:TensorFormsForDifferentFlowTypes}, together with the corresponding form of the strain rate [Eqs. (\ref{Eq:Form:StrainRates}) and (\ref{Eq:Form:Heaviside})], into the constitutive equation (\ref{Eq:Form:ConstitutiveEquation}). The independent components of the resulting tensor equation form a dynamical system of differential equations. Some equations of the system are then replaced with their linear combinations, so that the evolution equations for $\mathrm{tr}\,\bm{\tau}$ and for the normal stress differences of interest are obtained. Finally, the system is put into dimensionless form, which is done as described in Sec. \ref{Sec:Eqs:DimensionlessFormulation}.
\subsection{Dimensionless variables
\label{Sec:Eqs:DimensionlessFormulation}}
First, the time variable, $t$, is replaced with $\bar{t}=t/\lambda$, so that 
\begin{equation}
\der{}{t}=\dfrac{1}{\lambda}\der{}{\bar{t}},
\label{Eq:DF:DimensionlessTime}
\end{equation}
regardless of the flow type. Second, the strain rate $(\dot{\gamma}_0$ or $\dot{\epsilon}_0)$ is replaced with its dimensionless analogue, $\Lambda$ (definitions of $\Lambda$ for different flow types are given in the first row of Table \ref{Tab:DimensionlessVariables}). Third, the dimensionless stresses in startup ($+$) and cessation ($-$) flows ($\hT^\pm$, $\hnen^\pm$, $\hnto^\pm$, and $\hs^\pm$) and their steady-flow values ($\hT$, $\hnen$, $\hnto$, and $\hs$, respectively) are introduced according to the formulas provided in the second row of Table \ref{Tab:DimensionlessVariables}.
\begin{table*}[htp]
\caption{
\label{Tab:DimensionlessVariables}
Dimensionless strain rates (first row) and stresses (second row), introduced for  startup ($+$) and cessation ($-$) of steady shear and extensional flows. Corresponding definitions for steady flow regimes are obtained by omitting the superscripts ($\pm$).}
\begin{center}
\begin{ruledtabular}
\begin{tabular}{c c c}
Shear flow & Planar extension & Uniaxial and biaxial extension \\ 
\hline \rule{0pt}{1.2em}
$\Lambda = [(1-\xi) \epsilon]^{1/2}\lambda \dot{\gamma}_0$
& $\Lambda = (1-\xi) \lambda \dot{\epsilon}_0$ 
& $\Lambda = (1-\xi) \lambda \vert \dot{\epsilon}_0 \vert$
\\ \rule{0pt}{1.8em}
$\begin{bmatrix}
\hT^\pm \\
\hs^\pm \\
\hnen^\pm \\
\hnto^\pm
\end{bmatrix}
=-\dfrac{\lambda}{\eta_0}
\begin{bmatrix}
\epsilon \, \mathrm{tr} \boldsymbol{\tau} \\
[(1-\xi) \epsilon]^{1/2}\tau_{12} \\
(1-\xi) \epsilon (\tau_{11}-\tau_{22}) \\
(1-\xi) \epsilon (\tau_{22}-\tau_{33})
\end{bmatrix}$
& $ \begin{bmatrix}
\hT^\pm \\
\hnen^\pm \\
\hnto^\pm
\end{bmatrix}
=
-\dfrac{\epsilon \lambda}{\eta_0}
\begin{bmatrix}
\mathrm{tr} \boldsymbol{\tau} \\
\tau_{33}-\tau_{11} \\
\tau_{33}-\tau_{22}
\end{bmatrix}$
& $ \begin{bmatrix}
\hT^\pm \\
\hnen^\pm 
\end{bmatrix}
=
-\dfrac{\epsilon \lambda}{\eta_0}
\begin{bmatrix}
\mathrm{tr} \boldsymbol{\tau} \\
(\tau_{33}-\tau_{11})\mathrm{sgn}\,\dot{\epsilon}_0
\end{bmatrix}$
\end{tabular}
\end{ruledtabular}
\end{center}
\end{table*}
\par 
This procedure completely eliminates the model parameters $\lambda$ and $\eta_0$ from the equations and puts the equations in a much simpler form. Further simplification is made by effectively combining the two remaining model parameters, $\xi$ and $\epsilon$, into a single quantity in a way that depends on the flow type. We introduce the effective shear flow parameter, $\chi$, by
\begin{equation}
\chi = \dfrac{\xi(2-\xi)}{2(1-\xi)\epsilon},
\label{Eq:DF:Chi}
\end{equation}  
and the effective extensional flow parameter, $\alpha$, by 
\begin{equation}
\alpha=\dfrac{\epsilon}{1-\xi}.
\label{Eq:DF:Alpha}
\end{equation}
Thus, our dimensionless equations contain only one model parameter instead of the original four in their dimensionful form. Note that $\chi$ is non-negative (with $\chi=0$ only if $\xi=0$, i.e., for affine PTT fluids), while $\alpha$ is strictly positive. The values of $\xi$ and $\epsilon$ used in  rheological characterization\cite{White1988, Quinzani1995, Li1998, Burghardt1999, Langouche1999, Solovyov1999, Dietz2015, Hatzikiriakos1997, Christodoulou2000,Nasseri2004,CampoDeano2013, Ramiar2017, Gudino2019} most commonly result in $\chi$ and $\alpha$ in the ranges $0\leq \chi <10$ and $0<\alpha<2$, although much larger values ($\chi \sim 10^5$, $\alpha \sim 10$) are also relevant.\cite{Langouche1999, Hatzikiriakos1997, Dietz2015}
\par 
The definitions of dimensionless stresses and strain rates, as given in Table \ref{Tab:DimensionlessVariables}, ensure that these dimensionless quantities are always positive (the only exception being $\hnto^\pm$ in shear flows, which is always negative to match the sign of $\Psi_2^\pm$). This feature of positivity is crucial and shall be extensively used in the  analysis. An overview of the relations between the dimensionless stresses and the corresponding material functions is provided in Table \ref{Tab:MatFunctionsVsDimlessVars}. 
\begin{table*}[htp]
\caption{
\label{Tab:MatFunctionsVsDimlessVars}
Material functions describing startup ($+$) and cessation ($-$) regimes of steady shear and extensional flows, expressed in terms of the dimensionless stresses and strain rates. The corresponding relations for steady flow regimes are obtained by ignoring the time-dependence and omitting the superscripts ($\pm$).}
\begin{center}
\begin{ruledtabular}
\begin{tabular}{c c c}
Shear flow & Planar extension & Uniaxial and biaxial extension \\ 
\hline \rule{0pt}{5em} 
$\begin{bmatrix}
\eta^\pm(\bar{t},\Lambda) \\
\Psi_1^\pm(\bar{t},\Lambda) \\
\Psi_2^\pm(\bar{t},\Lambda)
\end{bmatrix}=\eta_0
\begin{bmatrix}
\hs^\pm / \Lambda \\
\lambda \hnen^\pm/\Lambda^2 \\
\lambda \hnto^\pm/\Lambda^2
\end{bmatrix}$
& $\begin{bmatrix}
\bar{\eta}_{1}^\pm(\bar{t},\Lambda) \\
\bar{\eta}_{2}^\pm(\bar{t},\Lambda)
\end{bmatrix}=\dfrac{\eta_0}{\alpha \Lambda}
\begin{bmatrix}
\hnen^\pm \\
\hnto^\pm
\end{bmatrix}$ 
& $ \bar{\eta}^\pm(\bar{t},\Lambda)=\dfrac{\eta_0}{\alpha \Lambda}
\hnen^\pm$  
\end{tabular}
\end{ruledtabular}
\end{center}
\end{table*}

\par 
If not directly stated otherwise, the transient dimensionless stresses will be considered as functions of $\bar{t}$ with two parameters ($\Lambda$ and $\chi$ for shear flows, $\Lambda$ and $\alpha$ for extensional flows), while the steady-flow dimensionless stresses will be treated as functions of $\Lambda$ with one parameter ($\chi$ or $\alpha$, depending on the flow type).

\subsection{Startup of steady shear flow
\label{Sec:Eqs:Shear}}
For startup of steady shear flow, the result of nondimensionalization can be written as
\begin{align}
\der{\hT^+}{\bar{t}} &= -(1 + \hT^+) \hT^+ +2\Lambda \hs^+, \label{Eq:Eqs:Shear:1TEvolution}\\
\der{\hs^+}{\bar{t}} &= -(1 + \hT^+) \hs^+ -\chi \Lambda \hT^+ + \Lambda, \label{Eq:Eqs:Shear:2SEvolution}
\end{align}
with initial conditions $\hT^+(0)=\hs^+(0)=0$. The dimensionless normal stress differences, $\hnen^+$ and $\hnto^+$, are proportional to $\hT^+$, with
\begin{align}
\hnen^+ &= \hT^+, \label{Eq:Eqs:Shear:N1} \\
\hnto^+ &= -\dfrac{\xi}{2} \hnen^+ \label{Eq:Eqs:Shear:N2}
\end{align}
at any $\bar{t}$, and therefore do not appear in the system as independent variables. From Eq. (\ref{Eq:Eqs:Shear:N2}), it immediately follows that 
\begin{equation}
\label{Eq:Eqs:Shear:Psi2Plus}
\Psi_2^+ = -\dfrac{\xi}{2}\Psi_1^+.
\end{equation}
\subsection{Startup of steady planar extensional flow
\label{Sec:Eqs:PlaEx}}
For startup of steady planar extensional flow, the same procedure leads to
\begin{align}
\der{\hT^+}{\bar{t}} &= - (1 + \hT^+) \hT^+ + 2 \Lambda \hnen^+, \label{Eq:Eqs:PlaEx:1TEvolution}\\
\der{\hnen^+}{\bar{t}} &= - (1 + \hT^+) \hnen^+ +2 \Lambda \hT^+ + 4\alpha \Lambda, \label{Eq:Eqs:PlaEx:2N1Evolution}
\end{align}
with initial conditions $\hT^+(0)=\hnen^+(0)=0$. The dimensionless second normal stress difference is simply a halfsum of $\hT^+$ and $\hnen^+$,
\begin{equation}
\hnto^+ = \dfrac{1}{2}(\hT^++\hnen^+) \label{Eq:Eqs:PlaEx:N2}
\end{equation}
at any $\bar{t}$.
\subsection{Startup of steady uniaxial and biaxial extensional flows
\label{Sec:Eqs:UBEx}}
For startup of steady uniaxial and biaxial extensional flows [with upper and lower sign chosen in Eq. (\ref{Eq:Eqs:UBEx:2N1Evolution}), respectively], one arrives at
\begin{align}
\der{\hT^+}{\bar{t}} &= - (1 + \hT^+) \hT^+ + 2 \Lambda \hnen^+, \label{Eq:Eqs:UBEx:1TEvolution}\\
\der{\hnen^+}{\bar{t}} &= - (1 + \hT^+) \hnen^+ +\Lambda (\hT^+ \pm \hnen^+)+3 \alpha \Lambda, \label{Eq:Eqs:UBEx:2N1Evolution}
\end{align}
with $\hT^+(0) = \hnen^+(0)=0$.
\subsection{Cessation of steady shear and extensional flows
\label{Sec:Eqs:Cessation}}
For cessation of steady shear and extensional flows,  nondimensionalization leads to a single independent differential equation,
\begin{align}
\label{Eq:Eqs:Cessation:1T}
\der{\hT^-}{\bar{t}} = -(1+\hT^-)\hT^-,
\end{align}
with $\hT^-(0)=\hT$. The other dimensionless stresses, when  relevant (note that for affine LPTT fluids in cessation of steady shear flow, $\hnto^-=0$ identically), are proportional to $\hT^-$,
\begin{equation}
\label{Eq:Eqs:Cessation:2Rest}
\left(\dfrac{\hs^-}{\hs}= \right)\dfrac{\hT^-}{\hT}=\dfrac{\hnen^-}{\hnen} \left(=\dfrac{\hnto^-}{\hnto} \right)
\end{equation}
at any $\bar{t}$.
\par 
Equations (\ref{Eq:Eqs:Cessation:1T}) and (\ref{Eq:Eqs:Cessation:2Rest}) contain neither $\chi$ nor $\alpha$ and thus are identical to those for affine LPTT fluids. The corresponding solving procedure can be found in our recent work\cite{Shogin2020SLPTT} and shall not be reproduced here. The final result, however, possesses some distinct features and shall be discussed in Sec. \ref{Sec:Transient:Cessation}.
\section{
\label{Sec:SteSol}
Solutions for steady flow regimes}
\par 
The equations governing the rheology of LPTT fluids in steady shear and extensional flows are algebraic and can be obtained from the corresponding equations for startup regimes by omitting the superscript ($+$) in the variables and setting all time derivatives to zero.
\par 
As pointed out in our recent work,\cite{Shogin2020SLPTT} in steady shear and extensional flows, the dimensionless equations for the affine linear PTT model are identical to those for the FENE-P dumbbell model\cite{Bird1980} of dilute polymer solutions, derived from the kinetic theory. Therefore, the results of Sec. \ref{Sec:SteSol} (with $\chi=0$ and $\alpha \equiv \epsilon$) are also valid for this fluid model. The dimensionful steady-flow material functions of the FENE-P dumbbell solution can be obtained using the formulas in Table \ref{Tab:MatFunctionsVsDimlessVars}, provided one performs the conversion according to the following rules:\cite{Shogin2020SLPTT}
\begin{align}
\eta_0 & \leftrightarrow \dfrac{b}{b+3}nk_\mathrm{B}T\lambda_H, \\
\lambda & \leftrightarrow \dfrac{b}{b+3} \lambda_H, \\
\epsilon & \leftrightarrow \dfrac{1}{b+3}, \label{Eq:SteSol:FENE:EpsilonB}
\end{align}
where $b$, $\lambda_H$, and $(nk_\mathrm{B}T)$ are the model parameters of the FENE-P dumbbells.\cite{Bird1980} As an important note, $b>0$, which implies $\epsilon=\alpha<1/3$ [see Eq. (\ref{Eq:SteSol:FENE:EpsilonB})]. Therefore, some rheological behaviors in steady extensional flows (see Secs. \ref{Sec:SteSol:PlaEx} and \ref{Sec:SteSol:UBEx}), common for LPTT fluids, are impossible within the FENE-P dumbbell framework.
\subsection{Shear flow \label{Sec:SteSol:Shear}}
The steady-flow versions of Eqs. (\ref{Eq:Eqs:Shear:1TEvolution}) and (\ref{Eq:Eqs:Shear:2SEvolution}) read
\begin{align}
\left(1+\hT \right) \hT &= 2\Lambda \hs, \label{Eq:SteSol:Shear:1T}\\
\left(1+\hT\right) \hs &= -\chi \Lambda \hT + \Lambda, \label{Eq:SteSol:Shear:2S}
\end{align}
the dimensionless normal stress differences being related to $\hT$ through the analogs of Eqs. (\ref{Eq:Eqs:Shear:N1}) and (\ref{Eq:Eqs:Shear:N2}),
\begin{align}
\hnen &= \hT, \label{Eq:SteSol:Shear:N1} \\
\hnto &=  -\dfrac{\xi}{2}\hnen. \label{Eq:SteSol:Shear:N2}
\end{align}
From Eq. (\ref{Eq:SteSol:Shear:1T}), one gets
\begin{equation}
\label{Eq:SteSol:Shear:MainResultS}
\hat{\mathbb{S}} = \dfrac{\left(1+\hat{\mathbb{T}}\right)\hat{\mathbb{T}}}{2\Lambda}.
\end{equation} 
Substituting this into Eq. (\ref{Eq:SteSol:Shear:2S}) and rearranging, one arrives at
\begin{equation}
\hT^3+2\hT^2+ (1+2\chi \Lambda^2)\hT-2\Lambda^2 =0.
\label{Eq:SteSol:Shear:Cubic}
\end{equation}
This equation is cubic in $\hT$ (for general solutions and some important features of cubic equations, see Appendix \ref{App:Cubics}). It has one real solution, which is positive (see Appendix \ref{App:CubicSolutions:Shear}) and can be written as
\begin{equation}
\label{Eq:SteSol:Shear:MainResultT}
\hat{\mathbb{T}}=\dfrac{1}{3}\left(-2+\sqrt[3]{Q+\sqrt{Q^2+P^3}}+\sqrt[3]{Q-\sqrt{Q^2+P^3}}\right),
\end{equation}
where
\begin{align}
P &= -1+6\chi \Lambda^2, \label{Eq:SteSol:Shear:P}\\
Q &= 1+9(3+2\chi)\Lambda^2.\label{Eq:SteSol:Shear:Q}
\end{align}
\par
\begin{figure}[htp]
\begin{center}
\includegraphics[width=3.37in]{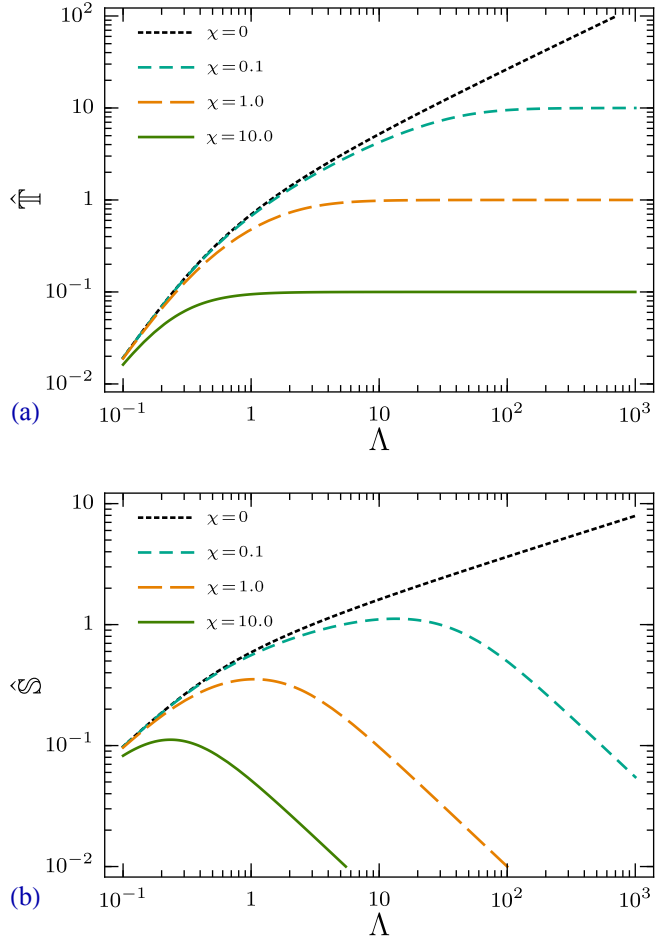}
\caption{\label{Fig:SteSol:Shear:Stresses} The dimensionless stresses in steady shear flow ($\hT$ and $\hs$) as functions of the dimensionless shear rate, $\Lambda$, at different values of the effective shear flow parameter, $\chi$. The rheology of affine ($\chi=0$) and non-affine ($\chi \neq 0$) LPTT fluids at large shear rates is qualitatively distinct, which is easily observed in the plots.}
\end{center}
\end{figure}

 
The absolute values of the normal stress differences are  proportional to $\hT$ [see Eqs. (\ref{Eq:SteSol:Shear:N1}) and (\ref{Eq:SteSol:Shear:N2})]. At $\chi \neq 0$, $\hT(\Lambda)$ is monotonically increasing and bounded (see Appendix \ref{App:Bijectivity}), while the dimensionless shear stress, $\hs(\Lambda)$, does not increase monotonically (see Appendix \ref{App:Shapes:Shear}) but goes through a maximum [see Figs. \ref{Fig:SteSol:Shear:Stresses}(a) and \ref{Fig:SteSol:Shear:Stresses}(b), respectively]. The maximal value of $\hs$,
\begin{equation}
\label{Eq:SteSol:Shear:MaximumHatS}
\hs_\mathrm{max} = \dfrac{1}{(8\chi)^{1/2}},
\end{equation}
is reached at a critical value of $\Lambda$, which we shall call the first critical shear rate,
\begin{equation}
\label{Eq:SteSol:Shear:DefinitionLambdaFirst}
\Lambda_I = \dfrac{1+2\chi}{(2\chi)^{3/2}}.
\end{equation}
In contrast, at $\chi = 0$, both $\hT(\Lambda)$ and $\hs(\Lambda)$ increase monotonically with no upper bounds [see the solid lines in Figs. \ref{Fig:SteSol:Shear:Stresses} (a) and \ref{Fig:SteSol:Shear:Stresses} (b)].
\par 
The steady shear flow material functions $\eta(\Lambda)$ and $\Psi_1(\Lambda)$ are bounded and monotonically decreasing (see Appendix \ref{App:Shapes:Shear}), as shown in Fig. \ref{Fig:SteSol:Shear:MaterialFunctions}. The second normal stress coefficient, $\Psi_2$, is negative when $\chi \neq 0$ and only differs from $\Psi_1$ by a constant factor of $-\xi/2$, as seen from Eq. (\ref{Eq:SteSol:Shear:N2}). At $\chi=0$, $\Psi_2 = 0$ identically. 
\begin{figure}[htp]
\begin{center}
\includegraphics[width=3.37in]{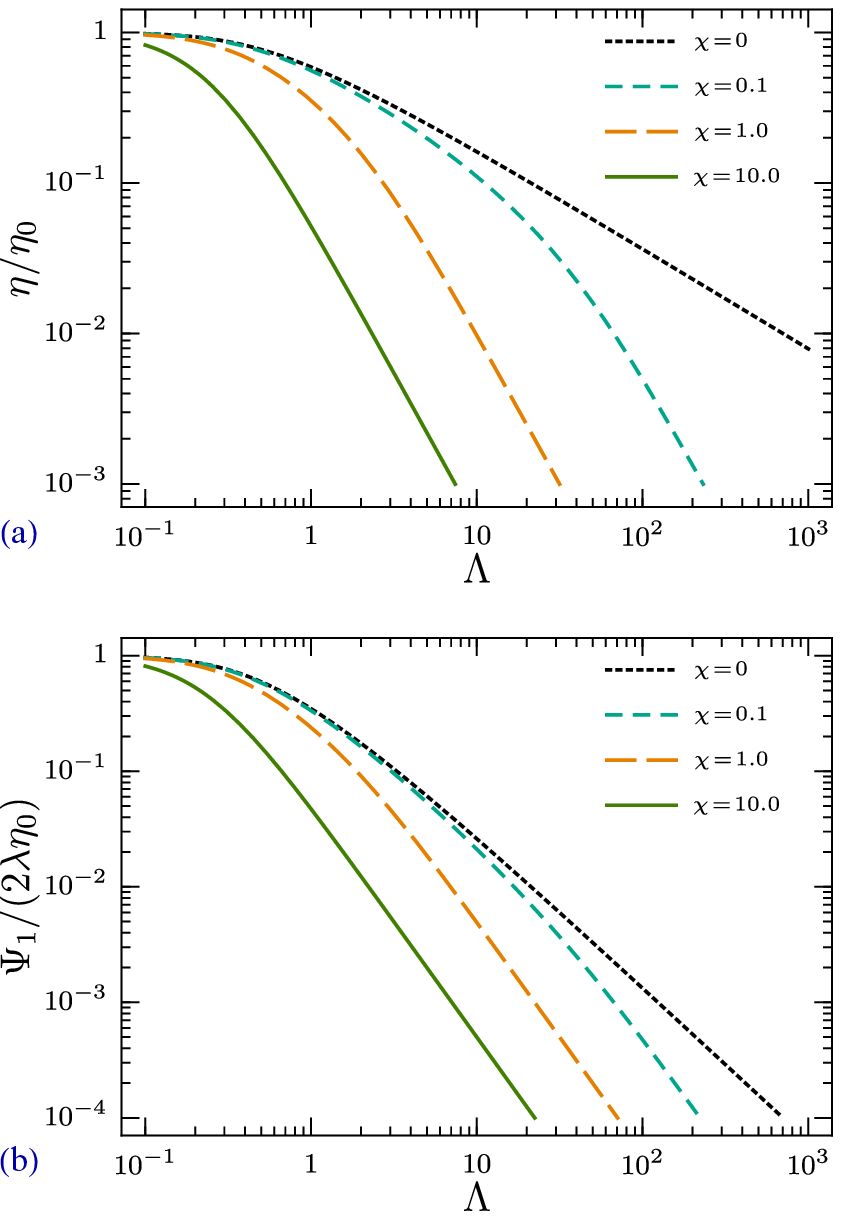}
\caption{\label{Fig:SteSol:Shear:MaterialFunctions} The steady-shear-flow material functions, normalized to their zero-shear-rate values: the non-Newtonian viscosity (a) and the first normal stress coefficient (b), as functions of the dimensionless shear rate, $\Lambda$. The plots corresponding to different values of the effective shear flow parameter, $\chi$, are shown.}
\end{center}
\end{figure}
\par 
Equations (\ref{Eq:SteSol:Shear:MainResultT})--(\ref{Eq:SteSol:Shear:Q}) and (\ref{Eq:SteSol:Shear:MainResultS}) can be used to obtain the asymptotic expansions for the dimensionless stresses at small and large shear rates. When $\Lambda $ is small,
\begin{align}
\label{Eq:SteSol:Shear:AsymptoticZeroHatT}
\hT &= 2\Lambda^2 -4 (2+\chi)\Lambda^4 
 +8(7+6\chi+\chi^2)\Lambda^6+O(\Lambda^8), \\
\label{Eq:SteSol:Shear:AsymptoticZeroHatS}
\hs &= \Lambda - 2(1+\chi)\Lambda^3 
+4(1+\chi)(3+\chi)\Lambda^5+ O(\Lambda^7).
\end{align}
At large $\Lambda$ and $\chi \neq 0$,
\begin{align}
\label{Eq:SteSol:Shear:AsymptoticInfHatT}
\hT &= 
 \dfrac{1}{\chi} - \dfrac{(1+\chi)^2}{2\chi^4\Lambda^2} 
 + \dfrac{(1+\chi)^3(3+\chi)}{
4\chi^7\Lambda^4} +O\left(\dfrac{1}{\Lambda^6}\right), \\
\label{Eq:SteSol:Shear:AsymptoticInfHatS}
\hs &= \dfrac{1+\chi}{2\chi^2 \Lambda}-\dfrac{(1+\chi)^2(2+\chi)}{4\chi^5\Lambda^3} 
+ \dfrac{(1+\chi)^3(7+6\chi+\chi^2)}{8\chi^8\Lambda^5} + O\left(\dfrac{1}{\Lambda^7}\right),
\end{align}
while at large $\Lambda $ and $\chi=0$,
\begin{align}
\label{Eq:SteSol:Shear:AsymptoticInfHatT-SLPTT}
\hT &= \sqrt[3]{2}\Lambda^{2/3}-\dfrac{2}{3}+\dfrac{1}{9\sqrt[3]{2}\Lambda^{2/3}}+O\left( \dfrac{1}{\Lambda^{4/3}} \right), \\
\label{Eq:SteSol:Shear:AsymptoticInfHatS-SLPTT}
\hs &=\dfrac{1}{\sqrt[3]{2}}\Lambda^{1/3}
-\dfrac{1}{3\sqrt[3]{4}} \dfrac{1}{\Lambda^{1/3}}+\dfrac{1}{162\sqrt[3]{2}\Lambda^{5/3}}+O\left(\dfrac{1}{\Lambda^{7/3}} \right).
\end{align}
Equations (\ref{Eq:SteSol:Shear:AsymptoticZeroHatT})--(\ref{Eq:SteSol:Shear:AsymptoticInfHatS-SLPTT}) can be  converted into the corresponding asymptotic expressions for the material functions using Eqs. (\ref{Eq:SteSol:Shear:N1}) and (\ref{Eq:SteSol:Shear:N2}) together with the first column of Table \ref{Tab:MatFunctionsVsDimlessVars}.
\par 
Expressions (\ref{Eq:SteSol:Shear:MainResultT})--(\ref{Eq:SteSol:Shear:Q}) and (\ref{Eq:SteSol:Shear:N1})--(\ref{Eq:SteSol:Shear:MainResultS}) provide a complete and fully explicit analytical description of the rheological properties of LPTT fluids in steady shear flows. Our equations  are different in form but equivalent to Eqs. (16)--(20) of Xue \textit{et al.} \cite{Xue1998} (which, in contrast to ours, are not fully explicit) and to Eqs. (12)--(15) of Alves \textit{et al.}\citep{Alves2001} (which are fully explicit but essentially more complicated than ours).
\par 
One should remark that the critical value of shear rate [Eq. (\ref{Eq:SteSol:Shear:DefinitionLambdaFirst})] is important for  shear flows of non-affine LPTT fluids. Alves \textit{et al.}\cite{Alves2001} have shown that the critical Weissenberg number [equivalent to $\Lambda_I$ defined by Eq. (\ref{Eq:SteSol:Shear:DefinitionLambdaFirst})] marks the onset of “constitutional instability” of steady pipe and slit flows of single-mode non-affine LPTT fluids. Even though the situation may be different for the multimode fluids (unless $\Lambda>\Lambda_I$ for each of the modes), this result demonstrates that non-affine LPTT fluid models should be applied to shear flows with care, especially at shear rates which are not small.
\subsection{Planar extension \label{Sec:SteSol:PlaEx}}
For steady state, Eqs. (\ref{Eq:Eqs:PlaEx:1TEvolution}) and (\ref{Eq:Eqs:PlaEx:2N1Evolution}) reduce to
\begin{align}
(1+\hT)\hT &= 2 \Lambda \hnen, \label{Eq:SteSol:PlaEx:1T}\\
(1+\hT)\hnen &= 2 \Lambda \hT + 4\alpha \Lambda, \label{Eq:SteSol:PlaEx:2N1}
\end{align}
while the algebraic relation for the dimensionless second normal stress difference, Eq. (\ref{Eq:Eqs:PlaEx:N2}), becomes
\begin{equation}
\hnto = \dfrac{1}{2}\left(\hT+\hnen\right). \label{Eq:SteSol:PlaEx:MainResultN2}
\end{equation}
From Eq. (\ref{Eq:SteSol:PlaEx:1T}), one gets
\begin{equation}
\hnen = \dfrac{\left(1+\hT\right)\hT}{2 \Lambda}. \label{Eq:SteSol:PlaEx:MainResultN1}
\end{equation}
Having used this to eliminate $\hnen$ from Eq. (\ref{Eq:SteSol:PlaEx:2N1}) and made simple rearrangements, one arrives at a cubic equation for $\hT$,
\begin{equation}
\label{Eq:SteSol:PlaEx:Cubic}
\hT^3 + 2\hT^2 + \left(1-4 \Lambda^2 \right) \hT - 8 \alpha \Lambda^2=0.
\end{equation}
Depending on $\alpha$ and $\Lambda$, Eq. (\ref{Eq:SteSol:PlaEx:Cubic}) can have either one real solution, which is positive, or three real solutions, only one of which is positive (see Appendix \ref{App:CubicSolutions:PlaEx}). In both cases, the positive solution can be written as
\begin{equation}
\label{Eq:SteSol:PlaEx:MainResultTExplicit}
\hT = \left\{ \renewcommand*{\arraystretch}{1.8} \begin{array}{ll}
-\dfrac{2}{3}+ \dfrac{2\sqrt{P}}{3}\cos\left(\dfrac{1}{3} \arccos \dfrac{Q}{P\sqrt{P}} \right), & \quad \text{if }Q^2\leq P^3, \\
-\dfrac{2}{3}+\dfrac{2\sqrt{P}}{3}\cosh \left(\dfrac{1}{3}\ln\dfrac{Q+\sqrt{Q^2-P^3}}{P\sqrt{P}} \right), & \quad \text{if } Q^2>P^3, \end{array} \right.
\end{equation}
where 
\begin{align}
P &= 1+12 \Lambda^2, \label{Eq:SteSol:PlaEx:P}\\
Q &= 1-36(1-3\alpha)\Lambda^2. \label{Eq:SteSol:PlaEx:Q}
\end{align}
\par
\begin{figure}[htp]
\begin{center}
\includegraphics[width=3.37in]{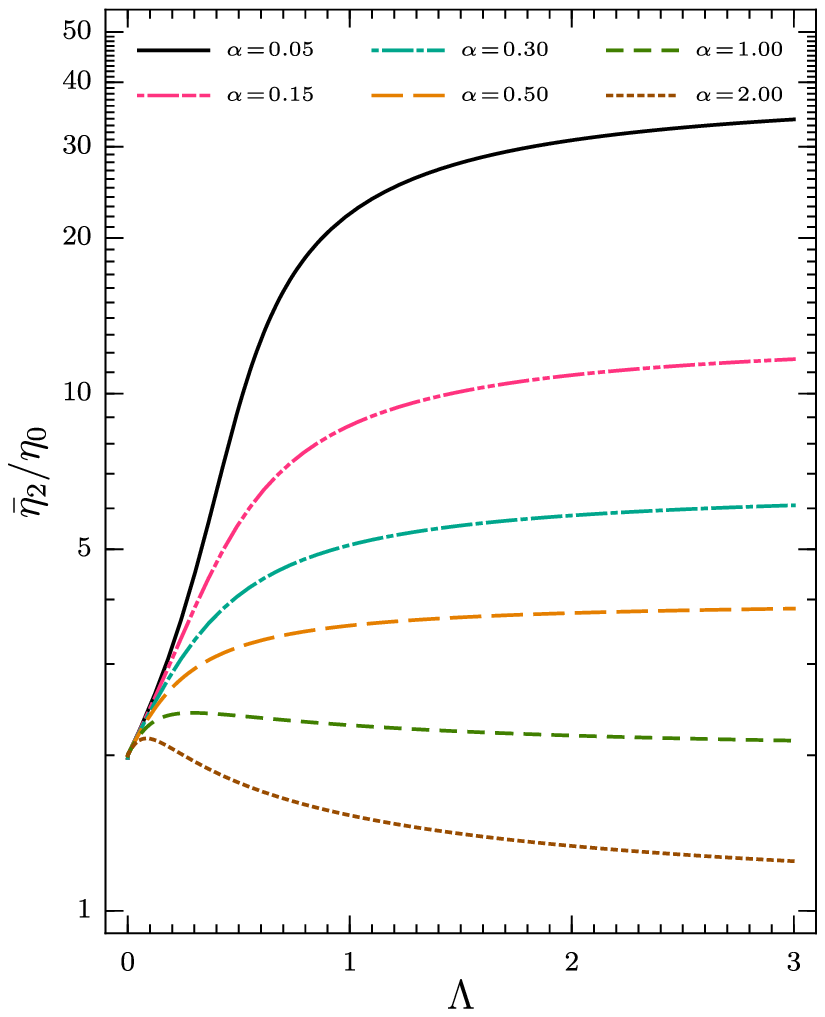}
\caption{\label{Fig:SteSol:PlaEx:SecondViscosity} The scaled second planar extensional viscosity as a function of the dimensionless elongation rate, $\Lambda$, at different values of the effective extensional flow parameter, $\alpha$.}
\end{center}
\end{figure}

\begin{figure}[htp]
\begin{center}
\includegraphics[width=3.37in]{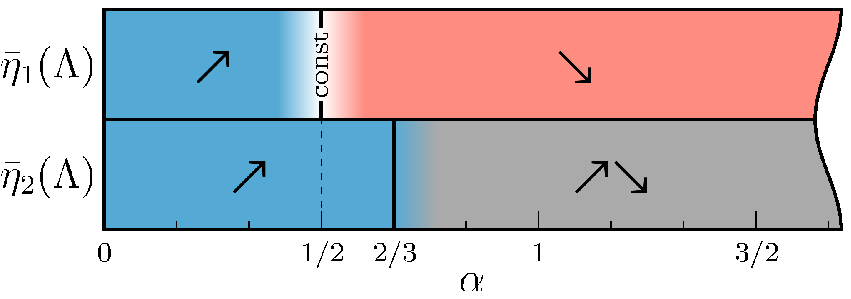}
\caption{\label{Fig:SteSol:PlaEx:Map} The possible shapes of the planar extensional viscosity curves. Color transitions are used to show the gradual character of changes between the different shapes: For example, when $\alpha \to 2/3$ from the right, the interval of decrease in $\bar{\eta}_2(\Lambda)$ is smoothed out, the maximum becomes less pronounced, and $\bar{\eta}_2(\Lambda)$ asymptotically approaches monotonic increase.}
\end{center}
\end{figure}

 
The dimensionless trace of the stress tensor, $\hT$, and both dimensionless normal stress differences, $\hnen$ and $\hnto$, are strictly increasing, unbounded functions of $\Lambda$ (see Appendices \ref{App:Bijectivity} and \ref{App:Shapes:PlaEx}).
In contrast, the extensional viscosities, $\bar{\eta}_1(\Lambda)$ and $\bar{\eta}_2(\Lambda)$, are bounded  (see Appendix \ref{App:Shapes:PlaEx}), as shown in Figs. \ref{Fig:SteSol:PlaEx:FirstViscosity} and \ref{Fig:SteSol:PlaEx:SecondViscosity}. Monotonicity of these material functions is controlled by the value of $\alpha$, which is summarized in Fig. \ref{Fig:SteSol:PlaEx:Map}. At $\alpha<1/2$, $\bar{\eta}_1(\Lambda)$ increases monotonically; at $\alpha=1/2$, $\bar{\eta}_1=4\eta_0$ and it is independent of $\Lambda$; and at $\alpha>1/2$, it decreases monotonically. At the same time, $\bar{\eta}_2(\Lambda)$ increases monotonically at $\alpha \leq 2/3$ and goes through a maximum at $\alpha > 2/3$. The maximum value,
\begin{equation}
\label{Eq:SteSol:PlaEx:maxN2-N2}
\bar{\eta}_{2,\mathrm{max}} = \dfrac{2\eta_0 (1+X)}{1 + (2\alpha-1)X^2},
\end{equation}
is achieved at 
\begin{equation}
\label{Eq:SteSol:PlaEx:maxN2-Lambda}
\Lambda_\mathrm{max} = \dfrac{X + (2\alpha-1)X^3}{2(1-X^2)},
\end{equation}
where
\begin{equation}
\label{Eq:SteSol:PlaEx:maxN2-A}
X = \sqrt{ \dfrac{3-4\alpha+\sqrt{2(2\alpha-1)^3/\alpha}}{-1+2\alpha+\sqrt{2(2\alpha-1)^3/\alpha}} }.
\end{equation}
\begin{figure}[htp]
\begin{center}
\includegraphics[width=3.37in]{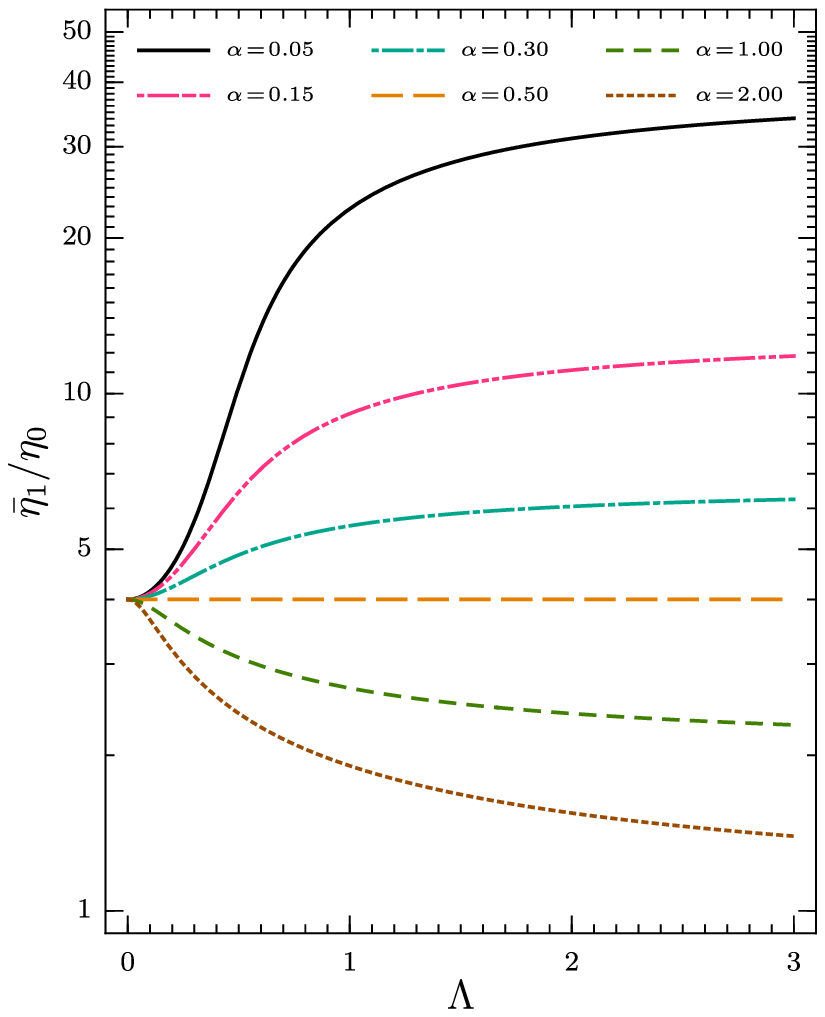}
\caption{\label{Fig:SteSol:PlaEx:FirstViscosity} The scaled first  planar extensional viscosity as a function of the dimensionless elongation rate, $\Lambda$, at different values of the effective extensional flow parameter, $\alpha$. }
\end{center}
\end{figure}
Furthermore, at $\alpha \to \infty$, $\bar{\eta}_{2,\mathrm{max}} \to 2\eta_0 = \bar{\eta}_2(0)$: Thus, the interval of increase “smoothes out” in this limit and the behavior of the second extensional viscosity asymptotically approaches monotonic decrease. Similarly, the interval of decrease gradually disappears as $\alpha \to 2/3$ from the right.
\par 
Asymptotic expressions for the dimensionless stresses at small and large elongation rates can be obtained from Eqs. (\ref{Eq:SteSol:PlaEx:MainResultTExplicit})--(\ref{Eq:SteSol:PlaEx:Q}), (\ref{Eq:SteSol:PlaEx:MainResultN1}), and (\ref{Eq:SteSol:PlaEx:MainResultN2}). At small $\Lambda$,
\begin{align}
\hT &= 8\alpha \Lambda^2 + 32 \alpha (1-4\alpha)\Lambda^4 + 128\alpha(1-12\alpha+28\alpha^2)\Lambda^6+O(\Lambda^8), \label{Eq:SteSol:PlaEx:AsymptoticZeroHatT}\\
\hnen &= 4\alpha \Lambda + 16\alpha (1-2\alpha)\Lambda^3 + 64\alpha(1-8\alpha+12\alpha^2)\Lambda^5+ O(\Lambda^7), \label{Eq:SteSol:PlaEx:AsymptoticZeroHatN1}\\
\hnto &= 2\alpha \Lambda + 4\alpha \Lambda^2 + 8\alpha(1-2\alpha)\Lambda^3+ O(\Lambda^4),\label{Eq:SteSol:PlaEx:AsymptoticZeroHatN2}
\end{align}
while at large $\Lambda$,
\begin{align}
\hT &= 2\Lambda - (1-\alpha) +\dfrac{\alpha(2-3\alpha)}{4\Lambda} + O\left(\dfrac{1}{\Lambda^2} \right), \label{Eq:SteSol:PlaEx:AsymptoticInfHatT}\\
\hnen &= 2\Lambda -(1-2\alpha) + \dfrac{\alpha(1-2\alpha)}{2\Lambda} + O\left( \dfrac{1}{\Lambda^2}\right), \label{Eq:SteSol:PlaEx:AsymptoticInfHatN1}\\
\hnto &= 2 \Lambda -\dfrac{2-3\alpha}{2}  + \dfrac{\alpha(4-7\alpha)}{8\Lambda}+ O\left( \dfrac{1}{\Lambda^2}\right). \label{Eq:SteSol:PlaEx:AsymptoticInfHatN2}
\end{align}
The corresponding expressions for the planar extensional viscosities can be obtained from Eqs. (\ref{Eq:SteSol:PlaEx:AsymptoticZeroHatN1}), (\ref{Eq:SteSol:PlaEx:AsymptoticZeroHatN2}), (\ref{Eq:SteSol:PlaEx:AsymptoticInfHatN1}), and (\ref{Eq:SteSol:PlaEx:AsymptoticInfHatN2}) using the second column of Table \ref{Tab:MatFunctionsVsDimlessVars}.
\par 
To our knowledge, the only existing alternative to Eqs. (\ref{Eq:SteSol:PlaEx:MainResultTExplicit})--(\ref{Eq:SteSol:PlaEx:Q}), (\ref{Eq:SteSol:PlaEx:MainResultN1}), and (\ref{Eq:SteSol:PlaEx:MainResultN2}) is Eqs. (23) and (24) of Xue \textit{et al.}\cite{Xue1998} (with $\beta=1$), which are incomplete (the authors have provided a formula only for the first planar extensional viscosity) and, in addition, may allow for  ambiguities: The expressions of Xue \textit{et al.} contain an implicit function defined as a solution to a cubic but do not involve any selection procedure which would allow one to choose the physically relevant value of this function. Our solution, being fully explicit, resolves the ambiguity issue completely. 
\par 
The asymptotic expansions for the normal stress differences  [Eqs. (\ref{Eq:SteSol:PlaEx:AsymptoticZeroHatN1}),  (\ref{Eq:SteSol:PlaEx:AsymptoticZeroHatN2}), (\ref{Eq:SteSol:PlaEx:AsymptoticInfHatN1}), and (\ref{Eq:SteSol:PlaEx:AsymptoticInfHatN2})] are, of course, in agreement with physical expectations. They are consistent with the numerical simulations of Xue \text{et al.}\cite{Xue1998} and with early asymptotic results for LPTT fluids [see Eqs. (30) and (31) of Petrie\cite{Petrie1990} but note that Eq. (30) has a missing factor of $(1-\xi)$].
\par 
An additional remarkable feature of our formulas is their purely real character, which we find very useful: Whenever complex values arise in intermediate calculations of real quantities, the accumulating numerical errors may cause the final result to have a small but nonvanishing imaginary part, thus leading to unnecessary complications.
\par 
To the best of our knowledge, we are the first to describe the variety of possible behaviors of the planar extensional viscosity curves for LPTT fluids (see Fig. \ref{Fig:SteSol:PlaEx:Map}).
\par 
Finally, all the results obtained in this section remain valid at $\xi=0$: The behavior of affine and non-affine LPTT fluids in extensional flows is qualitatively similar. Thus, this section provides a better alternative to the expressions proposed by Shogin\cite{Shogin2020SLPTT} for the affine case. Note that  different shapes of the planar extensional viscosity curves are also possible for affine LPTT fluids but were not described by Shogin\cite{Shogin2020SLPTT} because of the restriction adopted there: It was assumed that $\epsilon<1/4$, which for the affine case implies $\alpha<1/4$.
\subsection{Uniaxial and biaxial extension \label{Sec:SteSol:UBEx}}
For steady flow regime, Eqs. (\ref{Eq:Eqs:UBEx:1TEvolution}) and (\ref{Eq:Eqs:UBEx:2N1Evolution}) become
\begin{align}
(1+\hT)\hT &= 2\Lambda \hnen, \label{Eq:SteSol:UBEx:1T}\\
(1+\hT)\hnen &= \Lambda (\hT \pm \hnen) + 3 \alpha \Lambda.\label{Eq:SteSol:UBEx:2N1} 
\end{align}
From Eq. (\ref{Eq:SteSol:UBEx:1T}),
\begin{equation}
\hnen = \dfrac{(1+\hT)\hT}{2\Lambda}.
\label{Eq:SteSol:UBEx:MainResultN1}
\end{equation} 
Substituting this into Eq. (\ref{Eq:SteSol:UBEx:2N1}) and rearranging, one arrives at a cubic equation for $\hT$,
\begin{equation}
\label{Eq:SteSol:UBEx:Cubic}
\hT^3 + \left(2 \mp \Lambda \right)\hT^2 + \left(1 \mp \Lambda - 2 \Lambda^2 \right)\hT - 6\alpha \Lambda^2=0.
\end{equation}
Similarly to the case of planar extension, Eq. (\ref{Eq:SteSol:UBEx:Cubic}) can have one or three real solutions, depending on the values of $\alpha$ and $\Lambda$. Regardless of the signs chosen, only one of the solutions is both real and positive. This solution (see Appendix \ref{App:CubicSolutions:UBEx}) can be written as
\begin{equation}
\label{Eq:SteSol:UBEx:MainResultTExplicit}
\hT(\Lambda) = \left\{ \renewcommand*{\arraystretch}{1.8} \begin{array}{ll}
\dfrac{1}{3}\left[-2 \pm \Lambda + 2\sqrt{P}\cos\left( \dfrac{1}{3}\arccos\dfrac{Q}{2P\sqrt{P}} \right) \right], & \quad \text{if }Q^2\leq 4P^3, \\
\dfrac{1}{3}\left[-2 \pm \Lambda + 2\sqrt{P}\cosh\left( \dfrac{1}{3}\ln\dfrac{Q+\sqrt{Q^2-4P^3}}{2P\sqrt{P}} \right) \right],& \quad \text{if }Q^2>4P^3,
\end{array} \right.
\end{equation}
where
\begin{align}
P &= 1 \mp \Lambda + 7 \Lambda^2, \label{Eq:SteSol:UBEx:P}\\
Q &= 2 \mp 3\Lambda -3(13-54\alpha)\Lambda^2 \pm 20\Lambda^3. \label{Eq:SteSol:UBEx:Q}
\end{align}
\par
\begin{figure}[htp]
\begin{center}
\includegraphics[width=3.37in]{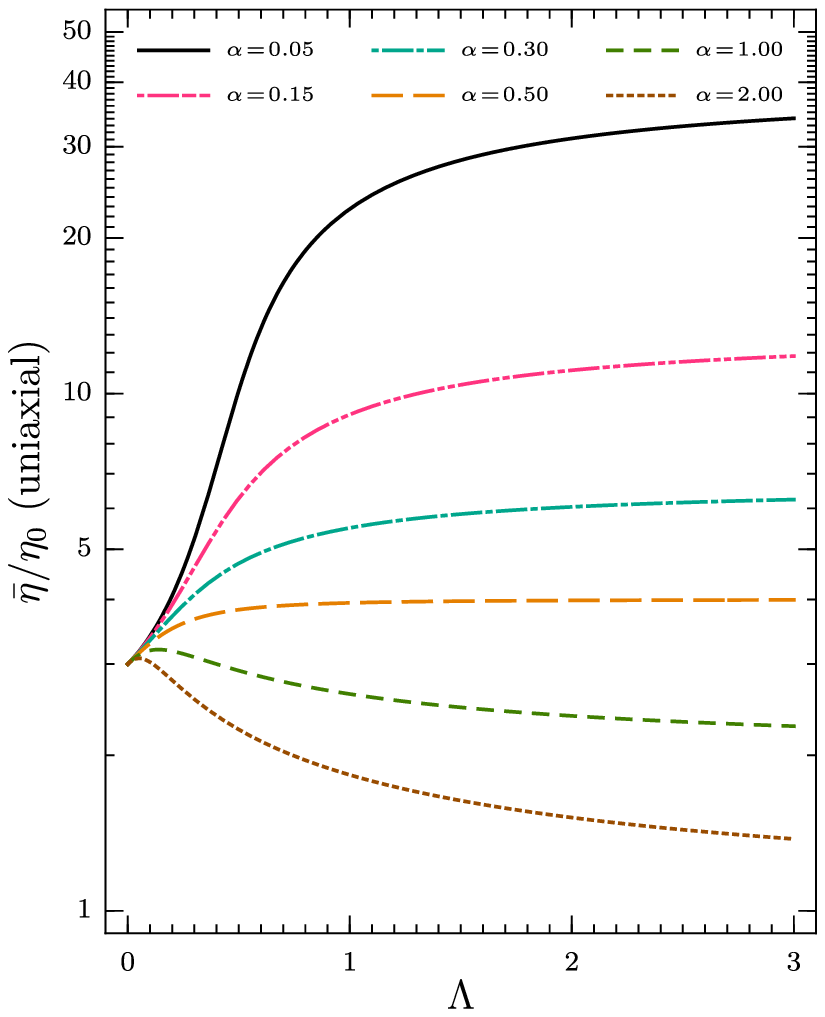}
\caption{\label{Fig:SteSol:UBEx:UniaxialViscosity} The scaled uniaxial extensional viscosity as a function of the dimensionless elongation rate, $\Lambda$, at different values of the effective extensional flow parameter, $\alpha$.}
\end{center}
\end{figure}

\begin{figure}[htp]
\begin{center}
\includegraphics[width=3.37in]{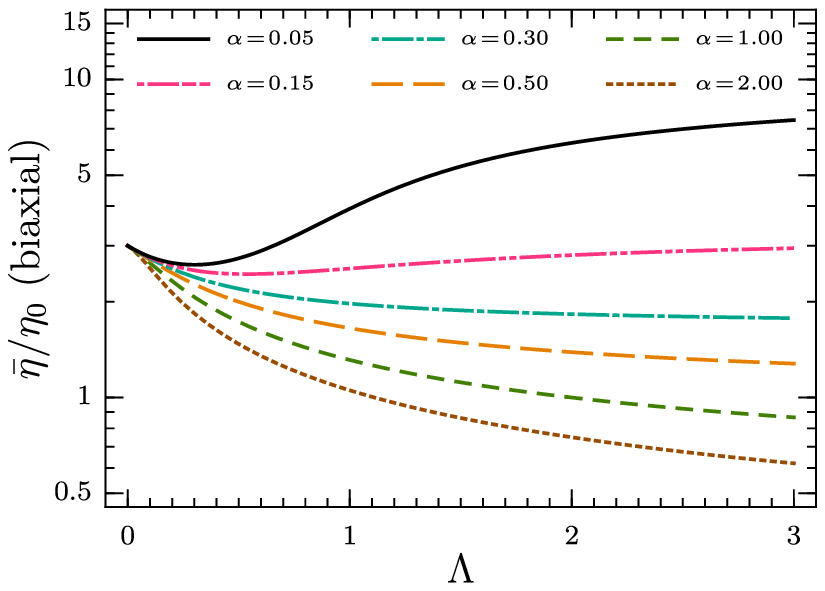}
\caption{\label{Fig:SteSol:UBEx:BiaxialViscosity} The scaled biaxial extensional viscosity as a function of the dimensionless elongation rate, $\Lambda$, at different values of the effective extensional flow parameter, $\alpha$.}
\end{center}
\end{figure}

\begin{figure}[htp]
\begin{center}
\includegraphics[width=3.37in]{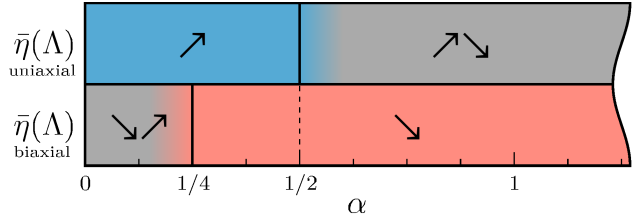}
\caption{\label{Fig:SteSol:UBEx:Map} The possible shapes of the uniaxial and biaxial extensional viscosity curves. Color transitions are used to show the gradual character of the shape changes with $\alpha$.}
\end{center}
\end{figure}

 
The dimensionless trace of the stress tensor, $\hT$, and the dimensionless normal stress difference, $\hnen$, are strictly increasing, unbounded functions of $\Lambda$ in both uniaxial and biaxial extensional flows (see Appendices \ref{App:Bijectivity} and \ref{App:Shapes:UBEx}), while uniaxial and biaxial extensional viscosities are bounded (see Appendix \ref{App:CubicSolutions:UBEx}), as shown in Figs. \ref{Fig:SteSol:UBEx:UniaxialViscosity} and \ref{Fig:SteSol:UBEx:BiaxialViscosity}. The monotonic properties of the extensional viscosities depend on $\alpha$, which is summarized in Fig. \ref{Fig:SteSol:UBEx:Map}. For uniaxial extension, $\bar{\eta}(\Lambda)$ is increasing monotonically at $\alpha \leq 1/2$ but goes through a maximum at $\alpha>1/2$. The maximum value, 
\begin{equation}
\bar{\eta}_\mathrm{max} = \dfrac{8\eta_0(3\alpha-1)}{8\alpha-3},
\label{Eq:SteSol:UBEx:N1UniMaxValue}
\end{equation}
is reached at 
\begin{equation}
\label{Eq:SteSol:UBEx:N1UniMaxLambda}
\Lambda_\mathrm{max} = \dfrac{8\alpha-3}{12(2\alpha-1)(4\alpha-1)}.
\end{equation} 
For biaxial extension, $\bar{\eta}(\Lambda)$ goes through a minimum at $\alpha <1/4$, with
\begin{equation}
\label{Eq:SteSol:UBEx:N1BiMinValue}
\bar{\eta}_\mathrm{min} = \dfrac{8\eta_0(1-3\alpha)}{3-8\alpha}
\end{equation}
reached at
\begin{equation}
\label{Eq:SteSol:UBEx:N1BiMinLambda}
\Lambda_\mathrm{min} = \dfrac{3-8\alpha}{12(1-2\alpha)(1-4\alpha)},
\end{equation}
and decreases monotonically if $\alpha \geq 4$.

\par 
Asymptotic expansions for the dimensionless stresses at small and large elongation rates can be obtained from Eqs. (\ref{Eq:SteSol:UBEx:MainResultTExplicit})--(\ref{Eq:SteSol:UBEx:Q}) and (\ref{Eq:SteSol:UBEx:MainResultN1}). At small $\Lambda$,
\begin{align}
\hT &= 6\alpha \Lambda^2 \pm 6\alpha \Lambda^3  +18\alpha(1-4\alpha)\Lambda^4+O(\Lambda^5), 
\label{Eq:SteSol:UBEx:AsymptoticZeroHatT}\\
\hnen &= 3\alpha \Lambda \pm 3\alpha \Lambda^2 + 9\alpha(1-2\alpha)\Lambda^3+O(\Lambda^4),
\label{Eq:SteSol:UBEx:AsymptoticZeroHatN1}
\end{align}
while at large $\Lambda$,
\begin{align}
\hT & = 2 \Lambda - (1-\alpha)+\dfrac{\alpha(3-5\alpha)}{6\Lambda}+O\left(\dfrac{1}{\Lambda^2}\right), \label{Eq:SteSol:UBEx:AsymptoticInfHatT-U}\\
\hnen &= 2\Lambda - (1-2\alpha) +\dfrac{\alpha(3-7\alpha)}{6\Lambda}+O\left( \dfrac{1}{\Lambda^2}\right) \label{Eq:SteSol:UBEx:AsymptoticInfHatN1-U}
\end{align}
for uniaxial extension and 
\begin{align}
\hT &= \Lambda - (1-2\alpha)+\dfrac{2\alpha(3-8\alpha)}{3\Lambda}+O\left(\dfrac{1}{\Lambda^2}\right),
\label{Eq:SteSol:UBEx:AsymptoticInfHatT-B}\\
\hnen &= \dfrac{1}{2}\Lambda - \dfrac{1-4\alpha}{2}+\dfrac{\alpha(3-10\alpha)}{3\Lambda}+O\left( \dfrac{1}{\Lambda^2}\right)
\label{Eq:SteSol:UBEx:AsymptoticInfHatN1-B}
\end{align}
for biaxial extension. The corresponding expansions for uniaxial and biaxial extensional viscosities can be obtained from Eqs. (\ref{Eq:SteSol:UBEx:AsymptoticZeroHatN1}), (\ref{Eq:SteSol:UBEx:AsymptoticInfHatN1-U}), and (\ref{Eq:SteSol:UBEx:AsymptoticInfHatN1-B}) using the third column of Table \ref{Tab:MatFunctionsVsDimlessVars}.  
\par 
Steady-flow rheology of both non-affine and affine LPTT fluids in uniaxial and biaxial extensional flows is fully described by Eqs. (\ref{Eq:SteSol:UBEx:MainResultN1}) and (\ref{Eq:SteSol:UBEx:MainResultTExplicit})--(\ref{Eq:SteSol:UBEx:Q}). We are not aware of any analogs of  these equations in the literature. At $\xi=0$ ($\alpha=\epsilon$), our formulas are equivalent to the recent analytical results for affine LPTT fluids,\cite{Shogin2020SLPTT} but, in contrast to the latter, are fully explicit.
\par 
The asymptotic expansions of $\hnen$ at large elongation rates in uniaxial and planar extension [Eqs. (\ref{Eq:SteSol:UBEx:AsymptoticInfHatN1-U}) and (\ref{Eq:SteSol:PlaEx:AsymptoticInfHatN1}), respectively] are very similar, with difference between them first appearing only in the third term. This is in agreement with the earlier theoretical predictions for LPTT fluids.\cite{Petrie1990}
\section{
\label{Sec:TransientSolutions}
Solutions for startup and cessation regimes}
\begin{table*}
\caption{\label{Tab:Forms} The trigonometric form ($\mathfrak{T}$), the hyperbolic form ($\mathfrak{H}$), the exponential forms ($\mathfrak{E}^+$ and $\mathfrak{E}^-$), and the transient flows described by these forms.}
\begin{center}
\begin{ruledtabular}
\begin{tabular}{c c c c}
Form & Expression & Describes & Applies to \\
\hline \rule{0pt}{3em}
$\mathfrak{T}$ 
&
$1-\dfrac{K \left( \cos \omega \bar{t} + \dfrac{a}{\omega}\sin \omega \bar{t}\right)}{C \mathrm{e}^{\Omega \bar{t}}+A \cos \omega \bar{t} + \dfrac{B}{\omega} \sin \omega \bar{t}}$
& stress growth
& shear and extensional flows
 \\ \rule{0pt}{3em}
$\mathfrak{H}$ 
&
$1-\dfrac{K \left(\cosh \omega \bar{t} + \dfrac{a}{\omega} \sinh \omega \bar{t}\right)}{C e^{\Omega \bar{t}}+A \cosh \omega \bar{t} + \dfrac{B}{\omega} \sinh \omega \bar{t}}$
& stress growth
& extensional flows \\ \rule{0pt}{2em}
$\mathfrak{E}^+$ 
&
$1-\dfrac{K \left(1 + a\bar{t}\right)}{C e^{\Omega \bar{t}}+A + B\bar{t}}$
& stress growth
& extensional flows \\ \rule{0pt}{2.5em}
$\mathfrak{E}^-$ 
&
$\dfrac{1}{(1+\hT) \exp \bar{t}-\hT}$ 
& stress relaxation 
& shear and extensional flows
\end{tabular}
\end{ruledtabular}
\end{center}
\end{table*}
\begin{table*}
\caption{
\label{Tab:FormFunctions}
Expressions for the functions encountered in forms $\mathfrak{T}$, $\mathfrak{H}$, and $\mathfrak{E}^+$ for startup regimes of shear flow, of planar extension, and of uniaxial and biaxial extension (with upper and lower signs, respectively).}
\begin{center}
\begin{ruledtabular}
\begin{tabular}{r c c c}
& Shear flow & Planar extension & Uniaxial and biaxial extension \\
 \hline \rule{0pt}{2em}
$\omega\,$ & 
$ \dfrac{1}{2}\sqrt{8\chi \Lambda^2 +4\hT+3\hT^2} $ & 
$\dfrac{1}{2}\sqrt{\left\vert-16\Lambda^2+4\hT+3\hT^2\right\vert}$ &
$ \dfrac{1}{2}\sqrt{\left \vert -9\Lambda^2+(4\mp 2\Lambda)\hT +3\hT^2 \right\vert }$ 
\\ \rule{0pt}{1.2em}
$\Omega\,$ & 
$1 + \dfrac{3}{2} \hT$ &
$1 + \dfrac{3}{2}\hT$ &
$1 \mp \dfrac{1}{2}\Lambda +\dfrac{3}{2}\hT$ \\ \rule{0pt}{1.2em}
$K\,$ &
$1 + 2\chi \Lambda^2 + 4\hT +3\hT^2$ &
$1 - 4\Lambda^2 +4\hT +3\hT^2$ &
$1 \mp \Lambda -2\Lambda^2 +2(2\mp\Lambda)\hT+3\hT^2$ 
 \\ \rule{0pt}{1.2em}
$C\,$ & 
$1 + 2\chi \Lambda^2 +2\hT +\hT^2 $ &
$1 -4\Lambda^2 + 2\hT + \hT^2$ &
$1\mp \Lambda-2\Lambda^2+(2\mp \Lambda)\hT+\hT^2$ 
 \\ \rule{0pt}{1.2em}
$A\,$ & 
$2\hT(1+\hT) $ &
$2\hT(1+\hT)$ &
$ \hT(2\mp\Lambda+2\hT)$ \\ \rule{0pt}{1.2em}
$B\,$ & 
$\hT(1-2\chi \Lambda^2+\hT)$ &
$\hT(1+4\Lambda^2+\hT)$ &
$\dfrac{1}{2} \hT \left[2 \mp 2\Lambda+5\Lambda^2+(2\mp\Lambda)\hT \right]$ 
\end{tabular}
\end{ruledtabular}
\end{center}
\end{table*}
The stress growth functions (normalized to their steady-flow values) in startup of shear and extensional flows can be written in one of the three compact forms $\mathfrak{T}$, $\mathfrak{H}$, and $\mathfrak{E}^+$ (see Table \ref{Tab:Forms}; for derivation of this result, see Appendix \ref{App:Transient}). These forms are specified by defining seven quantities: the frequency ($\omega$), the damping factor $(\Omega)$, and the five coefficients ($K$, $C$, $A$, $B$, and $a$), all of them being functions of  $\Lambda$ with parameter $\chi$ for shear flows or $\alpha$ for extensional flows. The particular expressions for $\omega$, $\Omega$, $K$, $C$, $A$, and $B$ depend only on the flow type; these expressions are provided in Table \ref{Tab:FormFunctions}. In contrast, the coefficient $a$ is defined uniquely for each material function it is associated with; the definitions of $a$ for the stress growth functions investigated in this work are given in Table \ref{Tab:a}. With the coefficients from Tables \ref{Tab:FormFunctions} and \ref{Tab:a}, form $\mathfrak{T}$ specifies a function that increases nonmonotonically from zero at $\bar{t}=0$ toward the steady-flow value of one, which is asymptotically approached through a series of nonharmonic damped oscillations; while functions defined by  forms $\mathfrak{H}$ and $\mathfrak{E}^+$ increase from zero toward unity monotonically.
\par 
It should be noted that from a mathematical point of view, the forms $\mathfrak{T}$, $\mathfrak{H}$, and $\mathfrak{E}^+$ are three purely real representations of the same complex (but real-valued) oscillatory function for the cases when the frequency of oscillations is real and positive and equals $\omega$ (form $\mathfrak{T}$), imaginary and has absolute value $\omega$ (form $\mathfrak{H}$), and zero (form $\mathfrak{E}^+$).  
\par 
Regardless of the flow type, all the stress relaxation functions considered in this work (normalized to their initial steady-flow values) take the form $\mathfrak{E}^-$ and decrease monotonically from unity at $\bar{t}=0$ toward zero.
\par 
\begin{table*}
\caption{
\label{Tab:a}
The expressions for the coefficient $a$ (encountered in the forms $\mathfrak{T}$, $\mathfrak{H}$, and $\mathfrak{E}^+$) for the normalized stress growth functions in startup of shear and extensional flows. Upper and lower signs in the expressions in the third column correspond to the uniaxial and biaxial extension, respectively.}
\begin{center}
\begin{ruledtabular}
\begin{tabular}{c c c c c c} 
\multicolumn{2}{c}{Shear flow}  & \multicolumn{2}{c}{Planar extension} & \multicolumn{2}{c}{Uniaxial and biaxial extension} \\ 
\hline \rule{0pt}{2em}
$\dfrac{\hT^+}{\hT}, \dfrac{\Psi_1^+}{\Psi_1}$ & 
$1+\dfrac{1}{2}\hT$ &
$\dfrac{\hT^+}{\hT}$ &
$1+\dfrac{1}{2}\hT$ &
$\dfrac{\hT^+}{\hT}$ &
$1\mp \dfrac{1}{2}\Lambda +\dfrac{1}{2}\hT$ 
\\ \rule{0pt}{2em}
$ \dfrac{\eta^+}{\eta}$ &
$-\dfrac{4\chi \Lambda^2+\hT+\hT^2}{2(1+\hT)}$ &
$\dfrac{\bar{\eta}_1^+}{\bar{\eta}_1}$ &
$\dfrac{8\Lambda^2-\hT-\hT^2}{2(1+\hT)}$ &
$\dfrac{\bar{\eta}^+}{\bar{\eta}}$  &
$\dfrac{\pm \Lambda +4\Lambda^2-(1\mp\Lambda)\hT-\hT^2}{2(1+\hT)}$ 
\\ \rule{0pt}{2em}
& & 
$\dfrac{\bar{\eta}_2^+}{\bar{\eta}_2}$  & 
$2\Lambda-\dfrac{1}{2}\hT$ &
&
\end{tabular}
\end{ruledtabular}
\end{center}
\end{table*}
\subsection{Start-up of steady shear flow
\label{Sec:Transient:Shear}}
\begin{figure}[htp]
\begin{center}
\includegraphics[width=6.45in]{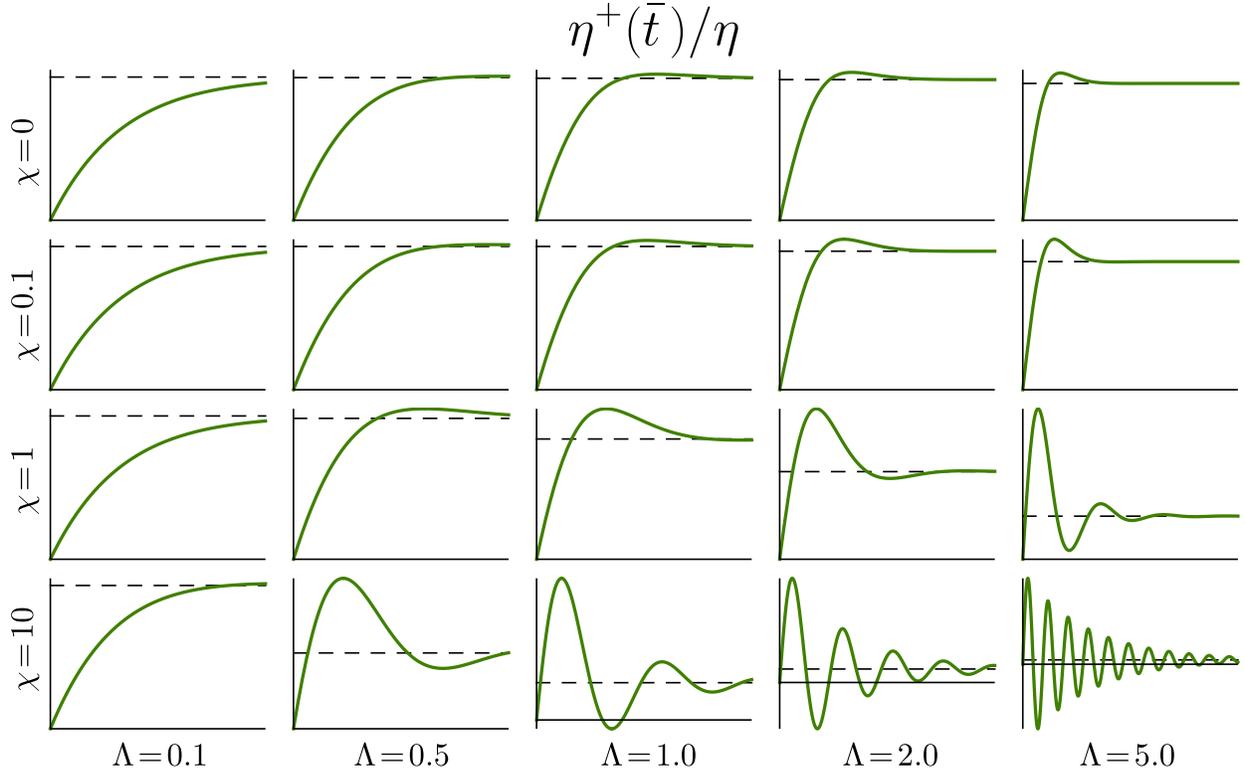}
\caption{\label{Fig:Transient:Shear:Shear} The normalized shear stress growth function, $\eta^+(\bar{t})/\eta$. The diagram shows a grid of miniplots made at different values of $\chi$ and $\Lambda$. The horizontal scale is the same for all miniplots (the range $0 \leq \bar{t} \leq 3$ is shown), while the vertical scale is not (the dashed horizontal lines in each miniplot mark the steady-flow value of unity). It is seen that the amplitude and the frequency of the oscillations increase with $\Lambda$ and $\chi$; and at large values of $\Lambda$ and/or $\chi$, it becomes possible for $\eta^+(\bar{t})$ to become negative.}
\end{center}
\end{figure}
\begin{figure}[htp]
\begin{center}
\includegraphics[width=6.45in]{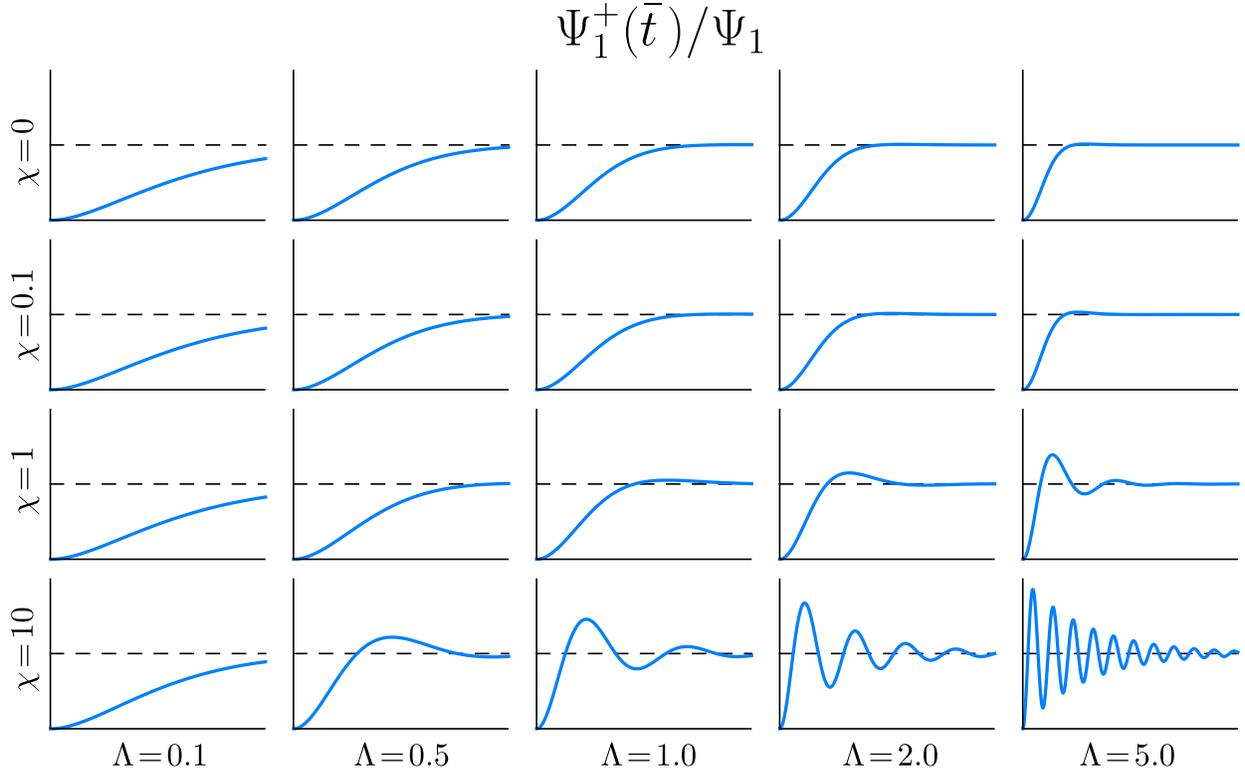}
\caption{\label{Fig:Transient:Shear:Normal} The normalized first normal stress difference growth function, $\Psi_1^+(\bar{t})/\Psi_1$, at  different values of $\chi$ and $\Lambda$. Each miniplot shows the region $[0,3]\times [0,2]$. The dashed horizontal line in each miniplot marks the steady-flow value of unity. In contrast to $\eta^+(\bar{t})$, the change of sign for $\Psi_1^+(\bar{t})$ is not possible.}
\end{center}
\end{figure}
\begin{figure}[htp]
\begin{center}
\includegraphics[width=3.37in]{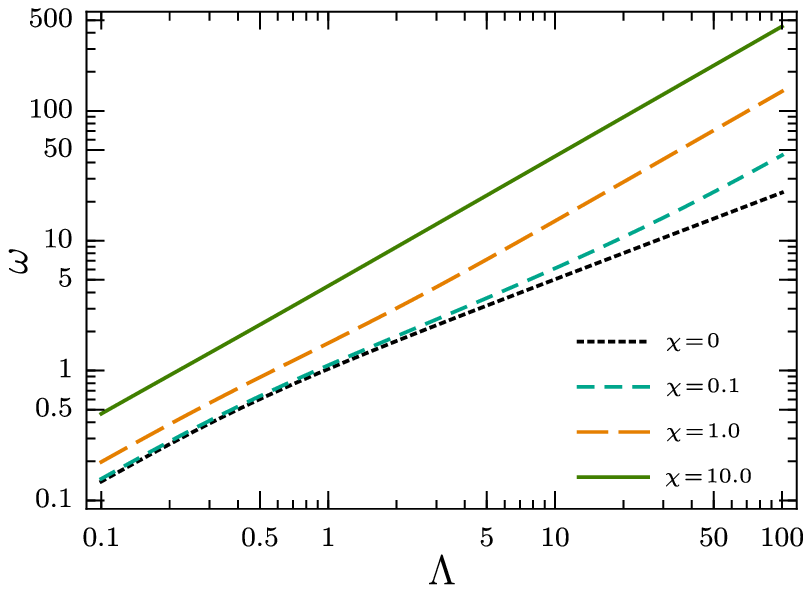}
\caption{\label{Fig:Transient:Shear:Frequency} The frequency of the stress oscillations in startup of steady shear flow, plotted as a function of the dimensionless shear rate, $\Lambda$, for different values of the effective shear flow parameter, $\chi$.}
\end{center}
\end{figure}
For any $\chi$ and $\Lambda$, the stress growth functions in startup of steady shear flow are of trigonometric form $\mathfrak{T}$ with $\omega$, $\Omega$, $K$, $A$, $B$, and $C$ given in the first column of Table \ref{Tab:FormFunctions} and the coefficients $a$ from the first column of Table \ref{Tab:a}. 
\par 
The stress growth functions $\eta^+(\bar{t})/\eta$ and $\Psi_1^+(\bar{t})/\Psi_1$ are shown in Figs. \ref{Fig:Transient:Shear:Shear} and \ref{Fig:Transient:Shear:Normal}, respectively [the second normal stress difference growth function is not shown, since $\Psi_2^+(\bar{t})=0$ if $\chi=0$ and $\Psi_2^+(\bar{t})/\Psi_2=\Psi_1^+(\bar{t})/\Psi_1$ if $\chi \neq 0$]. The solutions are always oscillatory, with a series of overshoots (maxima) and undershoots (minima) present, although not always easily observable. The frequency of oscillations, $\omega$, increases monotonically with $\Lambda$, as shown in Fig. \ref{Fig:Transient:Shear:Frequency}. At small $\Lambda$,
\begin{equation}
\omega \sim \sqrt{2(1+\chi)}\Lambda,
\end{equation}
while at large $\Lambda$,
\begin{equation}
\omega \sim 2\sqrt{\chi} \Lambda
\end{equation}
for non-affine LPTT fluids and 
\begin{equation}
\omega \sim \dfrac{\sqrt{3}}{\sqrt[3]{4}}\Lambda^{2/3} 
\end{equation}
for affine ones. Note that $\omega \to 0$ when 
$\Lambda \to 0$, so that the stresses build up monotonically in this limiting case, with
\begin{align}
\eta^+ & \to \eta_0\left[1-\exp(-\bar{t})\right], \label{Eq:Transient:Shear:ZeroLambdaS}\\
\Psi_1^+ & \to 2\eta_0 \lambda \left[1-(1+\bar{t})\exp(-\bar{t}) \right]. \label{Eq:Transient:Shear:ZeroLambdaN1} 
\end{align}
The response of shear stresses in this limit [Eq. (\ref{Eq:Transient:Shear:ZeroLambdaS})] is identical to the “linear viscoelastic response” of the Maxwell model.\cite{Bird1987a}
\par 
During the oscillations, the stresses go through their steady-flow values periodically, reaching them for the $k^\mathrm{th}$ time at 
\begin{equation}
\bar{t}_0^{\;[k]}=\dfrac{1}{\omega}\left[\phi+(k-1)\pi \right],
\label{Eq:Transient:Shear:EquilibriumPositions}
\end{equation}
where 
\begin{equation}
\label{Eq:Transient:Shear:Phi:Shear}
\phi = -\arctan \dfrac{\omega}{a}
\end{equation}
for the shear stresses (since $a$ corresponding to $\eta^+/\eta$ is negative) and 
\begin{equation}
\label{Eq:Transient:Shear:Phi:Normal}
\phi = \pi-\arctan \dfrac{\omega}{a}
\end{equation}
for the normal stress differences (since $a$ corresponding to $\Psi_1^+/\Psi_1$ is positive).
\par
\begin{figure}[htp]
\begin{center}
\includegraphics[width=6.45in]{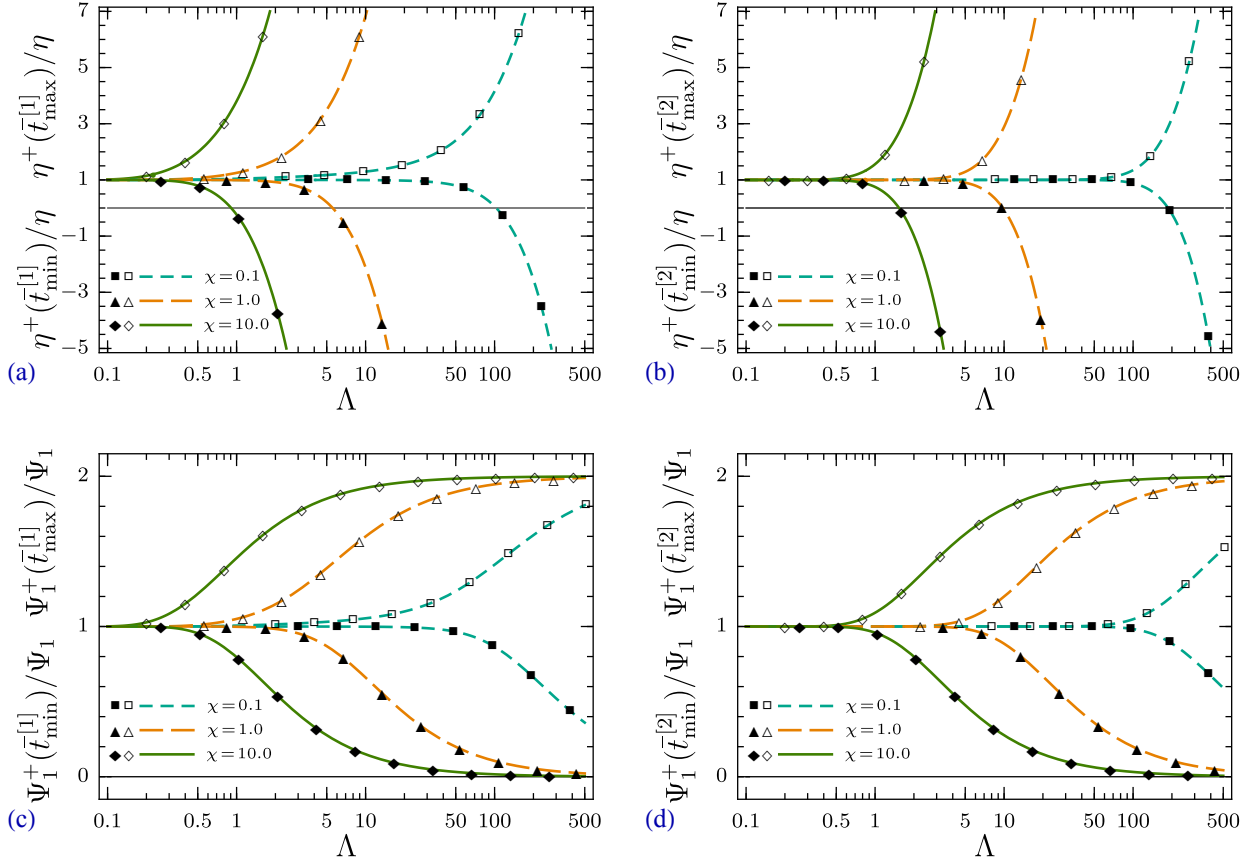}
\caption{\label{Fig:Transient:Shear:OverUnderShoots} Stress overshoots and undershoots during startup of steady shear flow. Top row: the values of $\eta^+(\bar{t})/\eta$ at the first (a) and the second (b) overshoot (inclining curves, hollow symbols) and undershoot (declining curves, black symbols). Bottom row: the values of $\Psi_1^+(\bar{t})/\Psi_1$ at the first (c) and the second (d) 	overshoot (inclining curves, hollow symbols) and undershoot (declining curves, black symbols). The accurate numerical values (shown by symbols) were calculated by applying functions \texttt{FindMaximum} and \texttt{FindMinimum} of Wolfram Mathematica to the exact analytical solutions, while the approximate values (shown by lines) were  obtained by substitution of Eqs. (\ref{Eq:Transient:Shear:OvershootPositions}) and (\ref{Eq:Transient:Shear:UndershootPositions}) into the form $\mathfrak{T}$. The difference between the numerical results and the approximation is seen to be negligible.}
\end{center}
\end{figure}

 
The occurrence of overshoots and undershoots is nearly periodic. As shown in Appendix \ref{App:Truncation}, the positions of $k^\mathrm{th}$ overshoot and undershoot are accurately described by the approximate formulas
\begin{align}
\bar{t}_\mathrm{max}^{\;[k]} & \approx \dfrac{1}{\omega}\left[\theta+(2k-2)\pi \right], \label{Eq:Transient:Shear:OvershootPositions} \\
\bar{t}_\mathrm{min}^{\;[k]} & \approx \dfrac{1}{\omega}\left[\theta+(2k-1) \pi \right],
\label{Eq:Transient:Shear:UndershootPositions}
\end{align}
respectively, where  
\begin{equation}
\label{Eq:Transient:Shear:Theta:Shear}
\theta =
\left\{ 
\renewcommand{\arraystretch}{1.5} \begin{array}{ll}
\pi + \arctan\dfrac{(a-\Omega)\omega}{a\Omega+\omega^2}  & \quad \text{if} \quad \Lambda<\sqrt[4]{\dfrac{1+\chi}{4\chi^3}}, \\
\arctan\dfrac{(a-\Omega)\omega}{a\Omega+\omega^2} & \quad \text{if} \quad \Lambda>\sqrt[4]{\dfrac{1+\chi}{4\chi^3}}, \\
\dfrac{\pi}{2} & \quad \text{if} \quad \Lambda=\sqrt[4]{\dfrac{1+\chi}{4\chi^3}},
\end{array} \right.
\end{equation}
with $a$ corresponding to $\eta^+/\eta$, for the shear stress growth function, and
\begin{equation}
\label{Eq:Transient:Shear:Theta:Normal}
\theta = \pi + \arctan\dfrac{(a-\Omega)\omega}{a\Omega+\omega^2},
\end{equation}
with $a$ corresponding to $\Psi_1^+/\Psi_1$, for the normal stress difference growth functions. Then, the  values of the material functions at the points of overshoots and undershoots can be evaluated by substituting Eqs. (\ref{Eq:Transient:Shear:OvershootPositions}) and (\ref{Eq:Transient:Shear:UndershootPositions}) into the form $\mathfrak{T}$. The result is shown in Fig. \ref{Fig:Transient:Shear:OverUnderShoots}.
\par 
The character of oscillations differs drastically between the full and the simplified versions of the LPTT model. For affine LPTT fluids, $\hT(\Lambda)$ is not bounded, which means that the damping factor, $\Omega$, also has no upper bound. While the frequency of oscillations increases with $\Lambda$, larger values of $\Lambda$ also lead to stronger damping; as a result, the first overshoot is the only pronounced extremum of the stress growth functions, as described previously by Shogin\cite{Shogin2020SLPTT} (see also the miniplots corresponding to $\chi=0$ in Figs. \ref{Fig:Transient:Shear:Shear} and \ref{Fig:Transient:Shear:Normal}). In contrast, for non-affine LPTT fluids, the frequency increases both with $\Lambda$ and with $\chi$, while the damping factor is restricted from above (furthermore, its upper bound decreases with increasing $\chi$). As a consequence, when $\chi$ and/or $\Lambda$ are large enough, multiple overshoots and undershoots can be observed (seen Figs. \ref{Fig:Transient:Shear:Shear} and \ref{Fig:Transient:Shear:Normal}).
\par 
The amplitude of the stress oscillations increases with $\Lambda$. For non-affine LPTT fluids at large values of $\Lambda$, 
\begin{align}
\dfrac{\eta^+(\bar{t}_\mathrm{max}^{\;[k]})}{\eta} & = \dfrac{(2\chi^3)^{1/2}}{1+\chi}\Lambda+O(1), \\
\dfrac{\eta^+(\bar{t}_\mathrm{min}^{\;[k]})}{\eta} & = -\dfrac{(2\chi^3)^{1/2}}{1+\chi}\Lambda+O(1), \label{Eq:Transient:Shear:ShearUndershootAsymptotics}\\ 
\dfrac{\Psi_1^+(\bar{t}_\mathrm{max}^{\;[k]})}{\Psi_1} & = 2+o(1), \\
\dfrac{\Psi_1^+(\bar{t}_\mathrm{min}^{\;[k]})}{\Psi_1} & = o(1).
\end{align}
Thus, the normalized first normal stress difference growth function, $\Psi_1^+(\bar{t})/\Psi_1$, always takes the values in the range $(0,2)$ [see Figs. \ref{Fig:Transient:Shear:Normal}, \ref{Fig:Transient:Shear:OverUnderShoots} (c), and \ref{Fig:Transient:Shear:OverUnderShoots} (d)], while the normalized shear stress growth function, $\eta^+(\bar{t})/\eta$, has no universal upper bound, as seen in Figs. \ref{Fig:Transient:Shear:Shear}, \ref{Fig:Transient:Shear:OverUnderShoots} (a), and \ref{Fig:Transient:Shear:OverUnderShoots} (b). In contrast, for affine LPTT fluids, the normalized stress growth functions are restricted from above by certain irrational numbers slightly greater than one.\cite{Shogin2020SLPTT}
\begin{figure}[htp]
\begin{center}
\includegraphics[width=3.37in]{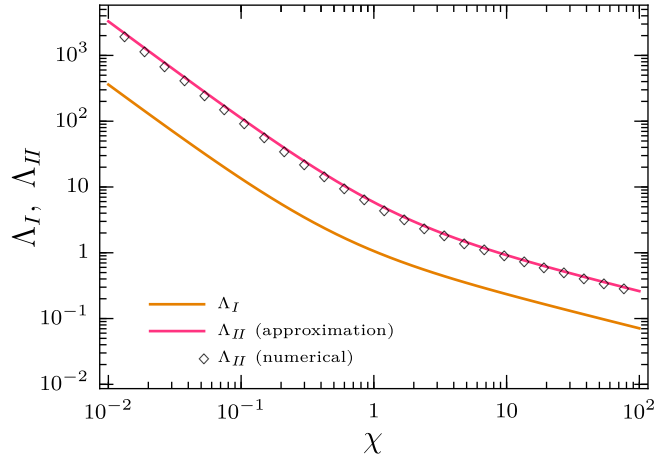}
\caption{\label{Fig:Transient:Shear:CriticalShearRates} The two critical shear rates, $\Lambda_I$ and $\Lambda_{II}$, as functions of the effective shear flow parameter, $\chi$. At $\Lambda>\Lambda_I$, the shear stress in steady shear flow decreases with increasing shear rate. At $\Lambda>\Lambda_{II}$, the shear stress growth function in startup of steady shear flow takes negative values during the undershoots. The numerical values of $\Lambda_{II}$ (hollow diamonds) were obtained by finding $\Lambda$ for which $\hs^+=0$ at the point of first undershoot using Wolfram Mathematica functions \texttt{FindMinimum} and \texttt{FindRoot}. The approximate values of $\Lambda_{II}$ (line) were calculated using Eq. (\ref{Eq:Transient:Shear:LambdaII}).}
\end{center}
\end{figure}
\par 
Furthermore, Eq. (\ref{Eq:Transient:Shear:ShearUndershootAsymptotics}) implies that for any $\chi>0$, there exists a critical value of the shear rate (which we shall call the second critical shear rate, $\Lambda_{II}$, in contrast to $\Lambda_I$ introduced in Sec. \ref{Sec:SteSol:Shear}), so that $\eta^+(\bar{t})$ takes negative values at some moment after startup when $\Lambda>\Lambda_{II}$. This feature of non-affine LPTT fluids is considered unphysical and was predicted by Stephenson,\cite{Stephenson1986} although he did not specify when exactly $\eta^+(\bar{t})$ can become negative. In Appendix \ref{App:LambdaII}, we derive an approximate  expression for $\Lambda_{II}$, 
\begin{equation}
\label{Eq:Transient:Shear:LambdaII}
\Lambda_{II} \approx \dfrac{3\sqrt{2}(3+2\chi)\pi}{8\chi^{3/2} W_0 \left[ \dfrac{3(3+2\chi)\pi}{4(1+\chi)} \right]},
\end{equation}
where $W_0(x)$ is the principal branch of the Lambert $W$ function (product logarithm).\cite{Corless1996} From Fig. \ref{Fig:Transient:Shear:CriticalShearRates}, where the two critical shear rates, $\Lambda_I$ and $\Lambda_{II}$, are shown as functions of $\chi$, it is seen that $\Lambda_{II}$ is significantly larger than $\Lambda_I$. Thus, the sign change in the shear stresses occurs in situations when the LPTT fluid model most probably should not be applied at all, and, therefore, shall unlikely cause problems in most applications.
\subsection{Startup of steady planar extensional flow
\label{Sec:Transient:PlaEx}}
\begin{figure}[htp]
\begin{center}
\includegraphics[width=3.37in]{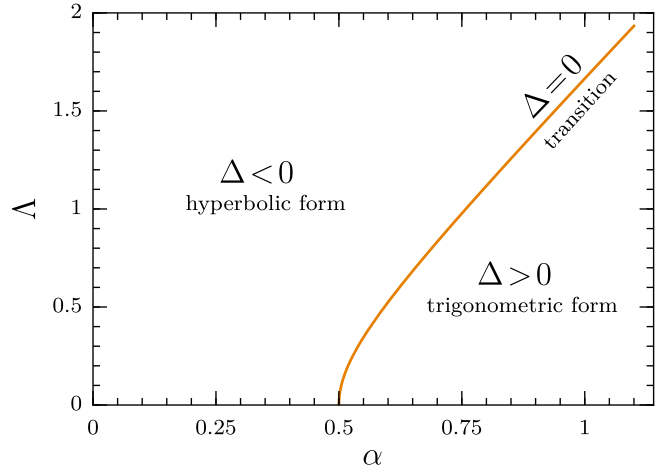}
\caption{\label{Fig:Transient:PlaEx:FormMap} The regions  of the $(\alpha,\Lambda)$-plane where the stress growth functions in startup of steady planar extensional flow take forms $\mathfrak{H}$, $\mathfrak{T}$, and $\mathfrak{E}^+$. Form $\mathfrak{E}^+$ applies at the boundary line where $\Delta=0$.}
\end{center}
\end{figure}

Depending on the sign of $\Delta$ (see Appendixes \ref{App:Transient:PlaEx} and \ref{App:FormShift:PlaEx} for details), which is determined  by the values of $\alpha$ and $\Lambda$, the stress  growth functions in startup of steady planar extensional flow can be described by any of the three forms $\mathfrak{H}$, $\mathfrak{T}$, and $\mathfrak{E}^+$, with $\omega$, $\Omega$, $K$, $A$, $B$, and $C$ provided in the second column of Table \ref{Tab:FormFunctions} and the coefficients $a$ given in the second column of Table \ref{Tab:a}. The conditions at which each of the form applies are formulated in Table \ref{Tab:FormsPlaEx}, where
\begin{equation}
\Lambda^\ast = \dfrac{1+\hT^\ast}{2} \sqrt{\dfrac{\hT^\ast}{ 2\alpha +\hT^\ast}},
\label{Eq:Transient:PlaEx:LambdaCrit}
\end{equation}
with 
\begin{equation}
\label{Eq:Transient:PlaEx:TCrit}
\hT^\ast = 3\alpha -2 +\sqrt{\alpha(9\alpha-4)}.
\end{equation}
The corresponding regions of the $(\alpha,\Lambda)$-plane are shown in Fig. \ref{Fig:Transient:PlaEx:FormMap}.
\begin{table*}[htp]
\caption{
\label{Tab:FormsPlaEx}
The conditions at which the stress growth functions in startup of steady planar extensional flow take the forms $\mathfrak{H}$, $\mathfrak{T}$, and $\mathfrak{E}^+$.}
\begin{center}
\begin{ruledtabular}
\begin{tabular}{c c c}
$\mathfrak{H}$ & $\mathfrak{T}$ & $\mathfrak{E}^+$  \\
 \hline \rule{0pt}{2em}
$\begin{matrix}
\alpha \leq 1/2 \\
\alpha>1/2\text{ and }\Lambda>\Lambda^\ast  
\end{matrix}$ & 
$\alpha>1/2$ and $\Lambda<\Lambda^\ast$ & 
$\alpha>1/2$ and $\Lambda=\Lambda^\ast$ 
\end{tabular}
\end{ruledtabular}
\end{center}
\end{table*}
\par
\begin{figure}[htp]
\begin{center}
\includegraphics[width=6.45in]{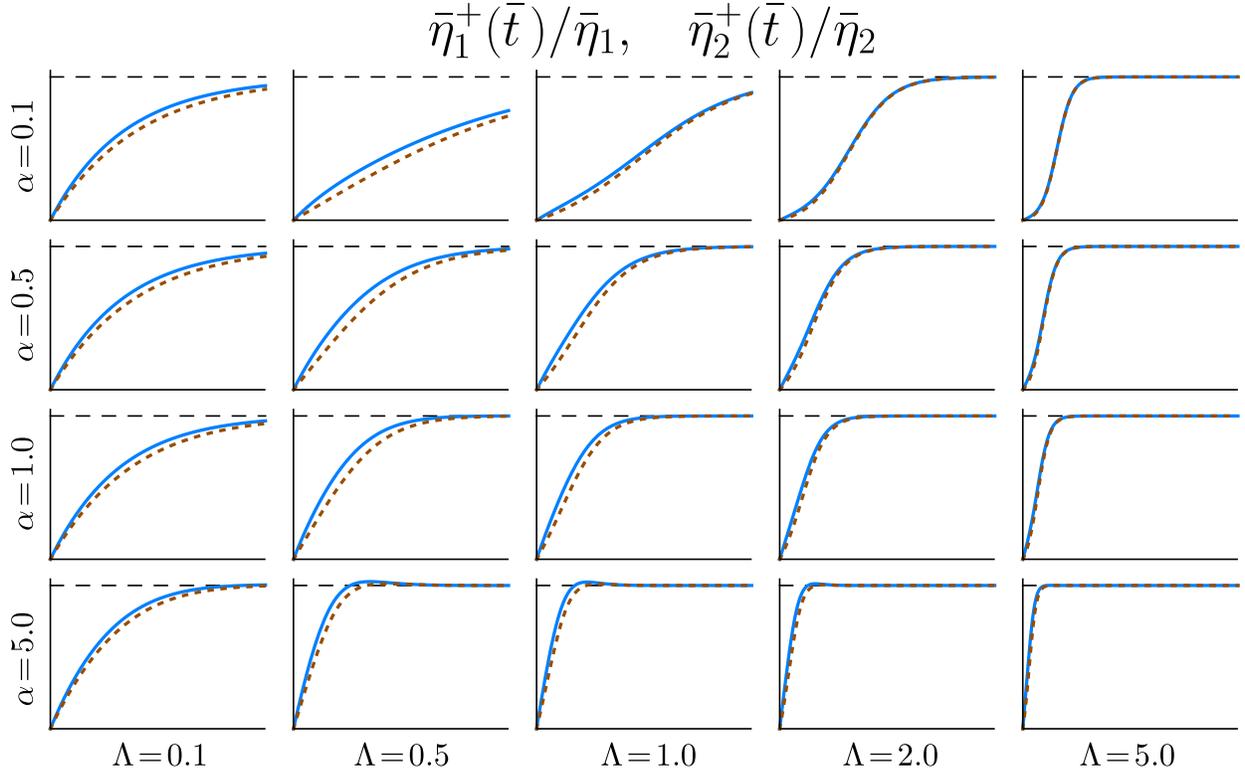}
\caption{\label{Fig:Transient:PlaEx:StressGrowthFunctions} The normalized stress growth functions in startup of planar extensional flow: $\bar{\eta}_1^+(\bar{t})/\bar{\eta}_1$ (solid lines) and $\bar{\eta}_2^+(\bar{t})/\bar{\eta}_2$ (dashed lines). The  miniplots have the same scale and show the region $[0,3]\times [0,1.05]$. The dashed horizontal line in each miniplot marks the steady-flow value of one.}
\end{center}
\end{figure}

\begin{figure}[htp]
\begin{center}
\includegraphics[width=3.37in]{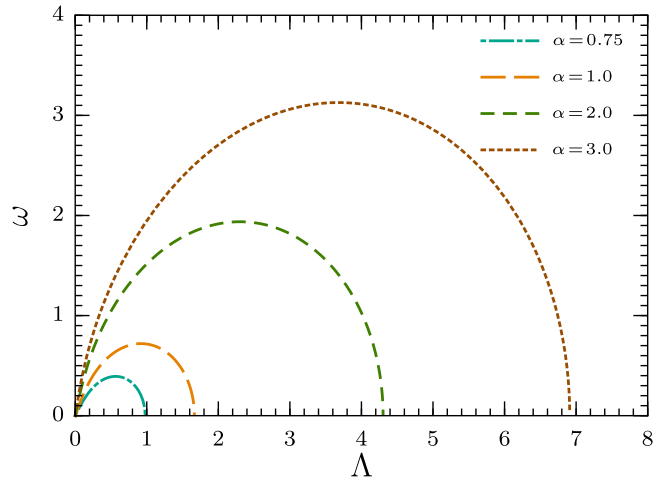}
\caption{\label{Fig:Transient:PlaEx:Frequency} The frequency of the stress oscillations at startup of steady planar extensional flow (in the case when the solution is described by the trigonometric form $\mathfrak{T}$). The frequency, $\omega$, is plotted as a function of the dimensionless shear rate, $\Lambda$, for different values of the effective extensional  flow parameter, $\alpha$.}
\end{center}
\end{figure}

 
The normalized stress growth functions $\bar{\eta}_1^+(\bar{t})/\bar{\eta}_1$ and $\bar{\eta}_2^+(\bar{t})/\bar{\eta}_2$ at different values of $\alpha$ and $\Lambda$ are plotted in Fig. \ref{Fig:Transient:PlaEx:StressGrowthFunctions}. It is seen that in the most cases of practical interest, the solutions are either  monotonic or nearly monotonic and qualitatively similar to those obtained for affine LPTT fluids,\cite{Shogin2020SLPTT} with
\begin{align}
\bar{\eta}^+_1 & \to 4\eta_0\left[1-\exp(-\bar{t})\right], \\
\bar{\eta}^+_2 & \to 2\eta_0\left[1-\exp(-\bar{t})\right]
\end{align}
at $\Lambda \to 0$. In the oscillatory regime, the frequency of the oscillations is bounded at any $\alpha$ and approaches zero at the edges of the region where it is real, as shown in Fig. \ref{Fig:Transient:PlaEx:Frequency}. Combined with the unbounded damping factor, $\Omega$, this leads to a strong damping of the oscillations. Of all the extrema, only the first maximum can be of observable magnitude, and even this is not always the case: The first overshoot is pronounced only when $\alpha$ is significantly large. 
\begin{figure}[htp]
\begin{center}
\includegraphics[width=3.37in]{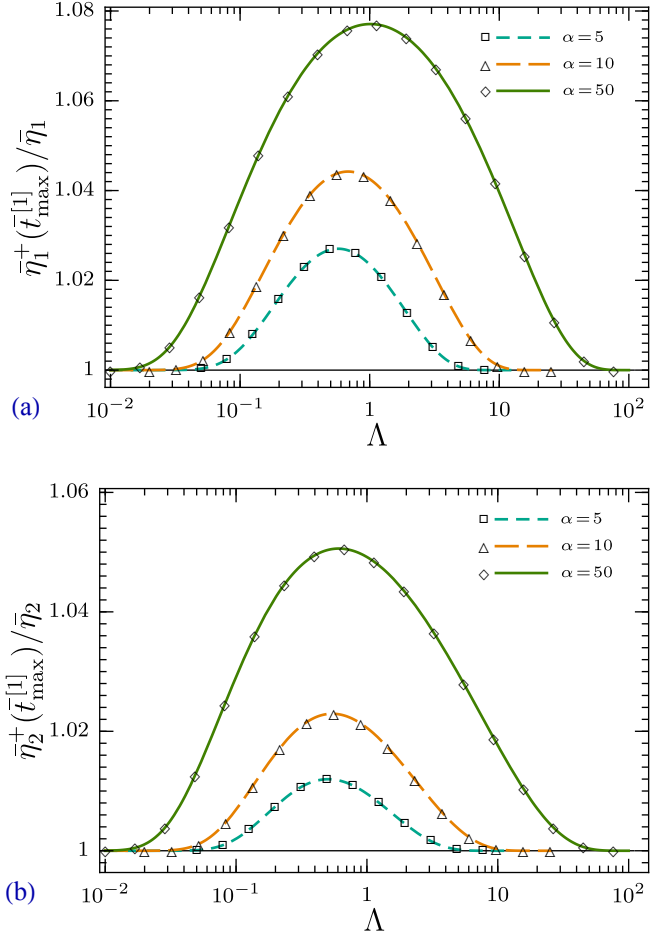}
\caption{\label{Fig:Transient:PlaEx:Overshoots} Stress overshoots at startup of steady planar extensional flow in the oscillatory regime, when form $\mathfrak{T}$ applies: the values of $\bar{\eta}_1^+(\bar{t})/\bar{\eta}_1$ (a) and $\bar{\eta}_2^+(\bar{t})/\bar{\eta}_2$ (b) at the moment of first overshoot. 
The numerical values (hollow symbols) were calculated using function \texttt{FindMaximum} in Wolfram Mathematica to the exact analytical solutions, while the approximate values (lines) were  obtained by substitution of Eq.  (\ref{Eq:Transient:PlaEx:FirstOvershootPosition}) into the form $\mathfrak{T}$. The approximate analytical results are in excellent agreement with the accurate numerical values. Large values of $\alpha$ are chosen for illustrative purposes.}
\end{center}
\end{figure}
Equations (\ref{Eq:Transient:Shear:EquilibriumPositions})  and (\ref{Eq:Transient:Shear:OvershootPositions}) still hold for the oscillatory regime, yielding
\begin{equation}
\label{Eq:Transient:PlaEx:FirstEquilibriumPosition}
\bar{t}_0^{\;[1]}  = \left\{ 
\renewcommand{\arraystretch}{1.5} \begin{array}{ll}
\dfrac{1}{\omega}\left[\pi - \arctan \dfrac{\omega}{a}\right]  & \quad \text{if} \quad a>0, \\
-\dfrac{1}{\omega}\arctan \dfrac{\omega}{a} & \quad \text{if} \quad a<0, \\
\dfrac{\pi}{2\omega} & \quad \text{if} \quad a=0
\end{array} \right.
\end{equation}
for the moment when the material functions reach their steady-flow values for the first time, and
\begin{equation}
\label{Eq:Transient:PlaEx:FirstOvershootPosition}
\bar{t}_\mathrm{max}^{\;[1]} \approx \dfrac{1}{\omega} \left[\pi + \arctan\dfrac{(a-\Omega)\omega}
{a\Omega+\omega^2}\right]
\end{equation}
for the first overshoots (for a particular stress growth function, the coefficient $a$ corresponding to this function must be chosen from the second column of Table \ref{Tab:a}). The approximate maximal values of the stress growth functions can be obtained by substituting Eq. (\ref{Eq:Transient:PlaEx:FirstOvershootPosition}) into the form $\mathfrak{T}$; the result is shown in Fig. \ref{Fig:Transient:PlaEx:Overshoots}.
\begin{table*}[htp]
\caption{
\label{Tab:FormsUBEx}
The conditions at which the stress growth functions describing startup of steady uniaxial and biaxial extensional flows take the forms $\mathfrak{H}$, $\mathfrak{E}^+$, and $\mathfrak{T}$.}
\begin{center}
\begin{ruledtabular}
\begin{tabular}{c c c}
 & Uniaxial extension & Biaxial extension  \\
 \hline \rule{0pt}{2em}
$\mathfrak{H}$ & 
$\begin{matrix}
\alpha < 1/3 \\
1/3\leq \alpha < 3/8 \text{ and } \Lambda<\Lambda_1^\ast \text{ or }\Lambda>\Lambda_2^\ast \\
\alpha \geq 1/3 \text{ and } \Lambda>\Lambda_2^\ast
\end{matrix}$ & 
$\begin{matrix}
\alpha \leq 3/8 \\
\alpha > 3/8 \text{ and } \Lambda>\Lambda_1^\ast
\end{matrix}$  \\ \hline \rule{0pt}{2em}
$\mathfrak{E}^+$ &
$\begin{matrix}
1/3 \leq \alpha < 3/8 \text{ and } \Lambda=\Lambda_1^\ast \text{ or }\Lambda=\Lambda_2^\ast\\
\alpha \geq 3/8 \text{ and } \Lambda=\Lambda_2^\ast 
\end{matrix}$ & 
$\alpha>3/8$ and $\Lambda=\Lambda_1^\ast$ \\ \hline \rule{0pt}{2em}
$\mathfrak{T}$ & 
$\begin{matrix}
1/3 < \alpha < 3/8 \text{ and } \Lambda_1^\ast<\Lambda<\Lambda_2^\ast \\
\alpha \geq 3/8 \text{ and } \Lambda<\Lambda_2^\ast 
\end{matrix}$ &
$\alpha>3/8$ and $\Lambda<\Lambda_1^\ast$
\end{tabular}
\end{ruledtabular}
\end{center}
\end{table*}
\par 
It should be noted that the amplitude of the oscillations increase with $\alpha$, but  at $\hT \to \infty$,
\begin{equation}
\left\{ \dfrac{\bar{\eta}_1^+(\bar{t}_\mathrm{max}^{\;[1]})}{\bar{\eta}_1},\dfrac{\bar{\eta}_2^+(\bar{t}_\mathrm{max}^{\;[1]})}{\bar{\eta}_2} \right\} \to 1+\sqrt{3}\exp\left[ -\dfrac{\sqrt{3}\pi}{2} \right] \approx 1.114.
\label{Eq:Transient:PlaEx:OvershootLimit}
\end{equation}
This number sets a restriction on the maximum values of the normalized stress growth functions. In a similar way, one can also show that for the values of the normalized material functions at the first undershoot,
\begin{equation}
 1-\sqrt{3}\exp\left[ -\dfrac{3\sqrt{3}\pi}{2} \right] <\left \{ \dfrac{\bar{\eta}_1^+(\bar{t}_\mathrm{min}^{\;[1]})}{\bar{\eta}_1}, \dfrac{\bar{\eta}_2^+(\bar{t}_\mathrm{min}^{\;[1]})}{\bar{\eta}_2} \right\} < 1,
\end{equation}
where the number on the left-hand side is approximately 0.9995. This confirms our earlier statement about the severely damped character of the oscillations.
\subsection{Startup of steady uniaxial and biaxial extensional flows
\label{Sec:Transient:UBEx}}
\begin{figure}[htp]
\begin{center}
\includegraphics[width=3.37in]{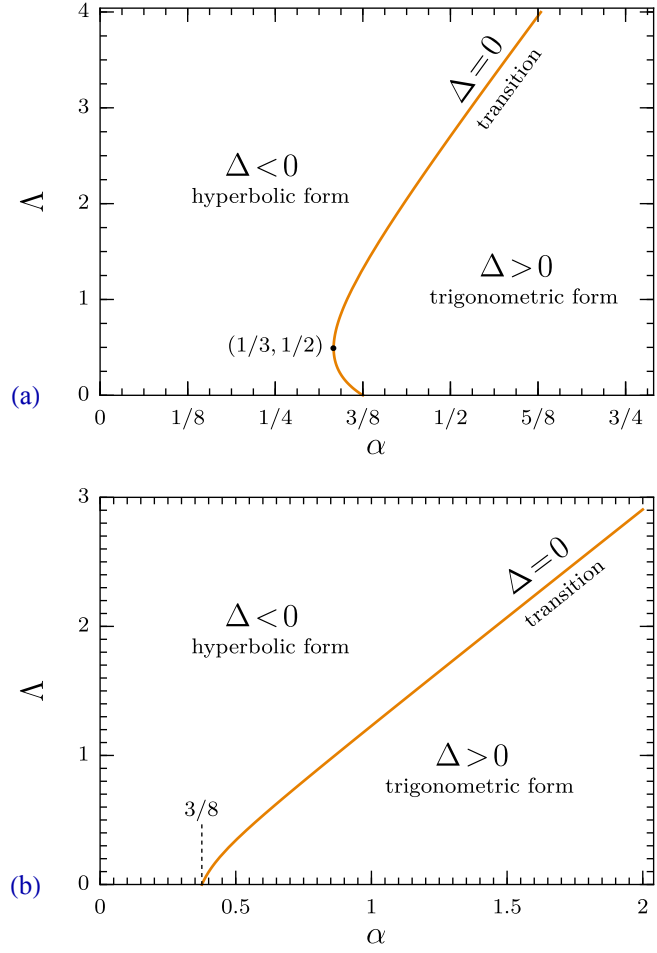}
\caption{\label{Fig:Transient:UBEx:FormMap} The regions  of the $(\alpha,\Lambda)$-plane where the stress growth functions describing startup of steady uniaxial (a) and biaxial (b) extensional flows take different forms. Form $\mathfrak{E}^+$ applies at the boundary lines where $\Delta=0$.}
\end{center}
\end{figure}
Similarly to the startup of steady planar extensional flow, the material functions in startup of steady uniaxial and biaxial extension can take any of the three forms $\mathfrak{H}$, $\mathfrak{E}^+$, and $\mathfrak{T}$, depending on the sign of $\Delta$ (see Appendixes \ref{App:Transient:UBEx} and \ref{App:FormShift:UBEx}). The functions $\omega$, $\Omega$, $K$, $A$, $B$, and $C$, needed to complete the forms, are found  in the third column of Table \ref{Tab:FormFunctions}, and the coefficients $a$ are provided in the third column of Table \ref{Tab:a}. The conditions at which each of the form applies are formulated in Table \ref{Tab:FormsUBEx}, where
\begin{equation}
\label{Eq:Transient:UBEx:LambdaStarUniaxial}
\Lambda^\ast_{1,2} = \dfrac{2\hT^\ast_{1,2}(1+\hT^\ast_{1,2})}{\sqrt{3\hT_{1,2}^\ast(8\alpha+3\hT^\ast_{1,2})} + \hT^\ast_{1,2}}
\end{equation}
\begin{figure}[htp]
\begin{center}
\includegraphics[width=6.45in]{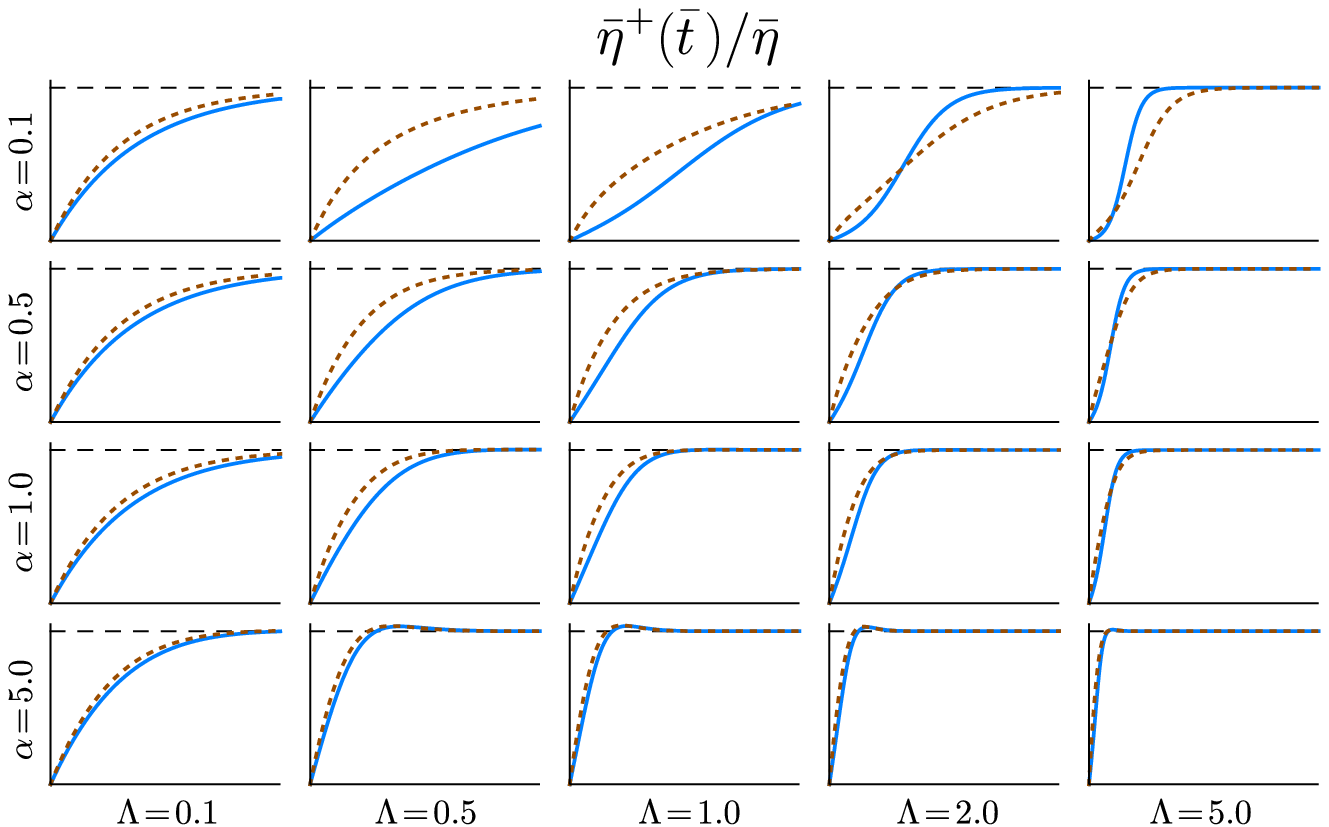}
\caption{\label{Fig:Transient:UBEx:StressGrowthFunctions} The normalized stress growth function $\bar{\eta}^+(\bar{t})/\bar{\eta}$, describing startup of steady uniaxial (solid lines) and biaxial (dashed lines) extensional flows. Each miniplot shows the region $[0,3]\times [0,1.05]$.}
\end{center}
\end{figure}
for uniaxial extension and 
\begin{equation}
\label{Eq:Transient:UBEx:LambdaStarBiaxial}
\Lambda^\ast_1 = \dfrac{2\hT^\ast_1(1+\hT^\ast_1)}{\sqrt{3\hT_1^\ast(8\alpha+3\hT^\ast_1)} - \hT^\ast_1}
\end{equation}
for biaxial extension, with 
\begin{align}
\hT^\ast_1 &= \dfrac{3}{4}(-3+8\alpha)+\dfrac{1}{2}\left[ \sqrt{l}-\sqrt{-l+2\left(p+\dfrac{q}{\sqrt{l}}\right)} \right], \\
\hT^\ast_2 &= \dfrac{3}{4}(-3+8\alpha)+\dfrac{1}{2}\left[ \sqrt{l}+\sqrt{-l+2\left(p+\dfrac{q}{\sqrt{l}}\right)} \right].
\end{align}
Here $p$ and $q$ are defined by
\begin{align}
p &= \dfrac{9}{8} (1-40\alpha+120\alpha^2 ), \\
q &= \dfrac{3}{2}\alpha(3\alpha-1)(168\alpha-31),
\end{align} 
while
\begin{equation}
l = \dfrac{2p}{3} 
+\sqrt[3]{Q+\sqrt{Q^2-P^3}}
+\sqrt[3]{Q-\sqrt{Q^2-P^3}},
\end{equation}
\begin{figure}[htp]
\begin{center}
\includegraphics[width=3.37in]{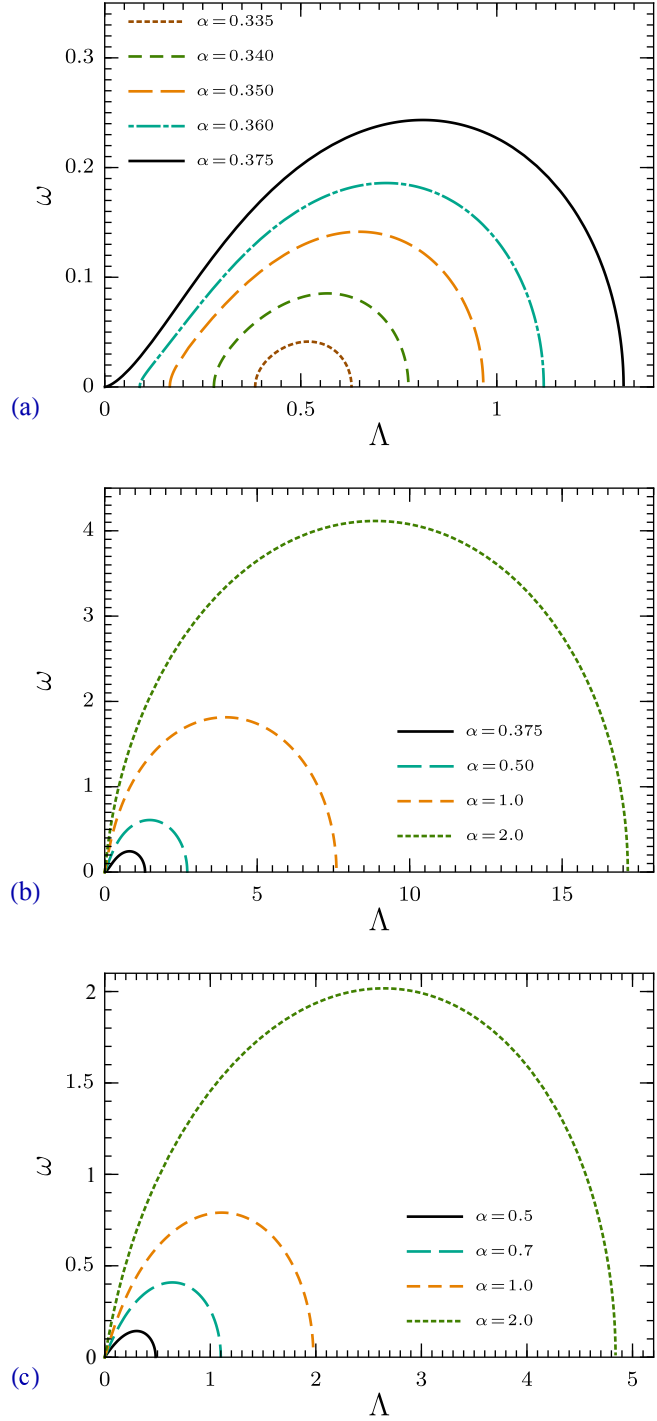}
\caption{\label{Fig:Transient:UBEx:Frequency} The frequency of the stress oscillations at startup of steady uniaxial [(a) and (b)]  and biaxial (c) extensional flows, when the stress growth functions take the trigonometric form $\mathfrak{T}$. The frequency, $\omega$, is
plotted as a function of the dimensionless shear rate, $\Lambda$, for different values of the effective extensional flow parameter, $\alpha$.}
\end{center}
\end{figure}
with
\begin{figure}[htp]
\begin{center}
\includegraphics[width=3.37in]{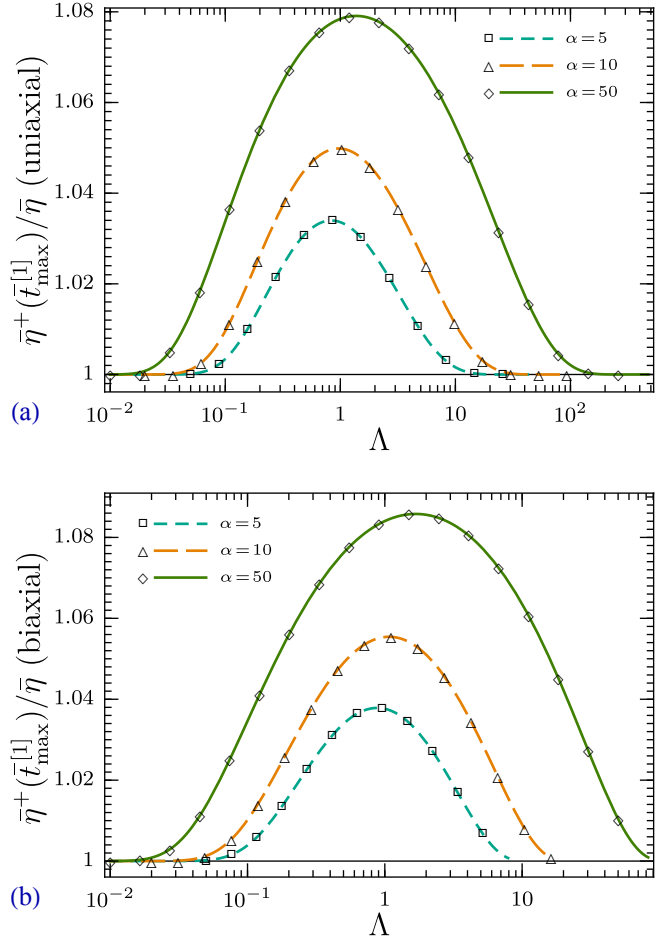}
\caption{\label{Fig:Transient:UBEx:Overshoots} Stress overshoots at startup of steady uniaxial (a) and biaxial (b) extensional flows in the oscillatory regime: the values of $\bar{\eta}^+(\bar{t})/\bar{\eta}$ at the moment of first overshoot. 
The numerical values (hollow symbols) were obtained using function \texttt{FindMaximum} of Wolfram Mathematica and are in a very good agreement with the approximate analytical values (lines). Large values of $\alpha$ are chosen for illustrative purposes.}
\end{center}
\end{figure}
\begin{align}
P &= \dfrac{9}{16}(1+8\alpha+56\alpha^2-672\alpha^3+1296\alpha^4), \\
Q &= \dfrac{9}{64}(3 + 36 \alpha - 700 \alpha^2 + 960 \alpha^3 + 23184 \alpha^4 - 108864 \alpha^5 + 139968 \alpha^6). \label{Eq:Transient:UBEx:Q}
\end{align}
The corresponding regions of the $(\alpha,\Lambda)$-plane are shown in Fig. \ref{Fig:Transient:UBEx:FormMap}.
\par 
The normalized stress growth functions $\bar{\eta}^+(\bar{t})/\bar{\eta}$ in startup of steady uniaxial and biaxial extensional flows are plotted in Fig. \ref{Fig:Transient:UBEx:StressGrowthFunctions} and share a lot of similarities with those related to startup of planar extension. At $\Lambda \to 0$,
\begin{equation}
\bar{\eta}^+ \to 3\eta_0 \left[1-\exp(-\bar{t})\right].
\end{equation}
When the solutions are oscillatory, the oscillations are damped very effectively:
At fixed $\alpha$, the frequency of the oscillations is bounded, as shown in Fig. \ref{Fig:Transient:UBEx:Frequency}, while the damping factor, $\Omega$, is not. As a result, the first overshoot is the only possibly pronounced extremum (in Fig. \ref{Fig:Transient:UBEx:StressGrowthFunctions}, it is seen  when $\alpha$ is significantly large). Note that the peculiar behavior of frequency at $1/3<\alpha<3/8$ in the uniaxial case [see Fig. \ref{Fig:Transient:UBEx:Frequency}(a)] does not lead to observable effects, since the frequency is very small at $\alpha$ in this range.
\par 
In the oscillatory regime, the moment when $\bar{\eta}^+(\bar{t})$ reaches its steady-flow value for the first time is still given by Eq. (\ref{Eq:Transient:PlaEx:FirstEquilibriumPosition}), which holds for both uniaxial and biaxial cases, provided that the corresponding values of $a$ are used (see the third column of Table \ref{Tab:a}). For the approximate positions of the first maximum, Eq. (\ref{Eq:Transient:PlaEx:FirstOvershootPosition}) holds for startup of uniaxial extension, while its analog for the biaxial case is
\begin{equation}
\label{Eq:Transient:UBEx:FirstOvershootPositionBiEX}
\bar{t}_\mathrm{max}^{\;[1]}  = \left\{ 
\renewcommand{\arraystretch}{1.5} \begin{array}{ll}
\dfrac{1}{\omega} \left[\arctan\dfrac{(a-\Omega)\omega}
{a\Omega+\omega^2}\right] & \quad \text{if} \quad a\Omega+\omega^2<0, \\

\dfrac{1}{\omega} \left[\pi + \arctan\dfrac{(a-\Omega)\omega}
{a\Omega+\omega^2}\right] & \quad \text{if} \quad a\Omega+\omega^2>0, \\
\dfrac{\pi}{2\omega} & \quad \text{if} \quad  a\Omega+\omega^2=0.
\end{array} \right.
\end{equation}
The approximate expressions for the values of the stress growth functions at the first maximum, obtained using Eqs. (\ref{Eq:Transient:PlaEx:FirstOvershootPosition}) and (\ref{Eq:Transient:UBEx:FirstOvershootPositionBiEX}) are very accurate, which is illustrated in Fig. \ref{Fig:Transient:UBEx:Overshoots}. It is seen that the amplitude of the oscillations increases with $\alpha$, but, as might be expected, the overshoot magnitude has an upper limit. This limit turns out to be the same for startup of uniaxial, biaxial, and planar extension [see the number on the right-hand side of Eq. (\ref{Eq:Transient:PlaEx:OvershootLimit})].
\subsection{Cessation of steady shear and extensional flows
\label{Sec:Transient:Cessation}}
All material functions in cessation of steady shear and extensional flows are similar to each other, take the form $\mathfrak{E}^-$ (see Table \ref{Tab:Forms}), and decrease monotonically, asymptotically approaching zero at late times. For cessation of shear flow, $\Psi_2^-(\bar{t})/\Psi_2=\Psi_1^-(\bar{t})/\Psi_1$ when $\chi \neq 0$ and $\Psi_2^-(\bar{t})=0$ identically when $\chi=0$.
\par 
At small strain rates, the decrease is asymptotically exponential,
\begin{equation}
\left\{
\dfrac { \eta^-} {\eta}, 
\dfrac { \Psi^-_1} {\Psi_1}, 
\dfrac {\bar{\eta}^-_1} {\bar{\eta}_1},
\dfrac {\bar{\eta}^-_2} {\bar{\eta}_2}, 
\dfrac {\bar{\eta}^-} {\bar{\eta}}
 \right\} \to \exp(-\bar{t}).
\end{equation}
\par
\begin{figure}[htp]
\begin{center}
\includegraphics[width=3.37in]{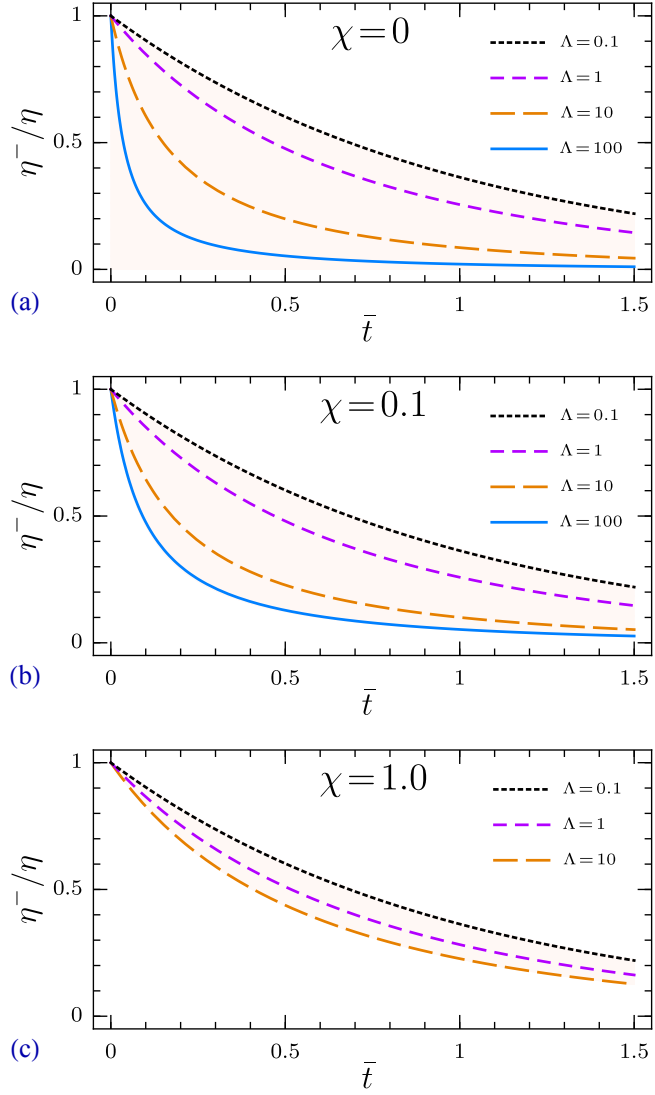}
\caption{\label{Fig:Transient:Cessation} (a) The normalized shear stress relaxation function $\eta^-(\bar{t})/\eta = \Psi^-_1(\bar{t})/\Psi_1$ of an affine LPTT fluid. It is seen that the graphs approach the coordinate axes as $\Lambda$ increases. (b) and (c) The normalized shear stress relaxation function $\eta^-(\bar{t})/\eta = \Psi^-_1(\bar{t})/\Psi_1=\Psi^-_2(\bar{t})/\Psi_2$ of non-affine LPTT fluids. As $\Lambda$ increases, the graphs first approach the attractor curve [the lower boundary of the colored region, see Eq. (\ref{Eq:Transient:Cessation:Attractor})] and then become insensitive to $\Lambda$.}
\end{center}
\end{figure}

 
When $\hT(\Lambda)$ is unbounded (extensional flows; shear flows of affine LPTT fluids), the decrease in the stress relaxation functions becomes steeper as $\Lambda$ increases; at large $\Lambda$, their graphs approach the coordinate axes. This can be seen in Fig. \ref{Fig:Transient:Cessation}(a); note that all plots lie in the colored region between the line $\exp (-\bar{t})$ and the coordinate axes.
\par 
The situation is different for cessation of shear flows of non-affine LPTT fluids, when $\hT(\Lambda)$ is bounded. In this case, the stress relaxation functions are less sensitive to $\Lambda$. At large $\Lambda$, their graphs approach the attractor curve, 
\begin{equation}
\label{Eq:Transient:Cessation:Attractor}
\left\{
\dfrac { \eta^-} {\eta}, 
\dfrac { \Psi^-_1} {\Psi_1},
\dfrac { \Psi^-_2} {\Psi_2}
 \right\}  \to \dfrac {\chi} {(1+\chi) \exp \bar{t}-1},
\end{equation}
as shown in Figs. \ref{Fig:Transient:Cessation}(b) and \ref{Fig:Transient:Cessation}(c); all plots lie within the colored region, restrained by the line $\exp(-\bar{t})$ and the attractor curve. Furthermore, this region becomes narrower as $\chi$ increases [compare Fig. \ref{Fig:Transient:Cessation}(b) to \ref{Fig:Transient:Cessation}(c)]; at large $\chi$,
\begin{equation}
\dfrac {\chi} {(1+\chi) \exp \bar{t}-1} \to \exp(-\bar{t}),
\end{equation}
and the region collapses into the curve $\exp(-\bar{t})$. In this limit, the stress relaxation functions become independent of $\Lambda$.
\section{Conclusion}
\label{Sec:Conclusion}
In this work, we have obtained exact analytical expressions description for the material functions of non-affine and affine  LPTT fluids in startup, cessation, and steady regimes of shear flows and of planar, uniaxial, and biaxial extension. Our formulas are simple, fully explicit, purely real, and as general as possible, since no additional restrictions on the model parameters were imposed. To the best of our knowledge, our results have no analogs in the literature.
\par 
Containing a detailed investigation of the properties of the aforementioned material functions, this work provides a complete analytical description of rheological properties of LPTT fluids in the most important simple rheometric flows. Given the classical status of the PTT models, we hope that this paper will be useful both to researchers and students in rheology and non-Newtonian fluid mechanics. 
\par 
Despite the relative simplicity of the mathematical methods implemented in this work, we believe that multiple technical aspects (e.g., reducing the number of the model parameters to one using a nonstandard scaling procedure, extensive application of cubic equations, and solving a system of coupled nonlinear differential equations by analytical means) deserve the attention of  mathematical physicists and, in general, of anyone interested in applied mathematics.
\section*{Supplementary material}
See the supplementary material for the Wolfram Mathematica codes verifying the key analytical results of this work. The interactive elements allow one to directly compare our explicit analytical expressions for the material functions to the numerical solutions of the corresponding systems of algebraic and differential equations at different values of $\chi$ (or $\alpha$), $\Lambda$, and $\bar{t}$.
\begin{acknowledgments}
This research has been funded by VISTA---a basic research program in collaboration between The Norwegian Academy of Science and Letters and Equinor. D.S. thanks Tamara and Mari Elisabeth Shogin for continuous support and numerous suggestions on the improvement of this paper.
\end{acknowledgments}

\begin{appendix}
\section{General solutions of cubic equations \label{App:Cubics}}
As seen from Sec. \ref{Sec:SteSol}, exact analytical solutions for steady flows rely heavily on finding positive real solutions of cubic equations of general form
\begin{equation}
x^3 + a_2 x^2 + a_1 x + a_0 = 0, \label{AEq:Cubics:Cubic}
\end{equation}
where the coefficients $a_0,a_1,a_2\in \mathbb{R}$ and $a_0 \neq 0$. The character of the roots of Eq. (\ref{AEq:Cubics:Cubic}) is determined by the sign of the discriminant,
\begin{equation}
D = 18 a_2 a_1 a_0 - 4 a_2^3 a_0 + a_2^2 a_1^2 - 4a_1^3-27a_0^2.
\label{AEq:Cubics:Discriminant}
\end{equation}
At $D \geq 0$, all three roots of Eq. (\ref{AEq:Cubics:Cubic}) are real (at $D \neq 0$, they are distinct), while at $D<0$, Eq. (\ref{AEq:Cubics:Cubic}) has only one real root, the other two being a couple of (non-real) complex conjugates. In either case, the roots of Eq. (\ref{AEq:Cubics:Cubic}) can be written  in compact form using two auxiliary quantities
\begin{align}
p &= \dfrac{1}{9}(3a_1-a_2^2), \label{AEq:Cubics:p}\\ 
q &= \dfrac{1}{54}(27a_0-9a_1a_2+2a_2^3), \label{AEq:Cubics:q}
\end{align}
with the sign of $q^2+p^3$ always being opposite to that of the discriminant.
\par 
In the case when Eq. (\ref{AEq:Cubics:Cubic}) has only one real root ($D<0$), this root can be expressed by Cardano's formula,
\begin{equation}
x = -\dfrac{a_2}{3}+\sqrt[3]{-q+\sqrt{q^2+p^3}}+\sqrt[3]{-q-\sqrt{q^2+p^3}}. \label{AEq:Cubics:RadicalForm}
\end{equation}
\par 
More generally, the three roots of Eq. (\ref{AEq:Cubics:Cubic}) can be written in trigonometric form,
\begin{equation}
x_j =-\dfrac{a_2}{3}+2\sqrt{-p}\cos \left[\dfrac{1}{3}\arccos \dfrac{q}{p\sqrt{-p}}-\dfrac{2\pi (j-1)}{3} \right],\quad j\in \{1,2,3 \}.
\label{AEq:Cubics:TrigSolution}
\end{equation}
If $D\geq 0$ (three real roots), expression (\ref{AEq:Cubics:TrigSolution}) is purely real. Furthermore, a straightforward evaluation of the cosine shows that $x_1 \geq x_2 \geq x_3$. If $D<0$, Eq. (\ref{AEq:Cubics:TrigSolution}) is still valid but involves complex functions; in calculations, the principal values of these functions should be used.
\par
From Vieta's formulas, it can be seen that the three roots of Eq. (\ref{AEq:Cubics:Cubic}) satisfy
\begin{align}
-a_0 &= x_1 x_2 x_3, \label{AEq:Cubics:VietaProduct}\\
a_1 &= x_1 x_2 + x_2 x_3 + x_3 x_1, \label{AEq:Cubics:VietaSumOfCrossProducts} \\
-a_2 &= x_1 + x_2 + x_3.\label{AEq:Cubics:VietaSum}
\end{align}
These relations are helpful when the signs of the (real) roots are to be determined. In particular, it follows immediately from Eq. (\ref{AEq:Cubics:VietaProduct}) that in the case of one real root, the sign of this root is identical to that of $(-a_0)$.
\section{Solution of cubic equations for $\hT$ \label{App:CubicSolutions}}
\subsection{Steady shear flow \label{App:CubicSolutions:Shear}}
Equation (\ref{Eq:SteSol:Shear:Cubic}) is of form (\ref{AEq:Cubics:Cubic}), with $a_0=-2\Lambda^2$, $a_1=1+2\chi \Lambda^2$, and $a_2=2$.
Its discriminant [see Eq. (\ref{AEq:Cubics:Discriminant})] can be written as
\begin{equation}
D = -4\Lambda^2\left[2(1+\chi) +(27+36\chi+8\chi^2)\Lambda^2+8\chi^3\Lambda^4\right]
\end{equation}
and is clearly negative. Equation (\ref{Eq:SteSol:Shear:Cubic}) thus has one real root, which is positive, since $a_0<0$. Having applied formulas (\ref{AEq:Cubics:p})-–(\ref{AEq:Cubics:RadicalForm}), one obtains Eqs. (\ref{Eq:SteSol:Shear:MainResultT})-–(\ref{Eq:SteSol:Shear:Q}).
\subsection{Steady planar extensional flow \label{App:CubicSolutions:PlaEx}}
Equation (\ref{Eq:SteSol:PlaEx:Cubic}) is a cubic equation of form (\ref{AEq:Cubics:Cubic}), with $a_0=-u$, $a_1=1-v$, and $a_2=2$, where
$u = 8\alpha \Lambda^2$ and $v = 4\Lambda^2$. Then, the discriminant of Eq. (\ref{Eq:SteSol:PlaEx:Cubic}) can be written as a quadratic polynomial in $u$,
\begin{equation}
D = -27 u^2 + 4u(9v-1)+4v(v-1)^2,
\end{equation}
which can take both positive and negative values at $u,v>0$. Therefore, either one or all three roots of Eq. (\ref{Eq:SteSol:PlaEx:Cubic}) are real, depending on the exact values of $\alpha$ and $\Lambda$. 
\par 
In the case of three real roots, only one of them is positive. This can be seen by applying Eqs. (\ref{AEq:Cubics:VietaProduct}) and (\ref{AEq:Cubics:VietaSum}): The product of the roots is positive (meaning that either one or all three roots are positive), while their sum is negative (which rules out the possibility of three positive roots). Then, the positive root is clearly the largest. Hence, it is given by Eq. (\ref{AEq:Cubics:TrigSolution}) with $j=1$, which together with Eqs. (\ref{AEq:Cubics:p}) and (\ref{AEq:Cubics:q}) leads to Eqs. (\ref{Eq:SteSol:PlaEx:MainResultTExplicit}) (first line),  (\ref{Eq:SteSol:PlaEx:P}), and (\ref{Eq:SteSol:PlaEx:Q}).
\par 
In the case of one real root, this root is positive, since $a_0<0$. It can still be found using Eq. (\ref{AEq:Cubics:TrigSolution}) with $j=1$ (using other values of $j$ leads to cosines of  complex numbers with nonvanishing real and imaginary parts, hence to complex results, while $j=1$ leads to a cosine of a purely imaginary quantity, thus to a real result). Therefore, the solution is still given by the first line of Eq. (\ref{Eq:SteSol:PlaEx:MainResultTExplicit}). Nevertheless, an alternative form of this result needs to be derived in order to eliminate complex functions from the final expression.
\par 
First, one demonstrates that
\begin{equation}
 P\sqrt{P}+Q=(1+3v)^{3/2}+(1-9v)+27u/2>(1+3v)^{3/2}+(1-9v)\geq 0. \label{AEq:CubicSolutions:PlaEx:Inequality}
\end{equation}
At $v\leq 1/9$, the last inequality is obvious. At $v>1/9$,
\begin{align}
(1+3v)^{3/2}+(1-9v) = \dfrac{(1+3v)^3-(9v-1)^2}{(1+3v)^{3/2}+(9v-1)}=\dfrac{27v(v-1)^2}{(1+3v)^{3/2}+(9v-1)}\geq 0,
\end{align}
which proves the inequality (\ref{AEq:CubicSolutions:PlaEx:Inequality}). Therefore, $Q/P\sqrt{P} > -1$, and the discriminant of Eq. (\ref{Eq:SteSol:PlaEx:MainResultTExplicit}) is negative only when $Q/P\sqrt{P}>1$. In this case, one proceeds with the following transformations (capitalized function names indicate  the principal values):
\begin{align}
\mathrm{Arccos}\,\dfrac{Q}{P\sqrt{P}} &= -i\,\mathrm{Ln}\, \left\{\dfrac{Q}{P\sqrt{P}}+i\left|1-\dfrac{Q^2}{P^3} \right|^{1/2}\exp\left[{\dfrac{i}{2}\mathrm{Arg}\left(1-\dfrac{Q^2}{P^3}\right)}\right] \right\} \nonumber \\
& = -i\, \mathrm{Ln}\,\left(\dfrac{Q}{P\sqrt{P}}+i\,\dfrac{\sqrt{Q^2-P^3}}{P\sqrt{P}}e^{i\pi/2} \right) \nonumber \\
&=-i \,\mathrm{Ln}\,\dfrac{Q-\sqrt{Q^2-P^3}}{P\sqrt{P}}=i \ln \dfrac{Q+\sqrt{Q^2-P^3}}{P\sqrt{P}}.
\label{AEq:CubicSolutions:PlaEx:FormConversion}
\end{align}
Finally, 
\begin{equation}
\cos \left( \dfrac{1}{3}\mathrm{Arccos}\,\dfrac{Q}{P\sqrt{P}}\right)=
\cosh \left( \dfrac{1}{3} \ln \dfrac{Q+\sqrt{Q^2-P^3}}{P\sqrt{P}} 
\right),
\end{equation}
which leads to a purely real expression for the real solution [Eq. (\ref{Eq:SteSol:PlaEx:MainResultTExplicit}), second line].
\subsection{Steady uniaxial and biaxial extensional flows \label{App:CubicSolutions:UBEx}}
Equation (\ref{Eq:SteSol:UBEx:Cubic}) is of form (\ref{AEq:Cubics:Cubic}), with $a_0=-w$, $a_1=1\mp\Lambda -2\Lambda^2$, and $a_2=2\mp \Lambda$, where $w=6\alpha \Lambda^2>0$. The discriminant of Eq. (\ref{Eq:SteSol:UBEx:Cubic}), calculated using Eq. (\ref{AEq:Cubics:Discriminant}), can be written as a quadratic polynomial in $w$,
\begin{equation}
D = -27w^2-2(2\mp\Lambda)(1\mp5\Lambda)(1\pm 4\Lambda)w+9\Lambda^2(1\pm\Lambda)^2(1\mp2\Lambda)^2,
\end{equation}
which can take both positive and negative values at $w,\Lambda>0$, depending on the values of $\alpha$ and $\Lambda$. Thus, Eq. (\ref{Eq:SteSol:UBEx:Cubic}) can have one or three real roots.
\par 
In the case of three real roots, their product is positive, since $a_0<0$ [see Eq. (\ref{AEq:Cubics:VietaProduct})]; therefore, either one or all three roots are positive. For uniaxial extension, either the sum of the roots [$-(2-\Lambda)$, see Eq. (\ref{AEq:Cubics:VietaSum})] or sum of their products [$1-\Lambda-2\Lambda^2$, see Eq. (\ref{AEq:Cubics:VietaSumOfCrossProducts})] is negative, which rules out the possibility of three real roots. For biaxial extension, the sum of the roots [$-(2+\Lambda)$, see Eq. (\ref{AEq:Cubics:VietaSum})] is negative, leading to the same result. Thus, if Eq. (\ref{Eq:SteSol:UBEx:Cubic}) has three real roots, only one of them is positive. Then, the positive root must be the largest of three; hence, it is calculated using Eq. (\ref{AEq:Cubics:TrigSolution}) with $j=1$. Having applied  Eqs. (\ref{AEq:Cubics:TrigSolution}), (\ref{AEq:Cubics:p}), and (\ref{AEq:Cubics:q}), one arrives at Eqs. (\ref{Eq:SteSol:UBEx:MainResultTExplicit}) (first line), (\ref{Eq:SteSol:UBEx:P}), and (\ref{Eq:SteSol:UBEx:Q}).
\par 
In case of one real root, this root is positive ($a_0<0$). Similarly to the case of planar extension, the root can also be found using the first line of Eq. (\ref{Eq:SteSol:UBEx:MainResultTExplicit}), but the process of calculation would involve complex functions. However, one observes that 
\begin{equation}
2P\sqrt{P}+Q > 2(1\mp \Lambda+7\Lambda^2)^{3/2}+2\mp3\Lambda-39\Lambda^2 
\pm20\Lambda^3 \geq 0.
\end{equation} 
This inequality is proven by showing that 
\begin{equation}
2(1\mp \Lambda+7\Lambda^2)^{3/2} \geq \vert 2\mp3\Lambda-39\Lambda^2 
\pm20\Lambda^3 \vert,
\end{equation}
which, is true, since
\begin{equation}
4(1\mp \Lambda +7\Lambda^2)^3-(2\mp3\Lambda-39\Lambda^2 
\pm20\Lambda^3)^2 = 243\Lambda^2(1\mp \Lambda -2\Lambda^2)^2 \geq 0.
\end{equation}
Therefore, the discriminant of Eq. (\ref{Eq:SteSol:UBEx:Cubic}) is negative only when $Q/2P\sqrt{P}>1$. Following the procedure described in Sec. \ref{App:CubicSolutions:PlaEx} [see Eq. (\ref{AEq:CubicSolutions:PlaEx:FormConversion})] step-by-step, one shows that
\begin{equation}
\cos\left(\dfrac{1}{3} \mathrm{Arccos}\,\dfrac{Q}{2P\sqrt{P}} \right) = \cosh \left( \dfrac{1}{3}\ln \dfrac{Q+\sqrt{Q^2-4P^3}}{2P\sqrt{P}}\right),
\end{equation}
which leads to Eq. (\ref{Eq:SteSol:UBEx:MainResultTExplicit}) (second line).
\section{Monotonic increase of $\hT(\Lambda)$ in steady flows
\label{App:Bijectivity}}
The purpose of this section is to prove that $\hT(\Lambda)$ is a strictly increasing function for all steady flows considered in this work. The proof shall be conducted as follows: Instead of investigating the explicit formulas [Eqs. (\ref{Eq:SteSol:Shear:MainResultT}),  (\ref{Eq:SteSol:PlaEx:MainResultTExplicit}), and (\ref{Eq:SteSol:UBEx:MainResultTExplicit})] directly, we shall demonstrate that the inverse functions, $\Lambda(\hT)$, are well-defined and monotonically increasing with $\hT$.
\par 
Equations (\ref{Eq:SteSol:Shear:Cubic}), (\ref{Eq:SteSol:PlaEx:Cubic}), and (\ref{Eq:SteSol:UBEx:Cubic}) are all quadratic with respect to the dimensionless strain rate and can be solved for $\Lambda$. For each of these equations, the solutions satisfying $\Lambda>0$ at $\hT>0$ must be chosen.
\par 
For shear flow [Eq. (\ref{Eq:SteSol:Shear:Cubic})], one gets
\begin{equation}
\label{AEq:Bijectivity:Shear:LambdaOfHatT}
\Lambda (\hT)=
(1+\hT)
\sqrt{
\dfrac{\hT}{2(1-\chi \hT)}}.
\end{equation}
One immediately observes that $\hT(\Lambda)$ is bounded at $\chi \neq 0$ [since the right-hand side of Eq. (\ref{AEq:Bijectivity:Shear:LambdaOfHatT}) implies $0<\chi\hT<1$] and unbounded at $\chi=0$. Differentiating Eq. (\ref{AEq:Bijectivity:Shear:LambdaOfHatT}) with respect to $\hT$, one obtains
\begin{equation}
\label{AEq:Bijectivity:Shear:DLambdaDHatT}
\der{\Lambda}{\hT}=\dfrac{1 +  \hT (3-2\chi\hT)}{2\sqrt{2\hT(1-\chi\hT)^3}},
\end{equation}
which is positive at $0\leq\chi\hT<1$.
\par  
For planar extensional flow [Eq. (\ref{Eq:SteSol:PlaEx:Cubic})], one gets
\begin{equation}
\label{AEq:Bijectivity:PlaEx:LambdaOfHatT}
\Lambda(\hT) = \dfrac{1+\hT}{2} \sqrt{\dfrac{\hT}{ 2\alpha +\hT}},
\end{equation}
the derivative of which,
\begin{equation}
\der{\Lambda}{\hT} = \dfrac{\alpha+3\alpha\hT+ \hT^2}{2 \sqrt{\hT (2\alpha+\hT)^3}},
\end{equation}
is clearly positive.
\par 
Finally, for uniaxial and biaxial extensional flows [Eq. (\ref{Eq:SteSol:UBEx:Cubic}), with upper and lower signs chosen, respectively],
\begin{equation}
\label{AEq:Bijectivity:UBEx:LambdaOfHatT}
\Lambda(\hT) = \dfrac{2\hT(1+\hT)}{Y \pm \hT},
\end{equation}
where 
\begin{equation}
Y = \sqrt{3\hT(8\alpha+3\hT)}>0.
\label{AEq:Bijectivity:UBEx:Y}
\end{equation}
Differentiating Eq. (\ref{AEq:Bijectivity:UBEx:LambdaOfHatT}) with respect to $\hT$, one arrives after some algebraic transformations at
\begin{equation}
\der{\Lambda}{\hT} = \dfrac{2\hT \left[12\alpha(1+3\hT)+\hT(9\hT \pm Y) \right]}{Y (Y \pm \hT )^2},
\label{AEq:Bijectivity:UBEx:DLambdaDHatTBiaxial}
\end{equation}
For uniaxial extensional flow [upper signs in Eq. (\ref{AEq:Bijectivity:UBEx:DLambdaDHatTBiaxial})], the positivity of $\lder{\Lambda}{\hT}$ is obvious. For biaxial extension [lower signs in Eq. (\ref{AEq:Bijectivity:UBEx:DLambdaDHatTBiaxial})], the sign of $\lder{\Lambda}{\hT}$ is identical to that of the expression in square brackets [see the numerator on the right-hand side of Eq. (\ref{AEq:Bijectivity:UBEx:DLambdaDHatTBiaxial})]. Having multiplied this expression by its positive conjugate and made simple rearrangements, one obtains
\begin{equation}
(12\alpha+36\alpha \hT+9 \hT^2)^2-(\hT Y)^2 = 24 \left[6\alpha^2(1+3\hT)^2+\alpha \hT^2(9+26\hT)+3\hT^4 \right],
\end{equation}
which is positive. Therefore, $\lder{\Lambda}{\hT}>0$ also for biaxial extension, which completes the proof of the original statement.
\section{Monotonic properties of stresses and material functions in steady flows{\label{App:Shapes}}}
As demonstrated in Appendix \ref{App:Bijectivity}, $\hT(\Lambda)$ is a bijective relation for all steady shear and extensional flows considered in this work. Then, with an appropriate variable change, an arbitrary function $f(\Lambda)$ can be rewritten as a function of $\hT$ instead. By the chain rule of differentiation, one obtains
\begin{equation}
\der{f}{\hT}
= \left( \der{\Lambda}{\hT} \right) \der{f}{\Lambda},
\end{equation}
and, since $\lder{\Lambda}{\hT}>0$ (see Appendix \ref{App:Bijectivity}), $f$ increases (decreases) with $\Lambda$ if and only if $f$ increases (decreases) with $\hT$. In the following, this property shall be used to investigate the monotonicity of stresses and material functions in steady shear and extensional flows.
\subsection{Shear flow \label{App:Shapes:Shear}}
For the dimensionless shear stress, 
\begin{equation}
\der{\hs}{\hT} = \der{}{\hT}\dfrac{( 1 + \hT ) \hT }{2\Lambda} =\dfrac{1-2\chi \hT}{2\sqrt{2\hT(1-\chi\hT)}}, \label{AEq:Shapes:Shear:DHatSDHatT}
\end{equation}
which is strictly positive when $\chi=0$ and changes its sign from “$+$” to “$-$” at
\begin{equation}
\label{AEq:Shapes:Shear:TatLambdaC} 
\chi\hat{\mathbb{T}} = 1/2
\end{equation}
when $\chi \neq 0$;
thus, shear stress maximizes at this point. Inserting Eq. (\ref{AEq:Shapes:Shear:TatLambdaC}) into Eq. (\ref{AEq:Bijectivity:Shear:LambdaOfHatT})
 yields Eq. (\ref{Eq:SteSol:Shear:DefinitionLambdaFirst}), while substituting Eqs. (\ref{AEq:Shapes:Shear:TatLambdaC}) and (\ref{Eq:SteSol:Shear:DefinitionLambdaFirst}) into Eq. (\ref{Eq:SteSol:Shear:MainResultS}) results in Eq. (\ref{Eq:SteSol:Shear:MaximumHatS}).
\par 
For the material functions, one gets
\begin{align}
\der{\eta}{\hT} & \propto \der{}{\hT}  \dfrac{( 1 + \hT ) \hT }{\Lambda^2(\hT)}= -\dfrac{2(1+\chi)}{(1+\hT)^2}, 
\label{AEq:Shapes:Shear:DSDHatT}\\
\der{\Psi_1}{\hT} & \propto \der{}{\hT} \dfrac{\hT}{\Lambda^2(\hT)} = -\dfrac{2(2+\chi-\chi \hT)}{(1+\hT)^3}.
\label{AEq:Shapes:Shear:DCheckTDHatT}
\end{align}
Both derivatives are clearly negative (recall that $0\leq \chi\hT<1$); therefore, $\eta$ and $\Psi_1$ are decreasing as $\Lambda$ increases.
\subsection{Planar extension \label{App:Shapes:PlaEx}}
For the dimensionless normal stress differences, one obtains
\begin{align}
\der{\hnen}{\hT} & = \der{}{\hT}  \dfrac{( 1 + \hT ) \hT }{2 \Lambda(\hT)}=  \dfrac{\alpha+ \hT}{\sqrt{\hT (2\alpha+ \hT)}}, 
\label{AEq:Shapes:PlaEx:DHatN1DHatT}\\
\der{\hnto}{\hT} &=\dfrac{1}{2}\der{}{\hT} 
\left(\hT + \dfrac{(1+\hT)\hT}{2\Lambda(\hT)} \right) = \dfrac{1}{2} \left[1+\dfrac{\alpha+\hT}{\sqrt{\hT(2\alpha+\hT)}} \right].
\label{AEq:Shapes:PlaEx:DHatN2DHatT}
\end{align}
These derivatives are obviously positive: Both dimensionless normal stress differences increase with $\Lambda$.
\par 
For the first extensional viscosity,
\begin{equation}
\der{\bar{\eta}_1}{\hT} \propto \der{}{\hT}  \dfrac{\hnen(\hT) }{\Lambda(\hT)}= \dfrac{ 2 (1-2\alpha) }{ (1+\hT)^2 }. 
\label{AEq:Shapes:PlaEx:DN1DHatT}
\end{equation}
This derivative changes its sign from “$+$” to “$-$” at $\alpha=1/2$. Therefore, $\bar{\eta}_1$ increases with $\Lambda$ at $\alpha<1/2$, is independent of $\Lambda$ at $\alpha = 1/2$, and decreases with $\Lambda$ at $\alpha>1/2$.
\par 
For the second extensional viscosity,
\begin{equation}
\der{\bar{\eta}_2}{\hT} \propto 
\der{}{\hT} \dfrac{\hnto(\hT)}{\Lambda(\hT)} = \dfrac{1}{(1+\hT)^2}\left[ 1-2\alpha+\dfrac{1+(1-\alpha)\hT}{\sqrt{\hT(2\alpha+\hT)}} \right].
\label{AEq:Shapes:PlaEx:DN2DHatT}
\end{equation}
At $\alpha\leq 1/2$, the expression in the square brackets [see the right-hand side of Eq. (\ref{AEq:Shapes:PlaEx:DN2DHatT})] is positive. At $\alpha>1/2$, it can change sign when
\begin{equation}
(2-3\alpha)\hT^2+2\alpha(3-4\alpha)\hT+\alpha=0,
\label{AEq:Shapes:PlaEx:DN2DHatTEqualsZero}
\end{equation}
provided that the root of Eq. (\ref{AEq:Shapes:PlaEx:DN2DHatTEqualsZero}) meets the requirement
\begin{equation} 
\label{AEq:Shapes:PlaEx:RootCheckCriterion}
1+(1-\alpha)\hT>0.
\end{equation}
Using the standard properties of quadratic polynomials, one can show that at Eq. (\ref{AEq:Shapes:PlaEx:DN2DHatTEqualsZero}) has no positive solutions at $\alpha \leq 2/3$, while at $\alpha>2/3$, it has one positive solution, 
\begin{equation}
\hT_\mathrm{max} = \dfrac{\alpha(3-4\alpha)+\sqrt{2\alpha(2\alpha-1)^3}}{\alpha(3\alpha-2)},
\label{AEq:Shapes:PlaEx:maxN2-HatTValue}
\end{equation}
which satisfies the condition (\ref{AEq:Shapes:PlaEx:RootCheckCriterion}). Then, a direct sign check shows that $\lder{\bar{\eta}_2}{\hT}>0$ at $\alpha \leq 2/3$, while at $\alpha>2/3$, this derivative changes its sign from “$+$” to “$-$” at $\hT=\hT_\mathrm{max}$. Inserting Eq. (\ref{AEq:Shapes:PlaEx:maxN2-HatTValue}) into Eq. (\ref{AEq:Bijectivity:PlaEx:LambdaOfHatT}) leads to Eqs. (\ref{Eq:SteSol:PlaEx:maxN2-Lambda}) and (\ref{Eq:SteSol:PlaEx:maxN2-A}).
The value of $\bar{\eta}_2$ at the point of maximum, Eq. (\ref{Eq:SteSol:PlaEx:maxN2-N2}), is obtained by subsequent substitution of Eqs. (\ref{Eq:SteSol:PlaEx:MainResultN1}), (\ref{AEq:Shapes:PlaEx:maxN2-HatTValue}), and (\ref{Eq:SteSol:PlaEx:maxN2-Lambda}) into Eq. (\ref{Eq:SteSol:PlaEx:MainResultN2}) using Eq. (\ref{Eq:SteSol:PlaEx:maxN2-A}) and Table \ref{Tab:MatFunctionsVsDimlessVars}.
\subsection{Uniaxial and biaxial extensional flows \label{App:Shapes:UBEx}}
For the dimensionless first normal stress difference,
\begin{equation}
\dfrac{\mathrm{d} \hnen}{\mathrm{d}\hT}  = \dfrac{\mathrm{d}}{\mathrm{d}\hT}  \dfrac{( 1 + \hT ) \hT }{2 \Lambda(\hT)}=  \dfrac{12\alpha+9\hT \pm Y}{4Y}, 
\label{AEq:Shapes:UBEx:DHatN1DHatT}
\end{equation}
where $Y>0$ is defined earlier by Eq. (\ref{AEq:Bijectivity:UBEx:Y}). For uniaxial extension (upper sign), it is easy to see that $\lder{\hnen}{\hT}>0$. For biaxial extension (lower sign), the same result can be shown by the following transformations:
\begin{equation}
\dfrac{\mathrm{d} \hnen}{\mathrm{d}\hT} = \dfrac{(12\alpha+9 \hT)^2-Y^2}{4Y(12\alpha+9 \hT+Y)} = \dfrac{6(6\alpha^2+8\alpha \hT+3\hT^2)}{Y(12\alpha+9 \hT + Y)}>0.
\end{equation}
Thus, the dimensionless first normal stress difference increases with $\Lambda$ in both uniaxial and biaxial extensional flows.
\par 
For the extensional viscosity, one gets
\begin{equation}
\der{\bar{\eta}}{\hT} \propto \der{}{\hT}  \dfrac{( 1 + \hT ) \hT }{2\Lambda^2(\hT)}= \dfrac{\left[3(3-8\alpha)\hT \pm Y\right](Y \pm \hT)}{8\hT (1+\hT)^2Y}. 
\label{AEq:Shapes:UBEx:DN1DHatT}
\end{equation}
Since $Y\pm \hT>0$, the sign of $\lder{\bar{\eta}}{\hT}$ is identical to that of $\left[3(3-8\alpha)\hT \pm Y\right]$.
\par 
For uniaxial extension (upper signs), this expression is clearly positive when $\alpha \leq 3/8$. If $\alpha>3/8$, one uses the transformation
\begin{equation}
\label{AEq:Shapes:UBEx:ConjugateMultiplication}
Y-3(8\alpha-3)\hT = \dfrac{Y^2-9(8\alpha-3)^2\hT^2}{Y+3(8\alpha-3)\hT} = \dfrac{24\hT\left[\alpha-3(2\alpha-1)(4\alpha-1)\hT\right]}{Y+3\hT(8\alpha-3)}.
\end{equation}
The numerator changes its sign from positive to negative at
\begin{equation}
\label{AEq:Shapes:UBEx:maxN1-hatT}
\hT_\mathrm{max} = \dfrac{\alpha}{3(2\alpha-1)(4\alpha-1)}
\end{equation}
at $\alpha >1/2$ and is strictly positive when $3/8 < \alpha \leq 1/2$. Therefore, uniaxial extensional viscosity is increasing monotonically with $\Lambda$ when $\alpha \leq 1/2$ and goes through a maximum when $\alpha>1/2$.
\par 
For biaxial extension [lower signs in Eq. (\ref{AEq:Shapes:UBEx:DN1DHatT})], it is easy to see that $(3-8\alpha)\hT-Y<0$ at $\alpha>3/8$. When $\alpha \leq 3/8$, the analog of Eq. (\ref{AEq:Shapes:UBEx:ConjugateMultiplication}) is 
\begin{equation}
3(3-8\alpha)\hT - Y = \dfrac{9(3-8\alpha)^2\hT^2-Y^2}{3(3-8\alpha)\hT+Y}=\dfrac{24\hT \left[ 3(1-2\alpha)(1-4\alpha)\hT-\alpha\right]}{3(3\alpha-8)\hT+Y}.
\end{equation}
The denominator of this expression is positive; the numerator is negative at $1/4 \leq \alpha \leq 3/8$ and changes its sign from negative to positive at 
\begin{equation}
\label{AEq:Shapes:UBEx:minN1-hatT}
\hT_\mathrm{min} = \dfrac{\alpha}{3(1-2\alpha)(1-4\alpha)}
\end{equation}
when $\alpha<1/4$. Thus, biaxial extensional viscosity decreases monotonically if $\alpha \geq 1/4$ and goes through a minimum if $\alpha <1/4$.
\par 
Substituting Eqs. (\ref{AEq:Shapes:UBEx:maxN1-hatT}) and (\ref{AEq:Shapes:UBEx:minN1-hatT}) into Eq. (\ref{AEq:Bijectivity:UBEx:LambdaOfHatT}) yields Eqs. (\ref{Eq:SteSol:UBEx:N1UniMaxLambda}) and (\ref{Eq:SteSol:UBEx:N1BiMinLambda}), respectively. Then, having substituted Eqs. (\ref{Eq:SteSol:UBEx:N1UniMaxLambda}) and (\ref{AEq:Shapes:UBEx:maxN1-hatT}) into Eq. (\ref{Eq:SteSol:UBEx:MainResultN1}) and applied Table \ref{Tab:MatFunctionsVsDimlessVars}, one arrives at Eq. (\ref{Eq:SteSol:UBEx:N1UniMaxValue}). Equation  (\ref{Eq:SteSol:UBEx:N1BiMinValue}) is obtained similarly by using Eqs. (\ref{Eq:SteSol:UBEx:N1BiMinLambda}) and (\ref{AEq:Shapes:UBEx:minN1-hatT}) in place of Eqs. (\ref{Eq:SteSol:UBEx:N1UniMaxLambda}) and (\ref{AEq:Shapes:UBEx:maxN1-hatT}).
\section{
\label{App:Transient}
Derivation of the analytical solutions for startup flows}
\subsection{
\label{App:Transient:Shear}
Start-up of steady shear flow}
The approach to solving Eqs. (\ref{Eq:Eqs:Shear:1TEvolution}) and (\ref{Eq:Eqs:Shear:2SEvolution}) is based on the method developed in our earlier work.\cite{Shogin2020SLPTT} The variables $\hT^+$ and $\hs^+$ are replaced with their deviations from steady-flow values, $\tilde{\mathbb{T}}$ and $\tilde{\mathbb{S}}$, respectively,
\begin{align}
\tilde{\mathbb{T}}(\bar{t}) &= \hT-\hT^+(\bar{t}), \\
\tilde{\mathbb{S}}(\bar{t})  &= \hs - \hs^+(\bar{t}).
\end{align}
Then, Eqs. (\ref{Eq:Eqs:Shear:1TEvolution}) and (\ref{Eq:Eqs:Shear:2SEvolution}) become
\begin{align}
\der{\tilde{\mathbb{T}}(\bar{t})}{\bar{t}} & = -(1 + 2 \hT)\tilde{\mathbb{T}}(\bar{t}) +2 \Lambda \tilde{\mathbb{S}}(\bar{t})   + \tilde{\mathbb{T}}^2(\bar{t}), \label{AEq:Transient:Shear:1TEvolution}\\
\der{\tilde{\mathbb{S}}(\bar{t})}{\bar{t}} &= - (1 + \hT)\tilde{\mathbb{S}}(\bar{t}) - \left(\hs+\chi\Lambda \right) \tilde{\mathbb{T}}(\bar{t}) + \tilde{\mathbb{S}}(\bar{t})\tilde{\mathbb{T}}(\bar{t}), \label{AEq:Transient:Shear:2SEvolution} 
\end{align}
with $\tilde{\mathbb{T}}(0) =\hT$ and $\tilde{\mathbb{S}}(0) =\hs$. Then, having subtracted Eq. (\ref{AEq:Transient:Shear:2SEvolution})  multiplied by $\tilde{\mathbb{T}}$ from Eq.  (\ref{AEq:Transient:Shear:1TEvolution}) multiplied by $\tilde{\mathbb{S}}$ and divided the result by $\tilde{\mathbb{S}}^2$, one obtains an ordinary differential equation,
\begin{align}
\der{V(\bar{t})}{\bar{t}} &= 2\Lambda - \hT V(\bar{t}) + (\hs+\chi \Lambda) V^2(\bar{t}), \label{AEq:Transient:Shear:V-ODE} \\
V(0) &= \hT/\hs,
\end{align}
for the function 
\begin{equation}
\label{AEq:Transient:Shear:V-Definition}
V(\bar{t})=\dfrac{\tilde{\mathbb{T}}(\bar{t})}{\tilde{\mathbb{S}}(\bar{t})}.
\end{equation} The variables in Eq. (\ref{AEq:Transient:Shear:V-ODE}) are separated by rewriting the equation as
\begin{equation}
\dfrac{\mathrm{d}V(\bar{t})}{\left[V(\bar{t})-\dfrac{\hT}{2(\hs+\chi \Lambda)} \right]^2+\dfrac{\omega^2}{(\hs+\chi \Lambda)^2}}=(\hs+\chi \Lambda)\mathrm{d}\bar{t},
\label{AEq:Transient:Shear:V-Integrable}
\end{equation}
where
\begin{equation}
\label{AEq:Transient:Shear:omega}
\omega=\dfrac{1}{2}\sqrt{8\Lambda(\hs+\chi\Lambda)-\hT^2}=\dfrac{1}{2}\sqrt{8\chi\Lambda^2+4\hT+3\hT^2}
\end{equation}
is a real quantity. Then, Eq. (\ref{AEq:Transient:Shear:V-Integrable}) is integrated directly; the result of integration can be written as
\begin{equation}
V(\bar{t}) =\dfrac{1}{2(\hs+\chi \Lambda)}\left[\hT+2\omega\dfrac{\sin \omega \bar{t}- \dfrac{a}{\omega} \cos \omega \bar{t}  }{\cos \omega \bar{t} +\dfrac{a}{\omega} \sin \omega \bar{t} }\right], \label{AEq:Transient:Shear:V-Solution}
\end{equation}
where 
\begin{equation}
a = -\dfrac{(\hs+2\chi \Lambda)\hT}{2 \hs}=-\dfrac{4\chi \Lambda^2+\hT+\hT^2}{2(1+\hT)}.
\label{AEq:Transient:Shear:aFirst}
\end{equation}
\par 
Then, $\tilde{\mathbb{T}}(\bar{t})$ is eliminated from Eq. (\ref{AEq:Transient:Shear:2SEvolution}) using Eq. (\ref{AEq:Transient:Shear:V-Definition}). The result is
\begin{align}
\der{\tilde{\mathbb{S}}(\bar{t})}{\bar{t}} &= - \left[1 + \hT+(\hs+\chi \Lambda) V(\bar{t})\right]\tilde{\mathbb{S}}(\bar{t}) + V(\bar{t}) \tilde{\mathbb{S}}^2(\bar{t}), \label{AEq:Transient:Shear:Bernoulli} 
\end{align}
with $\tilde{\mathbb{S}}(0) =\hs$, where $V(\bar{t})$ is now a known function specified by Eq. (\ref{AEq:Transient:Shear:V-Solution}). Equation (\ref{AEq:Transient:Shear:Bernoulli}) is a Bernoulli differential equation, which can be solved by standard methods.\cite{Ince2006} The solution can be written as
\begin{equation}
\dfrac{\tilde{\mathbb{S}}(\bar{t})}{\hs} = \dfrac{K \left[\cos \omega \bar{t} +\dfrac{a}{\omega}\sin \omega \bar{t}\right]}{C e^{\Omega \bar{t}}+A \cos \omega \bar{t} + \dfrac{B}{\omega} \sin \omega \bar{t}}, \label{AEq:Transient:Shear:TildeS}
\end{equation}
where the coefficients $K$, $A$, $B$, $C$ and the factor $\Omega$ are the functions in the first column of Table \ref{Tab:FormFunctions}.
\par 
Substituting Eqs. (\ref{AEq:Transient:Shear:V-Solution}) and (\ref{AEq:Transient:Shear:TildeS}) into Eq. (\ref{AEq:Transient:Shear:V-Definition}) and performing the multiplication, one obtains
\begin{equation}
\dfrac{\tilde{\mathbb{T}}(\bar{t})}{\hT} = \dfrac{K \left[\cos \omega \bar{t} + \dfrac{2+\hT}{2\omega}\sin \omega \bar{t}\right]}{C e^{\Omega \bar{t}}+A \cos \omega \bar{t} + \dfrac{B}{\omega} \sin \omega \bar{t}}. \label{AEq:Transient:Shear:TildeT}
\end{equation}
Note that it is identical in form to Eq. (\ref{AEq:Transient:Shear:TildeS}) but differs by the factor in front of $\sin \omega \bar{t}$ in the numerator. We denote this factor $a$ and allow it to be defined uniquely for each material function (see Table \ref{Tab:a}).
\par 
After going back to the original variables, ($\hT^+,\hnen^+$), one observes that the stress growth functions are expressed by the trigonometric form $\mathfrak{T}$ (see Table \ref{Tab:Forms}) with the coefficients provided in the first columns of Tables \ref{Tab:FormFunctions} and \ref{Tab:a}.
\subsection{
\label{App:Transient:PlaEx}
Startup of steady planar extensional flow
}
Having rewritten Eqs. (\ref{Eq:Eqs:PlaEx:1TEvolution}) and (\ref{Eq:Eqs:PlaEx:2N1Evolution}) in terms of the deviations of the hatted variables from their steady flow values,
\begin{align}
\tilde{\mathbb{T}}(\bar{t}) &= \hT-\hT^+(\bar{t}), \\
\tilde{\mathbb{N}}_1(\bar{t})   &= \hnen - \hnen^+(\bar{t}), \\
\tilde{\mathbb{N}}_2(\bar{t})   &= \hnto - \hnto^+(\bar{t}),
\end{align}
one obtains
\begin{align}
\der{\tilde{\mathbb{T}}(\bar{t})}{\bar{t}} &= - (1+2\hT)\tilde{\mathbb{T}}(\bar{t}) + 2\Lambda \tilde{\mathbb{N}}_1(\bar{t}) + \tilde{\mathbb{T}}^2(\bar{t}), \label{AEq:Transient:PlaEx:1TEvolution}\\
\der{\tilde{\mathbb{N}}_1(\bar{t})}{\bar{t}} &= - (1+\hT) \tilde{\mathbb{N}}_1(\bar{t}) +(2\Lambda-\hnen) \tilde{\mathbb{T}}(\bar{t}) + \tilde{\mathbb{N}}_1(\bar{t}) \tilde{\mathbb{T}}(\bar{t}), \label{AEq:Transient:PlaEx:2N1Evolution} \end{align}
with $\tilde{\mathbb{T}}(0) =\hT$ and $\tilde{\mathbb{N}}_1(0) =\hnen$, while the decoupled Eq. (\ref{Eq:Eqs:PlaEx:N2}) becomes simply
\begin{equation}
\tilde{\mathbb{N}}_2(\bar{t}) = \dfrac{1}{2} \left[ \tilde{\mathbb{N}}_1 (\bar{t}) +\tilde{\mathbb{T}}(\bar{t}) \right].
\label{AEq:Transient:PlaEx:3N2Evolution}
\end{equation}
Proceeding as described in Sec. \ref{App:Transient:Shear}, one constructs the evolution equation for
\begin{equation}
V(\bar{t})=\dfrac{\tilde{\mathbb{T}}(\bar{t})}{\tilde{\mathbb{N}}_1(\bar{t})},
\label{AEq:Transient:PlaEx:V-Definition}
\end{equation}
arriving at
\begin{equation}
\label{AEq:Transient:PlaEx:V-ODE}
\der{V(\bar{t})}{\bar{t}} = 2\Lambda - \hT V(\bar{t}) +  (\hnen-2\Lambda)V^2(\bar{t}), 
\end{equation}
with $V(0) = \hT/\hnen$. 
\par 
Then, one defines 
\begin{equation}
\label{AEq:Transient:PlaEx:Delta}
\Delta =-16\Lambda^2+8\Lambda \hnen -\hT^2 = -16\Lambda^2+4\hT+3\hT^2.
\end{equation}
One now considers three possible cases: $\Delta>0$, $\Delta<0$, and $\Delta=0$. First, let $\Delta>0$. 
Then, Eq. (\ref{AEq:Transient:PlaEx:V-ODE}) can be written as
\begin{equation}
\dfrac{\mathrm{d}V(\bar{t})}{\left[V(\bar{t})-\dfrac{\hT}{2(\hnen-2\Lambda)} \right]^2+\dfrac{\omega^2}{(\hnen-2 \Lambda)^2}}=(\hnen-2\Lambda)\mathrm{d}\bar{t},
\label{AEq:Transient:PlaEx:V-Integrable-T}
\end{equation}
where
\begin{equation}
\label{AEq:Transient:PlaEx:omega-Real}
\omega = \dfrac{1}{2}\sqrt{\Delta}
\end{equation}
is real. After integration, Eq. (\ref{AEq:Transient:PlaEx:V-Integrable-T}) yields
\begin{equation}
\label{AEq:Transient:PlaEx:V-Solution-T}
V(\bar{t})  = \dfrac{1}{2(\hnen-2\Lambda)} \left[\hT + 2\omega \dfrac{\sin \omega \bar{t} - \dfrac{a}{\omega} \cos \omega \bar{t}}{\cos \omega \bar{t} + \dfrac{a}{\omega} \sin \omega \bar{t}} \right],
\end{equation}
with
\begin{equation}
a = \dfrac{8\Lambda^2-\hT-\hT^2}{2(1+\hT)},
\end{equation}
which is similar in form to Eq. (\ref{AEq:Transient:Shear:V-Solution}). Having eliminated $\tilde{\mathbb{T}}$ from Eq. (\ref{AEq:Transient:PlaEx:2N1Evolution}) using Eq. (\ref{AEq:Transient:PlaEx:V-Definition}), 
one obtains a Bernoulli equation for $\tilde{\mathbb{N}}_1$,
\begin{equation}
\label{AEq:Transient:PlaEx:Bernoulli}
\der{\tilde{\mathbb{N}}_1(\bar{t})}{\bar{t}} = - \left[ 1+\hT + (\hnen-2\Lambda) V(\bar{t})\right] \tilde{\mathbb{N}}_1(\bar{t}) + V(\bar{t})\tilde{\mathbb{N}}_1^2(\bar{t}),
\end{equation}
with $\tilde{\mathbb{N}}_1(0)=\hnen$. The solution is
\begin{equation}
\label{AEq:Transient:PlaEx:SolutionTildeN1-T}
\dfrac{\tilde{\mathbb{N}}_1(\bar{t})}{\hnen} = \dfrac{K \left(\cos \omega \bar{t} +\dfrac{a}{\omega}\sin \omega \bar{t} \right)}{C  e^{\Omega \bar{t}}+A \cos \omega \bar{t} + \dfrac{B}{\omega} \sin \omega \bar{t}},
\end{equation}
where
$K$, $A$, $B$, $C$, and $\Omega$ are the functions found in the second column of Table \ref{Tab:FormFunctions}. Having used Eqs. (\ref{AEq:Transient:PlaEx:V-Definition}) and (\ref{AEq:Transient:PlaEx:3N2Evolution}) and reverted to the original variables, one obtains the trigonometric form $\mathfrak{T}$ (see Table \ref{Tab:Forms}) of $\hnen^+(\bar{t})$, $\hT^+(\bar{t})$, and $\hnto^+(\bar{t})$, the expressions for $a$ being provided in the second column of Table \ref{Tab:a}.
\par 
Now, let $\Delta<0$. Then, Eqs. (\ref{AEq:Transient:PlaEx:V-Integrable-T})--(\ref{AEq:Transient:PlaEx:SolutionTildeN1-T}) still hold but contain complex functions, since the “frequency” $\omega$, defined by Eq. (\ref{AEq:Transient:PlaEx:omega-Real}), is purely imaginary. Yet, having written $\omega = i \omega'$, where 
\begin{equation}
\omega' = \dfrac{1}{2}\sqrt{-\Delta},
\end{equation}
and applied the identities
\begin{align}
\cos i\omega' \bar{t} &= \cosh \omega' \bar{t}, \\
\sin i \omega' \bar{t} &= i \sinh \omega' \bar{t}, 
\end{align}
one obtains an alternative form of Eq. (\ref{AEq:Transient:PlaEx:SolutionTildeN1-T}), which is purely real at $\Delta<0$,
\begin{equation}
\label{AEq:Transient:PlaEx:SolutionTildeN1-H}
\dfrac{\tilde{\mathbb{N}}_1(\bar{t})}{\hnen} = \dfrac{K \left(\cosh \omega' \bar{t} +\dfrac{a}{\omega'}\sinh \omega' \bar{t}\right)}{C  e^{\Omega \bar{t}}+A \cosh \omega' \bar{t} + \dfrac{B}{\omega'} \sinh \omega' \bar{t}},
\end{equation}
where $K$, $A$, $B$, $C$, $\Omega$, and $a$ are identical to those in Eq. (\ref{AEq:Transient:PlaEx:SolutionTildeN1-T}). Equation (\ref{AEq:Transient:PlaEx:SolutionTildeN1-H}) leads to the hyperbolic form $\mathfrak{H}$ (see Table \ref{Tab:Forms}) of $\hnen^+(\bar{t})$, $\hT^+(\bar{t})$, and $\hnto^+(\bar{t})$. Note also that the definitions of $\omega$ and $\omega'$ are readily unified by redefining 
\begin{equation}
\omega = \dfrac{1}{2}\sqrt{\vert \Delta \vert}.
\end{equation}
\par 
Finally, in the case when $\Delta = 0$, the solution of Eq. (\ref{AEq:Transient:PlaEx:Bernoulli}) can be obtained by considering the limit $\omega \to 0$ at any fixed $\bar{t}$ in Eq. (\ref{AEq:Transient:PlaEx:SolutionTildeN1-T}). The result is 
\begin{equation}
\label{AEq:Transient:PlaEx:SolutionTildeN1-EPlus}
\dfrac{\tilde{\mathbb{N}}_1(\bar{t})}{\hnen} = \dfrac{K (1 +a\bar{t} )}{C  e^{\Omega \bar{t}}+A + B \bar{t}}.
\end{equation}
Equation (\ref{AEq:Transient:PlaEx:SolutionTildeN1-EPlus}) leads to the exponential form $\mathfrak{E}^+$ (see Table \ref{Tab:Forms}) of $\hnen^+(\bar{t})$, $\hT^+(\bar{t})$, and $\hnto^+(\bar{t})$.
\subsection{
\label{App:Transient:UBEx}
Startup of steady uniaxial and biaxial extensional flows}
Equations (\ref{Eq:Eqs:UBEx:1TEvolution}) and (\ref{Eq:Eqs:UBEx:2N1Evolution}), rewritten in terms of 
 \begin{align}
\tilde{\mathbb{T}}(\bar{t}) &= \hT-\hT^+(\bar{t}), \\
\tilde{\mathbb{N}}_1(\bar{t})   &= \hnen - \hnen^+(\bar{t}),
\end{align}
become
\begin{align}
\der{\tilde{\mathbb{T}}(\bar{t})}{\bar{t}} &= - (1+2\hT)\tilde{\mathbb{T}}(\bar{t}) + 2\Lambda \tilde{\mathbb{N}}_1(\bar{t}) + \tilde{\mathbb{T}}^2(\bar{t}), \label{AEq:Transient:UBEx:1TEvolution}\\
\der{\tilde{\mathbb{N}}_1(\bar{t})}{\bar{t}} &= - (1\mp \Lambda +\hT) \tilde{\mathbb{N}}_1(\bar{t}) + (\Lambda-\hnen) \tilde{\mathbb{T}}(\bar{t}) + \tilde{\mathbb{N}}_1(\bar{t}) \tilde{\mathbb{T}}, \label{AEq:Transient:UBEx:2N1Evolution}
\end{align}
with $\tilde{\mathbb{T}}(0) = \hT$ and $\tilde{\mathbb{N}}_1(0) = \hnen$.
The analog of Eq.  (\ref{AEq:Transient:PlaEx:V-ODE}) is
\begin{equation}
\label{AEq:Transient:UBEx:V-ODE}
\der{V(\bar{t})}{\bar{t}} = 2 \Lambda -(\hT \pm \Lambda)V(\bar{t}) + (\hnen-\Lambda)V^2(\bar{t}),
\end{equation}
with $V(0) = \hT/\hnen$, where
\begin{equation}
\label{AEq:Transient:UBEx:V-Definition}
V(\bar{t}) = \dfrac{\tilde{\mathbb{T}}(\bar{t})}{\tilde{\mathbb{N}}_1(\bar{t})}.
\end{equation}
Equation (\ref{AEq:Transient:UBEx:V-ODE}) can be rewritten as
\begin{equation}
\dfrac{\mathrm{d}V(\bar{t})}{\left[V(\bar{t})-\dfrac{\hT\pm \Lambda}{2(\hnen-\Lambda)} \right]^2+\dfrac{\Delta}{4(\hnen-2 \Lambda)^2}}=(\hnen-\Lambda)\mathrm{d}\bar{t},
\label{AEq:Transient:UBEx:V-Integrable-T}
\end{equation}
where
\begin{equation}
\label{AEq:Transient:UBEx:Delta}
\Delta = -9\Lambda^2\mp 2\Lambda \hT + 8 \Lambda \hnen - \hT^2 = -9\Lambda^2+(4\mp2\Lambda)\hT+3\hT^2
\end{equation}
can be positive, negative, or zero, depending on $\alpha$ and $\Lambda$. One therefore defines
\begin{equation}
\omega = \dfrac{1}{2}\sqrt{\vert \Delta \vert}.
\end{equation}
At $\Delta>0$, the result of integration of Eq. (\ref{AEq:Transient:UBEx:V-Integrable-T}) is
\begin{equation}
\label{AEq:Transient:UBEx:V-Solution-T}
V(\bar{t}) = \dfrac{1}{2(\hnen-\Lambda)}\left[\hT \pm \Lambda + 2\omega \dfrac{\sin \omega \bar{t} - \dfrac{a}{\omega} \cos \omega \bar{t}}{\cos \omega \bar{t} + \dfrac{a}{\omega} \sin \omega \bar{t}} \right],
\end{equation}
where 
\begin{equation}
a = \dfrac{\pm \Lambda +4\Lambda^2-(1\mp\Lambda)\hT-\hT^2}{2(1+\hT)}.
\end{equation}
Eliminating $\tilde{\mathbb{T}}(\bar{t})$ from Eq. (\ref{AEq:Transient:UBEx:2N1Evolution}) using Eq. (\ref{AEq:Transient:UBEx:V-Definition}) leads to a Bernoulli equation for $\tilde{\mathbb{N}}_1(\bar{t})$,
\begin{equation}
\der{\tilde{\mathbb{N}}_1(\bar{t})}{\bar{t}} = - \left[ 1\mp \Lambda+\hT + (\hnen-\Lambda) V(\bar{t})\right] \tilde{\mathbb{N}}_1(\bar{t}) + V(\bar{t})\tilde{\mathbb{N}}_1^2(\bar{t}),
\end{equation}
with $\tilde{\mathbb{N}}_1(0) = \hnen$ and $V(\bar{t})$ given by Eq. (\ref{AEq:Transient:UBEx:V-Solution-T}). Proceeding as in Sec. \ref{App:Transient:PlaEx}, one obtains the three forms $\mathfrak{T}$, $\mathfrak{H}$, and $\mathfrak{E}^+$ of the normalized material functions, with $K$, $A$, $B$, $C$, and $\Omega$ from the third column of Table \ref{Tab:FormFunctions} and $a$ from the third column of Table \ref{Tab:a}.
\section{Approximate expressions for the extrema of the transient material functions in oscillatory regime \label{App:Truncation}}
The extrema of the original form $\mathfrak{T}$ need to be determined from transcendental algebraic equations unsolvable by analytical methods. However, at $A=B=0$, the trigonometric form $\mathfrak{T}$ reduces to a simpler expression,
\begin{equation} 
\label{AEq:Truncation:FormT0}
1- \dfrac{K\left(\cos \omega \bar{t}+\dfrac{a}{\omega}\sin \omega \bar{t}\right)}{C \exp \Omega \bar{t}},
\end{equation}
which describes exponentially damped harmonic oscillations.
\par 
It is seen that the term $C \exp \Omega \bar{t}$ in the denominator of $\mathfrak{T}$ grows exponentially in time, while the sum of the two other terms is a bounded function of time,
\begin{equation}
\left\vert A \cos\omega \bar{t} +\dfrac{B}{\omega}\sin\omega \bar{t} \right \vert \leq \dfrac{\sqrt{A^2\omega^2+B^2}}{\omega}.
\end{equation}
Therefore, for any $(\chi,\Lambda)$ or $(\alpha,\Lambda)$, there exists a time point $\tilde{t}$ such that for all $\bar{t}\geq \tilde{t}$, 
\begin{equation}
\dfrac{\sqrt{A^2\omega^2+B^2}}{\omega}<<C \exp \Omega \bar{t}.
\end{equation}
\par 
Since the first overshoot occurs approximately at $\omega \bar{t} = \pi$ (this corresponds to half of the period of the oscillations), we shall consider $\tilde{t}=\pi/\omega$ and introduce the “truncation index,”
\begin{equation}
I_\mathrm{tr} = \dfrac{\sqrt{A^2\omega^2+B^2}}{C\omega \exp \dfrac{\pi \Omega}{\omega}}.
\label{AEq:Truncation:TruncationIndex}
\end{equation}
As $I_\mathrm{tr} \to 0$, the extrema of the form $\mathfrak{T}$ asymptotically approach those of the function in Eq. (\ref{AEq:Truncation:FormT0}).
\begin{figure}[htp]
\begin{center}
\includegraphics[width=6.45in]{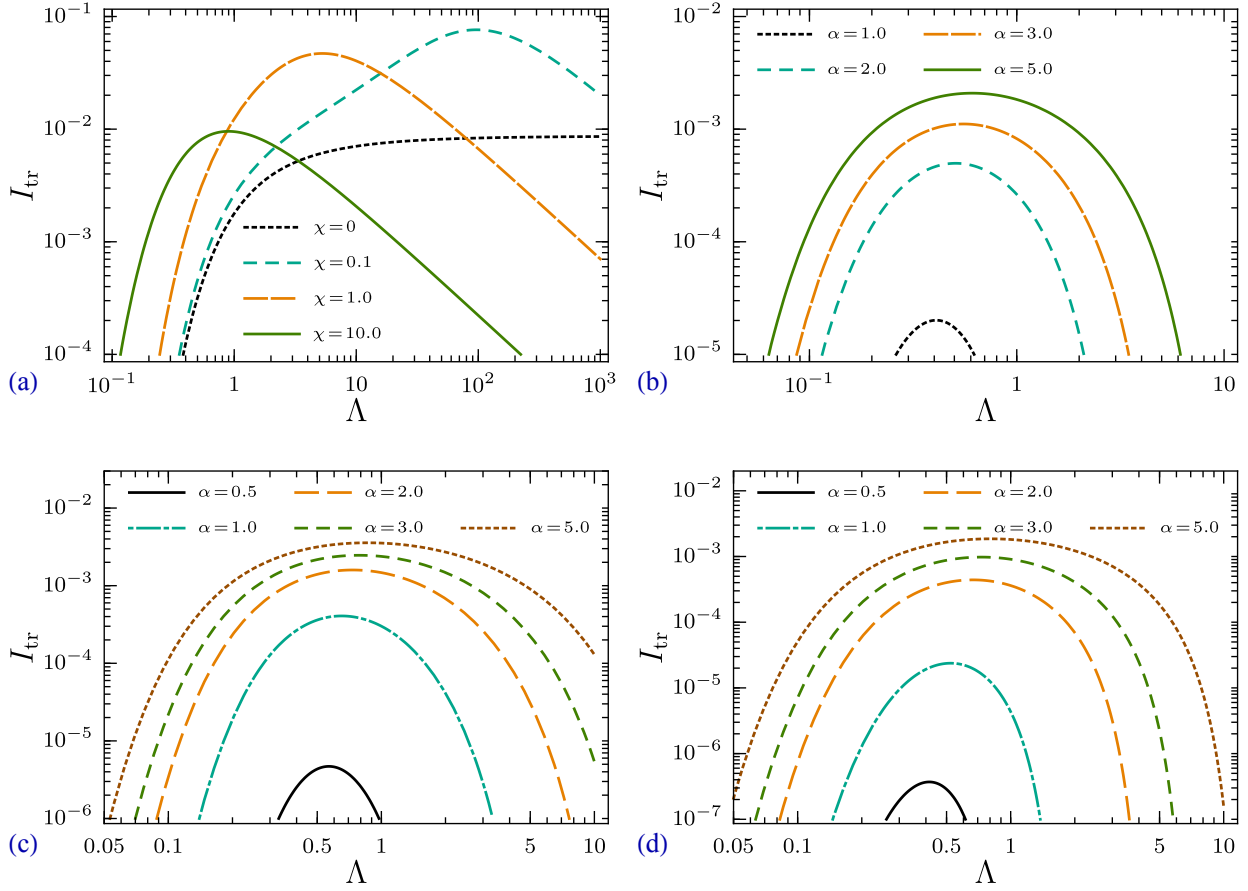}
\caption{\label{Fig:App:Truncation} The truncation index, $I_\mathrm{tr}$, defined by Eq. (\ref{AEq:Truncation:TruncationIndex}), for startup of shear flow (a), of planar extension (b), of uniaxial extension (c), and of biaxial extension (d), plotted as a function of the dimensionless strain rate, $\Lambda$. For startup of extensional flows, $I_\mathrm{tr}$ is evaluated only at those $\alpha$ and $\Lambda$ for which the material functions are oscillatory.}
\end{center}
\end{figure}
\par 
The truncation index for the stress growth functions related to the four startup flows considered in this work is shown in Fig. \ref{Fig:App:Truncation}. It is seen that in most situations, $I_\mathrm{tr}<10^{-2}$, the only exception being startup of steady shear flow at relatively large $\Lambda$ combined with very small $\chi$; in this case, $I_\mathrm{tr}\sim 10^{-1}$. Therefore, the overshoots and undershoots of the oscillatory stress growth functions can be accurately  approximated by the extrema of the function in Eq. (\ref{AEq:Truncation:FormT0}). The derivative of this function vanishes at 
\begin{equation}
\label{AEq:Truncation:EquationForTheta}
\tan \omega \bar{t} = \dfrac{(a-\Omega)\omega}{a\Omega+\omega^2}.
\end{equation}
Having defined $\theta$ by 
\begin{equation}
\theta =
\left\{ 
\renewcommand{\arraystretch}{1.5} \begin{array}{ll}
\pi + \arctan\dfrac{(a-\Omega)\omega}{a\Omega+\omega^2}  & \quad \text{if} \quad \dfrac{a-\Omega}{a\Omega+\omega^2} \leq 0, \\
\arctan\dfrac{(a-\Omega)\omega}{a\Omega+\omega^2} & \quad \text{if} \quad \dfrac{a-\Omega}{a\Omega+\omega^2} > 0, \\
\dfrac{\pi}{2} & \quad \text{if} \quad \ a\Omega+\omega^2=0,
\end{array} \right.
\label{AEq:Truncation:DefinitionOfTheta}
\end{equation}
 so that $\bar{t}=\theta/\omega$ is the smallest positive solution of Eq. (\ref{AEq:Truncation:EquationForTheta}), one immediately obtains Eqs. (\ref{Eq:Transient:Shear:OvershootPositions}) and (\ref{Eq:Transient:Shear:UndershootPositions}). 
\par 
For particular material functions, Eq. (\ref{AEq:Truncation:DefinitionOfTheta}) can be simplified. For instance, for the shear stress growth function, $\eta^+(\bar{t})/\eta$,
\begin{equation}
\dfrac{a-\Omega}{a\Omega+\omega^2} = -\dfrac{2(1+2\chi \Lambda^2+3\hT+2\hT^2)}{\hT(1-2\chi \Lambda +\hT)}=\dfrac{2(1-\chi \hT)(1+2\chi\Lambda^2+3\hT+2\hT^2)}{\hT(1+\hT)(-1+2\hT+\chi\hT^2)},
\label{AEq:Truncation:ThetaForShearGrowth}
\end{equation}
where we have used Eq. (\ref{AEq:Bijectivity:Shear:LambdaOfHatT}) to eliminate $\Lambda$. The right-hand side of Eq. (\ref{AEq:Truncation:ThetaForShearGrowth}) changes its sign from “$-$” to “$+$” at
\begin{equation}
\label{AEq:Truncation:Theta:Shear:Condition}
\hT = -1+\sqrt{1+1/\chi}.
\end{equation}
Using Eq. (\ref{AEq:Bijectivity:Shear:LambdaOfHatT}) once again to convert $\hT$ into $\Lambda$, one shows that condition (\ref{AEq:Truncation:Theta:Shear:Condition}) is equivalent to
\begin{equation}
\Lambda = \sqrt[4]{\dfrac{1+\chi}{4\chi^3}},
\end{equation}
which leads to Eq. (\ref{Eq:Transient:Shear:Theta:Shear}).
In contrast, for $\Psi_1^+(\bar{t})/\Psi_1$,
\begin{equation}
\dfrac{a-\Omega}{a\Omega+\omega^2} = -\dfrac{2\hT}{2+4\chi \Lambda^2 +6\hT +3\hT^2}<0,
\end{equation}
so that Eq. (\ref{AEq:Truncation:DefinitionOfTheta}) simplifies to Eq. (\ref{Eq:Transient:Shear:Theta:Normal}). The corresponding expressions related to startup of extensional flows are obtained in a similar way.
\section{The approximate expression for the second critical shear rate \label{App:LambdaII}}
Based on the numerical calculations shown in Fig. \ref{Fig:Transient:Shear:CriticalShearRates}, $\Lambda_{II}$ significantly exceeds $\Lambda_I$. Therefore, we assume $\Lambda  >> \Lambda_I$, which implies
\begin{equation}
\chi \Lambda^2  >>\left\{ \dfrac{1}{\chi^2},\dfrac{1}{\chi}, 1 \right\}.
\end{equation}
Under this assumption, 
\begin{equation}
\omega \bar{t}_{\mathrm{min}}^{\;[1]} \approx \pi + \arctan{(2\chi)^{3/2}\Lambda} \approx 3\pi/2,
\end{equation}
so that $\cos \omega\bar{t}_{\mathrm{min}}^{\;[1]} \approx 0$ and $\sin \omega\bar{t}_{\mathrm{min}}^{\;[1]} \approx -1$.
Then, for the first shear stress undershoot,
\begin{equation}
\dfrac{\hs^+(\bar{t}_{\mathrm{min}}^{\;[1]})}{\hs} \approx 1-\dfrac{\sqrt{2\chi^3}}{1+\chi}\Lambda \exp \left[-\dfrac{3(3+2\chi)\pi}{4\sqrt{2\chi^3}\Lambda} \right],
\end{equation}
which leads to an equation for $\Lambda_{II}$,
\begin{equation}
\dfrac{\sqrt{2\chi^3}}{1+\chi}\Lambda_{II} \exp \left[-\dfrac{3(3+2\chi)\pi}{4\sqrt{2\chi^3}\Lambda_{II}} \right]=1,
\end{equation}
the solution of which is expressed using the principal branch of the Lambert $W$ function [see Eq. (\ref{Eq:Transient:Shear:LambdaII})].

\section{Different forms of the stress growth functions describing startup of extensional flows
\label{App:FormShift}}
In this section, one shall derive the results presented in Tables \ref{Tab:FormsPlaEx} and \ref{Tab:FormsUBEx}, that is, the conditions at which the material functions in startup of steady extensional flows take the forms $\mathfrak{H}$, $\mathfrak{E}^+$, and $\mathfrak{T}$. 
\par 
As shown in Secs. \ref{App:Transient:PlaEx} and \ref{App:Transient:UBEx}, the form of the material functions is dictated by the sign of $\Delta$ [see Eqs. (\ref{AEq:Transient:PlaEx:Delta}) and (\ref{AEq:Transient:UBEx:Delta})]: The material functions  take the trigonometric form $\mathfrak{T}$ when $\Delta>0$, the hyperbolic form $\mathfrak{H}$ when $\Delta<0$, and the exponential form $\mathfrak{E}^+$ when $\Delta = 0$.
\par 
One shall start by formulating the conditions at which $\Delta=0$. These conditions specify a curve on the $(\alpha,\Lambda)$-plane on which the form $\mathfrak{E}^+$ applies. In each of the regions this curve divides the plane into, $\Delta$ has a constant sign, which can be found by a direct check.
\subsection{Startup of planar extensional flow
\label{App:FormShift:PlaEx}}
Having eliminated $\Lambda$ from Eq. (\ref{AEq:Transient:PlaEx:Delta}) using Eq. (\ref{AEq:Bijectivity:PlaEx:LambdaOfHatT}) and set the resulting expression equal zero, one obtains a quadratic equation for $\hT$,
\begin{equation}
\label{AEq:FormShift:PlaEx:ZeroDeltaEqn}
\hT^2 + 2 (2-3\alpha)\hT+4(1-2\alpha) =0.
\end{equation}
At $\alpha \leq 1/2$, this equation has no real positive roots. At $\alpha > 1/2$, Eq. (\ref{AEq:FormShift:PlaEx:ZeroDeltaEqn}) has two real roots, only one of which, expressed by Eq. (\ref{Eq:Transient:PlaEx:TCrit}), is positive. Substituting Eq. (\ref{Eq:Transient:PlaEx:TCrit}) into Eq. (\ref{AEq:Bijectivity:PlaEx:LambdaOfHatT}) yields Eq. (\ref{Eq:Transient:PlaEx:LambdaCrit}).
\subsection{Startup of uniaxial and biaxial extensional flows
\label{App:FormShift:UBEx}}
Eliminating $\Lambda$ from Eq. (\ref{AEq:Transient:UBEx:Delta}) using Eqs. (\ref{AEq:Bijectivity:UBEx:LambdaOfHatT}) and (\ref{AEq:Bijectivity:UBEx:Y}) leads after some algebraic transformations to a quartic equation for $\hT$ of form
\begin{equation}
\label{AEq:FormShift:UBEx:Quartic}
\hT^4 + b_3 \hT^3 + b_2 \hT^2 + b_1 \hT + b_0 =0,
\end{equation}
where one needs to look after the real positive roots satisfying $R(\alpha, \hT) \geq 0$ for uniaxial extension and $R(\alpha, \hT)\leq 0$ for biaxial extension. The coefficients of the quartic Eq. (\ref{AEq:FormShift:UBEx:Quartic}) are given by
\begin{align}
b_0 &= \dfrac{9}{4}(3-8\alpha)^2, \\
b_1 &= \dfrac{3}{2}(27-130\alpha+144\alpha^2), \\
b_2 &= \dfrac{9}{4}(13-52\alpha+36\alpha^2),\\
b_3 &= 3(3-8\alpha), \label{AEq:FormShift:UBEx:b3}
\end{align}
while 
\begin{equation}
\label{AEq:FormShift:UBEx:R}
R(\alpha, \hT) = 5\hT^2 + 18(1-2\alpha)\hT+6(3-8\alpha).
\end{equation}
Then, Eq. (\ref{AEq:FormShift:UBEx:Quartic}) is solved using the method described by Auckley.\cite{Auckley2007}
\par 
If one defines $p$, $q$, and $r$ by
\begin{align}
p &= b_2-\dfrac{3}{8}b_3^2 = -\dfrac{9}{8}(1-40\alpha+120\alpha^2), \label{AEq:FormShift:UBEx:p}\\
q &= b_1-\dfrac{1}{2}b_2 b_3 = -\dfrac{3}{2}\alpha(1-3\alpha)(31-168\alpha), \label{AEq:FormShift:UBEx:q}\\
r &= b_0-\dfrac{1}{4}b_1 b_3+\dfrac{1}{16}b_2 b_3^2-\dfrac{3}{256}b_3^4 \nonumber \\
&= \dfrac{27}{256}(3-8\alpha)(1+40\alpha-432\alpha^2+1152\alpha^3),
\end{align}
then the four solutions of Eq. (\ref{AEq:FormShift:UBEx:Quartic}) are given by\cite{Auckley2007}
\begin{align}
\label{AEq:FormShift:UBEx:QuarticSolution12}
\hT^\ast_{1,2} &=-\dfrac{b_3}{4}+
\dfrac{1}{2}\left[ \sqrt{l} \mp \sqrt{-l-2\left(p+\dfrac{q}{\sqrt{l}}\right)} \right], \\
\label{AEq:FormShift:UBEx:QuarticSolution34}
\hT^\ast_{3,4} &=-\dfrac{b_3}{4}+
\dfrac{1}{2}\left[- \sqrt{l} \mp \sqrt{-l-2\left(p-\dfrac{q}{\sqrt{l}}\right)} \right],
\end{align}
where $l$ is an arbitrary nonzero solution of the resolvent cubic equation of form (\ref{AEq:Cubics:Cubic}), with the coefficients
\begin{align}
a_0 &= -q^2, \\
a_1 &= p^2-4r, \\
a_2 &= 2p.
\end{align}
The discriminant of the quartic Eq. (\ref{AEq:FormShift:UBEx:Quartic}), which can be written as
\begin{equation}
D = 2^{13}3^7 \alpha^2 (1-3 \alpha)^3 (1 + 12 \alpha)^2 (3-18 \alpha + 28 \alpha^2),
\end{equation}
changes its sign from “$+$” to “$-$” at $\alpha=1/3$. Therefore, at $\alpha<1/3$, the four roots of Eq. (\ref{AEq:FormShift:UBEx:Quartic}) are either all real or all nonreal; at $\alpha>1/3$, two of the roots are real, while the other two are a couple of nonreal complex conjugates.
\par 
One observes that at $\alpha = 31/168$, which belongs to the half-interval $(-\infty,1/3)$, $q=a_0=0$, and $l=0$ is one of the roots of the resolvent cubic, which takes the form
\begin{equation}
l\left(l^2+\dfrac{8097}{1568}l+\dfrac{174375}{28672} \right)=0.
\label{AEq:FormShift:UBEx:ResolventPartialCase}
\end{equation}
The two nonzero roots of Eq. (\ref{AEq:FormShift:UBEx:ResolventPartialCase}) are
\begin{equation}
l = -\dfrac{3(2699\pm 128\sqrt{39})}{3136}.
\label{AEq:FormShift:UBEx:ResolventPartialSolution}
\end{equation}
Calculating the roots of Eq. (\ref{AEq:FormShift:UBEx:Quartic}) with Eqs. (\ref{AEq:FormShift:UBEx:QuarticSolution12}) and (\ref{AEq:FormShift:UBEx:QuarticSolution34}) using $\alpha=31/168$ and any value of $l$ from Eq. (\ref{AEq:FormShift:UBEx:ResolventPartialSolution}) yields four nonreal numbers. Therefore, at $\alpha<1/3$, all the roots of Eq. (\ref{AEq:FormShift:UBEx:Quartic}) are nonreal. It also means that at $\alpha=1/3$, two of the roots are nonreal, while the other two are real and equal to each other. Thus, only the values $\alpha \geq 1/3$ are of further interest.
\par 
Then, one calculates the discriminant of the resolvent cubic using Eq. (\ref{AEq:Cubics:Discriminant}). The result is
\begin{equation}
 D = 2^1 3^7 (1-3\alpha)^3(1+12\alpha)^2(3-18\alpha+28\alpha^2),
 \end{equation}
which is negative at $\alpha>1/3$. Thus, the resolvent cubic has one real root at $\alpha>1/3$, which is positive, since $a_0<0$. This root is given by 
\begin{equation}
\label{AEq:FormShift:UBEx:l}
l = -\dfrac{2p}{3} 
+\sqrt[3]{-Q+\sqrt{Q^2-P^3}}
+\sqrt[3]{-Q-\sqrt{Q^2-P^3}},
\end{equation}
with
\begin{align}
P &= -\dfrac{9}{16}(1+8\alpha+56\alpha^2-672\alpha^3+1296\alpha^4), \\
Q &= -\dfrac{9}{64}(3 + 36 \alpha - 700 \alpha^2 + 960 \alpha^3 + 23184 \alpha^4 - 108864 \alpha^5 + 139968 \alpha^6). \label{AEq:FormShift:UBEx:Q}
\end{align}
Note that Eq. (\ref{AEq:FormShift:UBEx:l}) is also applicable at $\alpha=1/3$ (when $q=0$), still returning a nonzero value $l=9/4$.
\par 
Thus, the four roots of Eq. (\ref{AEq:FormShift:UBEx:Quartic}) at $\alpha \geq 1/3$ are found using Eqs. (\ref{AEq:FormShift:UBEx:QuarticSolution12}) and (\ref{AEq:FormShift:UBEx:QuarticSolution34}), with all the quantities in these equations specified by Eqs. (\ref{AEq:FormShift:UBEx:b3}), (\ref{AEq:FormShift:UBEx:p}), (\ref{AEq:FormShift:UBEx:q}), and (\ref{AEq:FormShift:UBEx:l})--(\ref{AEq:FormShift:UBEx:Q}). Two of these solutions, $\hT^\ast_1$ and $\hT^\ast_2$, are real.
\par
\begin{figure}[htp]
\begin{center}
\includegraphics[width=3.37in]{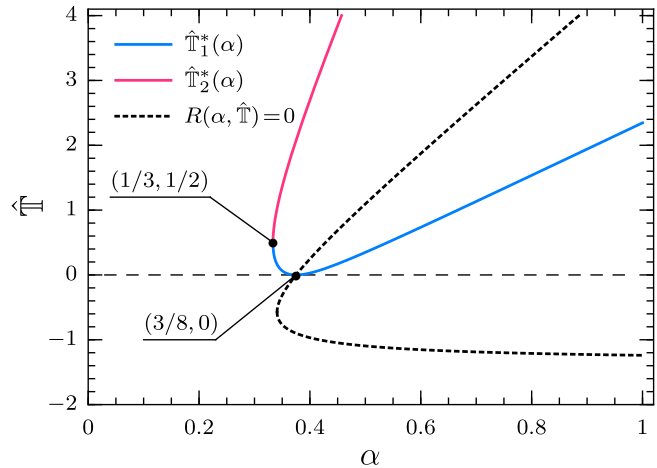}
\caption{\label{Fig:App:FormShift:UBEx} The two real solutions of Eq. (\ref{AEq:FormShift:UBEx:Quartic}) as functions of $\alpha$, plotted together with the curve $R(\alpha,\hT)=0$.}
\end{center}
\end{figure}

 
The real solutions of Eq. (\ref{AEq:FormShift:UBEx:Quartic}) are plotted in Fig. \ref{Fig:App:FormShift:UBEx} as functions of $\alpha$. The curve $R(\alpha,\hT)=0$, intersecting $\hT=\hT_1^\ast(\alpha)$ at $(3/8,0)$, is also shown in the plot. It is easy to see that $R(0,0)>0$ and $R(+\infty,0)<0$; therefore, the finite part of the curve $\hT=\hT_1^\ast(\alpha)$ and the entire curve $\hT=\hT_2^\ast(\alpha)$ satisfy $R(\alpha,\hT)>0$, while the infinite part of the curve $\hT=\hT_1^\ast(\alpha)$ satisfies $R(\alpha,\hT)<0$.
\par 
Finally, $\hT^\ast_1$ and $\hT^\ast_2$ are converted into $\Lambda^\ast_1$ and $\Lambda^\ast_2$, respectively, using Eq. (\ref{AEq:Bijectivity:UBEx:LambdaOfHatT}). This yields Table \ref{Tab:FormsUBEx} and, with a minor change of notations, Eqs. (\ref{Eq:Transient:UBEx:LambdaStarUniaxial})--(\ref{Eq:Transient:UBEx:Q}). 
\end{appendix}

\bibliography{FLPTT}

\end{document}